\documentclass[a4paper,fleqn]{cas-dc}

\usepackage[numbers]{natbib}

\def\tsc#1{\csdef{#1}{\textsc{\lowercase{#1}}\xspace}}
\tsc{WGM}
\tsc{QE}
\tsc{EP}
\tsc{PMS}
\tsc{BEC}
\tsc{DE}

\usepackage {hyperref}
\usepackage{booktabs}
\usepackage{colortbl}
\usepackage{rotating}
\usepackage{multirow}

\usepackage{algorithm}
\usepackage{algpseudocode}
\usepackage{lscape}
\usepackage[caption=false,font=scriptsize,labelfont=sf,textfont=sf]{subfig}
\usepackage{longtable} 
\begin{document}
\let\WriteBookmarks\relax
\def\floatpagepagefraction{1}
\def\textpagefraction{.001}
\shorttitle{IoTGeM: Generalizable Models for Behaviour-Based 
	IoT Attack Detection}
\shortauthors{Kostas et~al.}

\title [mode = title]{IoTGeM: Generalizable Models for Behaviour-Based 
	IoT Attack Detection}                      



\author[1]{Kahraman~Kostas}[type=editor,
                        auid=001,bioid=1,
                        orcid=0000-0002-4696-1857]
\cormark[1]


\affiliation[1]{organization={Ministry of
		National Education},
                city={Anakara},
                country={Türkiye}}

\author[2]{Mike~Just}[orcid=0000-0002-9669-5067]

\author[2]{Michael~A.~Lones}[type=editor,
auid=000,bioid=1,
orcid=0000-0002-2745-9896]


\affiliation[2]{organization={Department of Computer Science, Heriot-Watt University},
                postcode={EH14 4AS}, 
                city={Edinburgh},
                country={UK}}

\cortext[cor1]{Corresponding author}


\begin{abstract}
Previous research on behavior-based attack detection for networks of IoT devices has resulted in machine learning models whose ability to adapt to unseen data is limited and often not demonstrated. This paper presents IoTGeM, an approach for modeling IoT network attacks that focuses on generalizability, yet also leads to better detection and performance. We first introduce an improved rolling window approach for feature extraction. To reduce overfitting, we then apply a multi-step feature selection process where a Genetic Algorithm (GA) is uniquely guided by exogenous feedback from a separate, independent dataset. To prevent common data leaks that have limited previous models, we build and test our models using strictly isolated train and test datasets. The resulting models are rigorously evaluated using a diverse portfolio of machine learning algorithms and datasets. Our window-based models demonstrate superior generalization compared to traditional flow-based models, particularly when tested on unseen datasets. On these stringent, cross-dataset tests, IoTGeM achieves F1 scores of 99\% for ACK, HTTP, SYN, MHD, and PS attacks, as well as a 94\% F1 score for UDP attacks. Finally, we build confidence in the models by using the SHAP (SHapley Additive exPlanations) explainable AI technique, allowing us to identify the specific features that underlie the accurate detection of attacks. 

\end{abstract}


\maketitle

\section{Introduction}
The ever-growing number of IoT (internet of things) devices creates an ever-larger attack surface. Reports suggest that a new IoT device included in a network receives its first attack within five hours and becomes the target of a specific attack within 24 hours~\cite{modi_2019}. 
IoT devices can be particularly challenging to secure as they can vary widely in terms of their hardware, software, and interfaces, and they typically have more limited resources compared to conventional computing devices. Thus, classical security solutions 
often need to be tailored for IoT devices~\cite{zarpelao2017survey}.

Behaviour-based attack detection using machine learning (ML) 
is a common approach to securing IoT devices
by detecting anomalous 
network attacks.
However, 
common mistakes in some ML studies can raise serious doubts about the reliability of 
results~\cite{kapoor2022leakage,lones2021avoid,arp2020and}, with examples such as data leakage and feature overfitting in previous IoT attack detection studies (see Section~\ref{LR}).  
In this paper
we focus on automating the behaviour-based discrimination of benign and malicious network data with ML algorithms
while avoiding such pitfalls. 
To do this, we introduce an approach (\textit{IoTGeM}) for creating generalizable models for behaviour-based network attack detection for IoT devices, with the following 
contributions:
\begin{enumerate}
	\item  \textcolor{black}{A rolling window method for feature extraction that, unlike simple time-series analysis which often focuses only on packet timing and rates, generates a rich set of multi-dimensional statistical features (including packet size, payload entropy, and TCP flag statistics) to capture more complex and subtle attack behaviors.} 
	\item  \textcolor{black}{A novel multi-step feature selection process centered on a genetic algorithm. This GA is uniquely designed to optimize for generalizability by using exogenous feedback, where the fitness of each feature combination is evaluated on an entirely separate and unseen dataset, a key distinction from conventional feature selection methods.} 
	\item \textcolor{black}{A detailed examination of feature effectiveness by applying the SHAP (SHapley Additive exPlanations) explainable AI technique to interpret our trained models, thereby building confidence in their predictions by identifying the most influential features for attack detection.}
\end{enumerate}

We further adopt a strict methodology for ensuring the generalizability of our models, and the ease of validating, replicating, or extending our approach~\cite{lones2021avoid}:
\begin{itemize}
	\item We build and test our models using isolated train and test datasets to  
	avoid
	common data leaks.
	\item We rigorously evaluate our methodology with a diverse portfolio of machine learning models, evaluation metrics and datasets, and we avoid using inappropriate metrics for certain situations, e.g., erroneously using only accuracy with unbalanced data sets.  
	\item 
	We 
	use
	publicly available datasets and 
	make our scripts available to the public\footnote{Materials available at: \href{https://github.com/kahramankostas/IoTGeM}{github.com/kahramankostas/IoTGeM}\label{f1}}. 
\end{itemize}

\textcolor{black}{To achieve these contributions, our research considers a specific threat model and deployment scenario. We assume an attacker is located outside the local network attempting to compromise or disrupt IoT devices within it. Our proposed system, IoTGeM, is designed as a Network-based Intrusion Detection System (NIDS) to be deployed on a central network access point (``chokepoint''), like a gateway or router, where it can monitor all relevant traffic. The effectiveness of our approach is evaluated under increasingly stringent scenarios, moving beyond simple cross-validation to test model performance against data from entirely different sessions and, most importantly, from completely separate and unseen datasets to provide a robust measure of real-world generalizability.} 

This article is organized as follows. Section~\ref{LR} gives an overview of
related work. 
Section~\ref{Methodology}  describes the materials and methods we use for data selection, feature extraction and feature selection. Section~\ref{Performance}  evaluates the results of the models, compares them with other methods, and 
provides a detailed summary analysis relating feature effectiveness to different attacks.
Limitations are discussed in Section~\ref{Limitations}, and conclusions are given in Section~\ref{Conclusions}.

\section{Related Work}\label{LR}

\begin{figure*}[htbp]
	\subfloat[ Datasets ]{{\includegraphics[width=45mm]{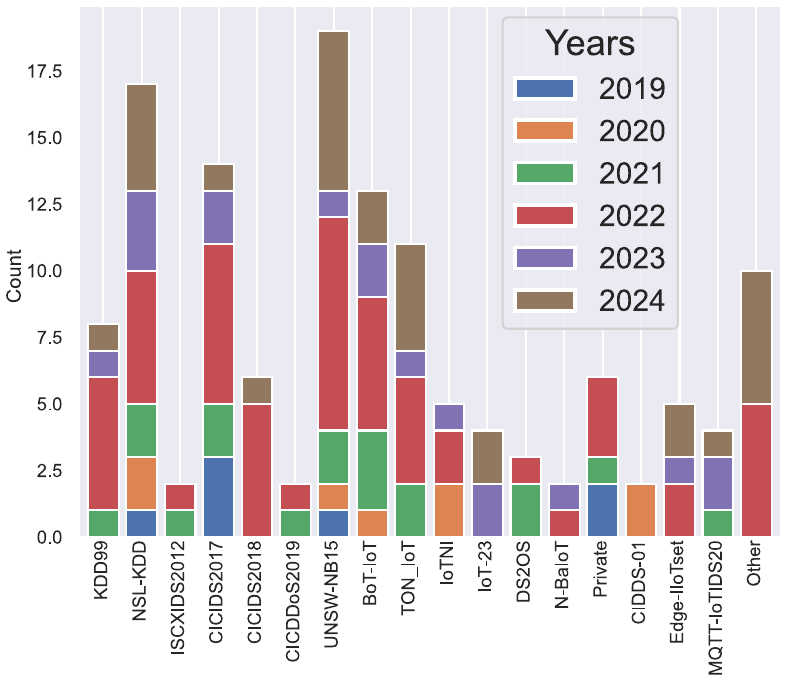}}\label{fig:comp_data1}}%
	\subfloat[ Feature Extraction Methods]{{\includegraphics[width=45mm]{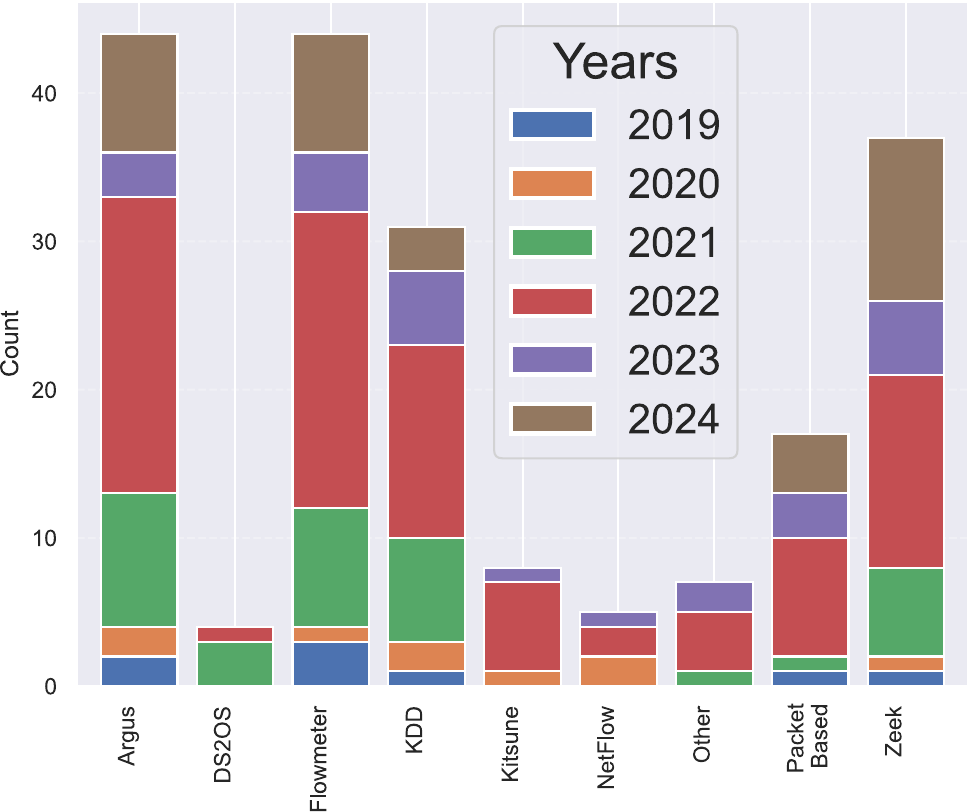}}\label{fig:comp_data3}}%
	\subfloat[ ML Algorithms]{{\includegraphics[width=45mm]{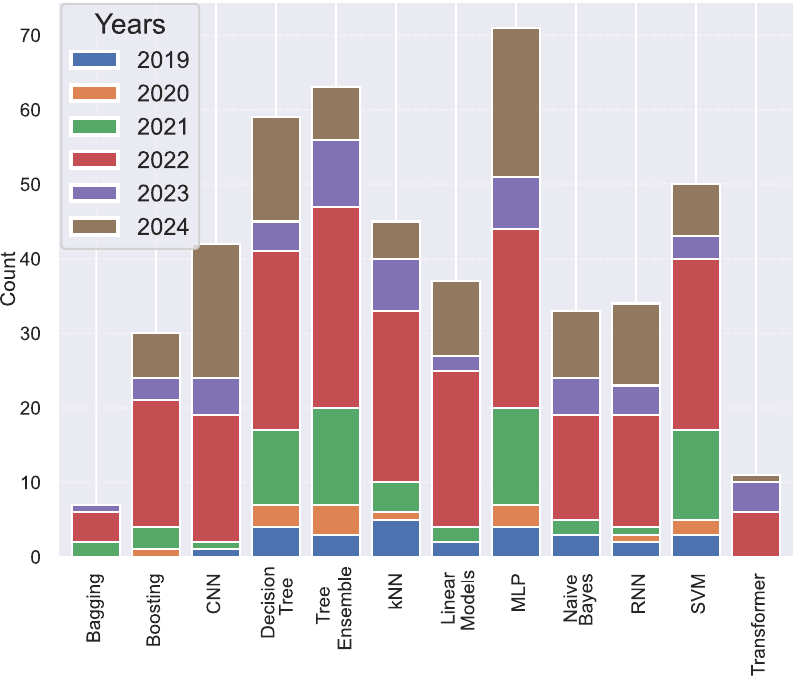}}\label{fig:comp_data2}} %
	\subfloat[ Evaluation Metrics]{{\includegraphics[width=45mm]{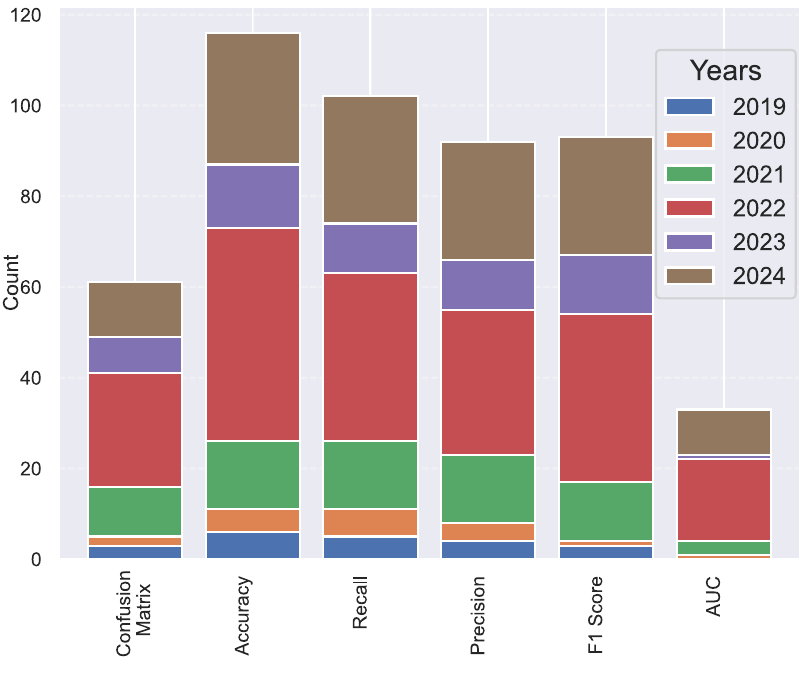}}\label{fig:comp_data4}}%
	\caption{Usage of datasets, machine learning (ML)  algorithms, feature types, and evaluation metrics in the literature.}\label{fig:bir}
\end{figure*}

Below we review related literature from the past several years that focuses on anomaly or intrusion detection for IoT devices using supervised ML models. We categorise this research into four areas, based on the data, features, ML models and evaluation metrics used in each study. Fig.~\ref{fig:bir} summarises the changes in these areas over time and frames our discussion in each corresponding subsection. 
A detailed summary and comparison of the works discussed in this section is available in Table~\ref{tab:LR1} in the Appendix.

\subsection{Data}\label{data}
The use of datasets in related work published from 2019 is summarised in Fig.~\ref{fig:comp_data1}.
Although a number of studies have been published in recent years, it is common practice to use old datasets. For example \cite{Bagaa2020,raghuvanshi2022intrusion,Saba2021,Manhas2021,ahmad2022supervised,zhang2022comparative,Mliki2020,el2022anomaly} used the ~\href{http://kdd.ics.uci.edu/databases/kddcup99/kddcup99.html}{KDD99}\footnote{
	See Table~\ref{tab:datasetlist} (Appendix) for listing of open-sources datasets.}
and \href{http://www.unb.ca/cic/datasets/nsl.html}{NSL-KDD} datasets. KDD99 dates from 1999, and NSL-KDD is an error-corrected version of KDD99. These are large, trusted datasets, and their use in a large number of studies makes them a useful benchmark. However, their age means they lack up-to-date attacks and modern devices. This seriously limits their applicability within contemporary studies of network security.

Furthermore, these datasets do not contain any data related to IoT devices. In fact, it is common practice for IoT-oriented studies of network security to use datasets which do not contain IoT data. For example, ~\cite{Roopak2019a,Jan2019a,gaur2022analysis,Amangele2019,ren2022id,dang2022using,okey2022boostedenml,singh2022machine,Ahmad2021,Hanif2019,saheed2022machine,mohmand2022machine,basharat2022machine,tiwari2022detecting} all use recent datasets, but these datasets do not contain any IoT data. Consequently, it is unclear how these studies yield results that are specific to IoT-based networks or attacks.
.

It is also common for studies to use private datasets. Although the datasets used in \cite{Zolanvari2019,gupta2022smart,adeleye2022security,Anthi2019a,DeCarvalhoBertoli2021,Anthi2021,mondal2022comparative} do contain IoT devices, as far as we are aware they are not publicly accessible, and consequently it is difficult to confirm their data quality. Whilst it is understandable that some datasets cannot be shared, reproducibility is a serious problem in ML-based studies~\cite{kapoor2022leakage}, and it is therefore important that studies which use private data are externally validated on a publicly-accessible dataset. 

Another issue with datasets  is that many (such as  KDD99, NSL-KDD, CIDDS-01, DS2OS, MQTTset, N-BaIoT, and InSDN) contain only pre-extracted features, and do not make the raw network data (pcap --packet capture-- file) available. Without raw data, it is not possible to investigate new features or extract features using different tools.

On the other hand, many datasets with IoT devices
with labelled and raw data for intrusion detection, both simulation-based datasets such as \href{https://research.unsw.edu.au/projects/bot-iot-dataset}{BoT-IoT}~\cite{koroniotis2019towards},
\href{https://ieee-dataport.org/documents/edge-iiotset-new-comprehensive-realistic-cyber-security-dataset-iot-and-iiot-applications}{Edge-IIoT}~\cite{ferrag2022edge}, 
\href{https://research.unsw.edu.au/projects/toniot-datasets}{TON-IoT}~\cite{MOUSTAFA2021TON}, 
\href{https://ieee-dataport.org/open-access/mqtt-iot-ids2020-mqtt-internet-things-intrusion-detection-dataset}{MQTT-IoT-IDS2020}~\cite{hindy2020MQTT},  and real device-based dataset such as \href{https://www.unb.ca/cic/datasets/iotdataset-2022.html}{CIC-IoT-2022}~\cite{CIC},
\href{https://ocslab.hksecurity.net/Datasets/iot-environment-dataset}{IoT-ENV}~\cite{IoT-ENV},
\href{https://ocslab.hksecurity.net/Datasets/iot-network-intrusion-dataset}{IoT-NID}~\cite{IoT-NID},
\href{https://www.kaggle.com/ymirsky/network-attack-dataset-kitsune}{Kitsune}~\cite{mirsky2018kitsune},
\href{https://www.stratosphereips.org/datasets-iot23}{IoT-23}~\cite{IoT-23} are widely used in the literature. These datasets are analysed in Section~\ref{Data Selection}.

\subsection{Features}\label{Feature}

The use of different feature extraction methods in related work published from 2019 is summarised in Fig.~\ref{fig:comp_data3}.
Most of the datasets used for intrusion detection are obtained with common methods and tools. Zeek (formerly Bro)~\cite{Bro}, Argus~\cite{argus}, and CICFlowMeter~\cite{flowmeter} (formerly ISCXFlowMeter) are examples of commonly used tools. For example, the UNSW-NB15 dataset used in \cite{Ahmad2021,Hanif2019,saheed2022machine,mohmand2022machine,basharat2022machine,tiwari2022detecting}  was generated using Zeek and Argus and contains 49 features. ToN\_IoT 
and Bot-IoT datasets were generated using Argus and contain 42 and 29 features respectively. The CICIDS2017-18-19 datasets were all generated by CICFlowMeter and contain 83 features. The common characteristics of these tools are that all of the generated features are flow-based. 
Unlike these tools, Kitsune, used in \cite{han2022machine,abu2022elba,faysal2022xgb,Desai2020} uses a sliding window approach, converting pcap files into a dataset with 115 features.

In the use of packet-based features, individual features obtained from network packets are used. This approach has been used in various studies~\cite{Anthi2019a,ferrag2022edge,DeCarvalhoBertoli2021,Anthi2021,saran2023comparative,mondal2022comparative,gebrye2023traffic,tomar2023cyber,kikissagbe2024machine,dos2024enhancing,cherfi2024exploring,gupta2024sustainable}. We can also list the following datasets that use individual packet features: MQTTset, Edge-IIoTset and MQTT-IoT-IDS20 (also contains flow-based). Apart from this, some studies~\cite{marin2018rawpower, de2021machine} detect attacks by training deep learning models on raw network data without first converting it into features. However, since the raw data (pcap file) consists of network packets, these studies face similar challenges to those which use individual packet features, such as the unintentional use of identifying features. Identifying features are those that directly or indirectly give clues about the label information, but are not generalizable. For example, in most cases, IP addresses are identifying because they uniquely identify the attacker or target. However, since these IP attributes may change in the event of another attack, this attribute is unique to that particular dataset and cannot be generalized. Therefore, using this feature will cause an overfitting problem, tying the selected features to one dataset. 
An analysis of the characteristics, advantages and disadvantages of the flow, window, and packet methods can be found in Section~\ref{FE-AD}.

It is also possible to create new feature sets by using different tools if the dataset includes a pcap file. For example, the IoTID20~\cite{albulayhi2022iot} dataset was created using CICFlowMeter from IoT-NID pcap files. So, both flow~\cite{albulayhi2022iot} and sliding window~\cite{Desai2020} features were extracted from the IoT-NID dataset. On the other hand, KDD99, NSL-KDD, CIDDS-01, DS2OS, MQTTset, N-BaIoT and InSDN datasets do not include pcap files. In this respect, feature extraction and modification studies that can be done with these datasets are quite limited.

\subsection{Machine Learning and Evaluation Metrics}\label{section:ml}

The use of different ML algorithms and metrics in related work published from 2019 is summarised in Fig.~\ref{fig:comp_data2}--\ref{fig:comp_data4}. The most widely used algorithms in the literature are tree ensembles (especially random forest), decision trees, MLP, SVM, boosting and kNN.  In addition to classical ML algorithms, the use of deep learning methods has been increasing recently. 
For example, the use of CNN and RNN has increased, and transformers have started to be used intensively since 2022. Researchers have often evaluated their approach using more than one ML model. In many studies with multiple comparisons, ensemble models (such as RF and XGB) gave the best results~\cite{zhang2022comparative,Mliki2020,el2022anomaly,gaur2022analysis,dang2022using,okey2022boostedenml,singh2022machine,Ahmad2021,mohmand2022machine,basharat2022machine,tiwari2022detecting,Zolanvari2019,gupta2022smart,abu2022elba,faysal2022xgb,Desai2020,yao2022cnn,Bhuvaneswari2020}.

Whilst deep learning (such as CNN, RNN and transformers) has become popular in recent years, classical ML algorithms remain widely used. In part, this can be attributed to the good performance of classical approaches  in cases where feature extraction has already taken place
~\cite{lundberg2020local}, something that is true of most datasets and studies in the area of network security.

Accuracy is the most commonly used  
evaluation metric.
In the majority of studies, accuracy is supported by other metrics such as recall, precision, and F1 score, but there are also studies~\cite{leon2022comparative,vitorino2023towards,raghuvanshi2022intrusion,manzano2022design} where only accuracy is reported. However, in an area such as attack detection, which suffers from unbalanced data distribution,  accuracy alone is not a reliable metric~\cite{unbalanced}.
Although the accuracies reported vary from 0.77-1, many studies report a success rate of greater than 0.99 (see Table~\ref{tab:LR1}). In cases of unbalanced data distribution and where only accuracy is reported, these scores may not accurately reflect model success. More generally, the reliability of  results reported in the literature is questionable due to 
common errors made in network security and ML-based studies (e.g., data leakage, overfitting)~\cite{arp2020and}. In this context, it is essential that studies are not only focused on high metric scores, but are also free from common errors, transparent and reproducible.

\subsection{Summary}

\textcolor{black}{This review of the field reveals several critical gaps that motivate the IoTGeM framework. Firstly, many studies rely on outdated or non-IoT datasets, limiting their applicability to modern security challenges, while the use of private data hinders reproducibility and validation. Secondly, a significant number of public datasets do not provide raw network captures, which prevents researchers from exploring novel feature extraction methods beyond pre-extracted flow-based features. Most critically, common methodological errors such as data leakage and feature overfitting are prevalent, raising serious doubts about the reported high-performance scores and the real-world generalizability of the resulting models. IoTGeM is therefore proposed to directly address these shortcomings by focusing on a strict evaluation methodology, novel feature engineering from raw data, and a feature selection process explicitly designed to produce robust and generalizable models.}

\textcolor{black}{While this section provides a broad overview of the field, a direct numerical comparison between IoTGeM and the results published in these studies is challenging. Our methodology employs a strict separation of datasets for training, validation, and testing to ensure a robust measure of generalizability, a practice not uniformly adopted in the literature. Therefore, to provide a fair and meaningful evaluation, this paper presents a rigorous internal benchmark in Section 4.1. There, we compare our proposed window-based features against the widely-used flow-based features under the exact same data pipeline and selection methodology, offering a controlled assessment of our approach's contribution.}

\section{Materials and Methods}\label{Methodology}

\subsection{System Model and Design Goals}\label{System Model and Design Goals}

\begin{figure}[htbp]
	\centering
	
	\centering{\includegraphics[width=82mm]{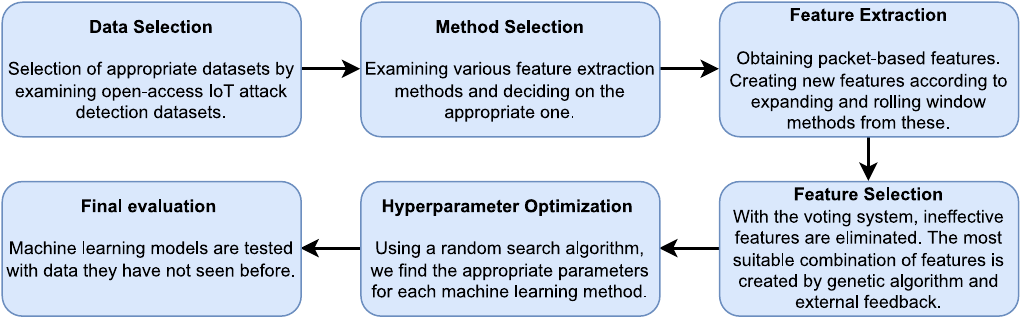}}%
	\caption{Steps in system implementation.}
	\label{fig:system}
\end{figure}

\begin{figure}[htbp]
	\centering
	
	\centering{\includegraphics[width=82mm]{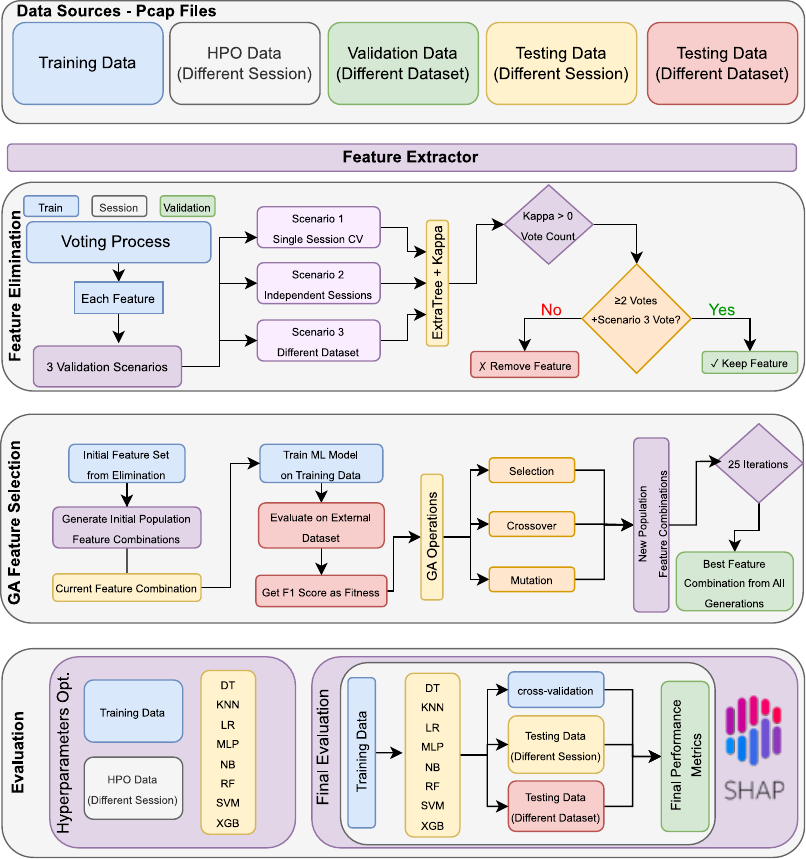}}%
	\caption{\textcolor{black}{IoTGeM Implementation Pipeline. This diagram illustrates the workflow of the IoTGeM framework, starting with labeled pcap-based network data. The process involves a three-step feature elimination using ExtraTree and Cohen’s Kappa-based validation, followed by genetic algorithm-driven feature selection with external feedback. Selected features train eight machine learning models, evaluated through cross-validation, independent sessions, and diverse datasets to ensure robust generalization. SHAP-based explainability enhances model interpretability.}} 
	\label{fig:pipiline}
\end{figure}

To build a generalizable ML model that detects attacks
we first analyse  
and decide on the most suitable datasets for our study. Then, we  
introduce our features by analysing the advantages and disadvantages of feature extraction methods. After feature extraction, we perform feature selection by analysing and eliminating some extracted features 
by analysing these features under different conditions. From the remaining features, we create the appropriate feature combination using a genetic algorithm (GA) and exogenous feedback. In the next stage, we discover the most suitable hyperparameters for each ML model using hyperparameter optimisation. Finally, we train our models with the selected features and hyperparameters, and then test these models with previously unseen data and obtain our final results. These steps are summarized in Fig.~\ref{fig:system}. This entire process, from data handling to final evaluation, is visualized in the IoTGeM implementation pipeline shown in Fig.~\ref{fig:pipiline}.

\subsection{Threat Model and System Placement}

\textcolor{black}{Our approach is based on a set of assumptions regarding the threat model and system placement. We assume an attacker is situated outside the local network and is attempting to compromise or disrupt specific IoT devices within it. The attacks range from flooding-based Denial of Service to reconnaissance (scanning) and unauthorized access (brute-force). Our proposed system, IoTGeM, is designed as a Network-based Intrusion Detection System (NIDS). It is intended for deployment on a central network chokepoint, such as a gateway, router, or a dedicated monitoring server with access to a SPAN port. This placement provides the necessary visibility into network traffic flowing to and from the protected IoT devices to perform its analysis.} 

\subsection{Data Selection}\label{Data Selection}

We examined the datasets used in IoT attack detection studies. We searched the ones that are publicly available, contain raw data (pcap files) and are labelled. The list of these is given in Table~\ref{tab:datasets}, where we also measure them in terms of the number of devices and attack types they contain and classify them in terms of whether they contain real IoT data and multiple sessions. IoT devices differ from non-IoT devices in particular by their heterogeneous nature, in that IoT devices often contain unique software and hardware. This diversity is very difficult to simulate, and consequently it is preferable that datasets include real rather than simulated IoT devices. Repeated attacks in different sessions are also desirable since it allows comparison of each attack across sessions and a deeper analysis of the nature of the attacks. It can also be used to prevent overfitting of session-specific features.

\begin{table}[htbp]
	\centering
	\caption{Datasets that can be used for IoT attack detection}
	\resizebox{0.475\textwidth}{!}{\begin{tabular}{@{}lccccc@{}}
			\toprule
			Dataset & Year  & Real IoT & Session & Devices & Attacks \\
			\midrule
			\href{https://research.unsw.edu.au/projects/bot-iot-dataset}{BoT-IoT}~\cite{koroniotis2019towards}& 2018  & ×     & ×     & 5     & 10 \\
			\href{https://www.unb.ca/cic/datasets/iotdataset-2022.html}{CIC-IoT-2022}~\cite{CIC}& 2022  &  \checkmark      &  \checkmark      & 40    & 4 \\
			\href{https://ieee-dataport.org/documents/edge-iiotset-new-comprehensive-realistic-cyber-security-dataset-iot-and-iiot-applications}{Edge-IIoT}~\cite{ferrag2022edge} & 2022  & ×     & ×     & 9     & 13 \\
			\href{https://ocslab.hksecurity.net/Datasets/iot-environment-dataset}{IoT-ENV}~\cite{IoT-ENV} & 2021  &  \checkmark      & ×     & 5     & 3 \\
			\href{https://ocslab.hksecurity.net/Datasets/iot-network-intrusion-dataset}{IoT-NID}~\cite{IoT-NID} & 2019  &  \checkmark      &  \checkmark      & 2     & 10 \\
			\href{https://www.kaggle.com/ymirsky/network-attack-dataset-kitsune}{Kitsune}~\cite{mirsky2018kitsune} & 2018  &  \checkmark      & ×     & 9     & 9 \\
			\href{https://research.unsw.edu.au/projects/toniot-datasets}{TON-IoT}~\cite{MOUSTAFA2021TON} & 2019  & ×     & ×     & 7     & 9 \\
			
			\href{https://www.stratosphereips.org/datasets-iot23}{IoT-23}~\cite{IoT-23} & 2020  &  \checkmark      & ×     & 3     & 8 \\
			\href{https://ieee-dataport.org/open-access/mqtt-iot-ids2020-mqtt-internet-things-intrusion-detection-dataset}{MQTT-IoT-IDS2020}~\cite{hindy2020MQTT} & 2020  & ×     & ×     & 13     & 4 \\
			\bottomrule
		\end{tabular}%
		\label{tab:datasets}}%
\end{table}%

The IoT-NID and CIC-IoT-22 datasets are the only ones which contain real, multi-session IoT data. While CIC-IoT-22 has a large number of devices, it contains relatively few attack types. Furthermore, attack situations are isolated in this dataset, in that only one device is involved in each session in which an attack occurs. IoT-NID, on the other hand, contains few devices but many attacks, which makes it particularly useful for training different attack detection models. 
Thus, IoT-NID is used as the main dataset for training models.

\begin{figure}[htbp]
	\centering
	
	\centering{\includegraphics[width=84mm]{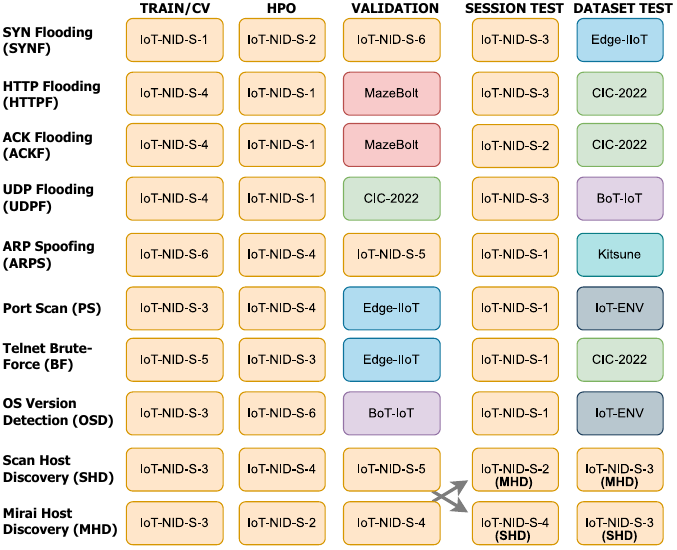}}%
	\caption{Use of data in this study. Train/CV=Training and cross-validation. HPO=Feature reduction and hyperparameter selection. Validation=Data validation and feature selection. 
		S-n= Session n.}
	\label{fig:datasets}
\end{figure}

It is important to completely isolate training and testing data. To ensure this, we always measure the success of a model on data that the model has never seen before. Fig.~\ref{fig:datasets} depicts how data is used in this study. The same colour occurring in different columns for the same attack represents different sessions from the IoT-NID dataset, i.e., where an attack was carried out more than once, in different sessions. 
The data in the first column are 
used for training the models. Data in the Train/CV, HPO and Validation columns are used to refine and select models, while data in the Session and Dataset Test columns are used to measure model generality. More specifically, data in the HPO column is used both for feature reduction and hyperparameter optimisation, and data in the validation column are used in the feature selection step. 

Data is quite limited for some of the attacks. In the case of HTTPS and ACK flood attacks, there was insufficient data 
to use a separate IoT-NID session for validation, 
hence we used
extra data from \href{https://kb.mazebolt.com/}{kb.mazebolt.com}   for these attacks as their structure of these two attacks closely resembled their versions in IoT-NID.
For BF, we addressed the paucity of data for specific attacks by using slightly different variants of the attack in the training, validation and test datasets. Specifically, the training data involves telnet BF, the validation data involves password attacks and the test data involves a RTSP BF attack. For SHD and MHD, there were no suitable equivalent data in other datasets. So that we could still measure the generality of these attack models, we used an MHD session to test SHD models, and vice versa. This is feasible  as SHD and MHD are very similar variants of the same attack (host discovery) made using different devices and tools (scan and Mirai).

\subsection{ML Algorithm Selection}\label{ML-Selec}

Since there is no one-size-fits-all approach for every problem, it is important to consider performance across a range of ML models \cite{lones2021avoid}. 
We selected one model from each of the following model categories~\cite{gollapudi2016practical}; 
linear model: logistic regression (LR),
tree-based: decision tree (DT),
Bayesian-based: naïve Bayes (NB),
kernel-based: support vector machine (SVM),
ensemble-based: random forest (RF),
instance-based: k-nearest neighbours (KNN), and 
artificial neural networks (ANN): multilayer perceptron (MLP).
This set of models provides a good intersection with those used in previous studies. Additionally, we also included XGB, which, although infrequently encountered in the attack detection literature, is known to work well on tabular datasets~\cite{shwartz2022tabular}.

We have made the code for our work public\textsuperscript{\ref{f1}} and we have used a  modular code structure
so that only minor code changes are required to support other algorithms. 
\textcolor{black}{It is important to note that our framework trains a separate binary classification model for each of the ten attack types. Each model is specialized in distinguishing one specific attack category from benign network traffic.}

\subsection{Feature Extraction Methods}\label{FE-AD}

In the literature, it is common to see flow-based or individual packet-based approaches to feature extraction. In this section, we describe these two approaches and propose a window-based approach as an alternative.

\subsubsection{Individual Packet-Based Features}\label{ind-AD}

Individual packet features can be obtained from the packet headers or packet payloads of the individual network packets. As discussed in Section~\ref{Feature}, whilst this approach is commonly used \cite{ferrag2022edge,DeCarvalhoBertoli2021,Anthi2021,saran2023comparative,prazeres2023engineering,mondal2022comparative}, it has a number of issues that can impair generalizability\cite{kostas2025ind}. For this reason, we do not use it in our main study. However, given its prevalence in the literature, in Appendix~\ref{use-ind} we further discuss the limitations of this approach and experimentally show how even simple features such as packet size or timestamp can become identifiers and lead to models with apparently high accuracy but very poor generality. For this, individual features were extracted from raw data using \href{https://www.python.org/}{Python},
\href{https://scapy.net/}{Scapy}, and \href{https://www.wireshark.org/}{Wireshark} tools, mostly from headers, but we also extracted payload-based features such as payload-entropy or payload-bytes.

\subsubsection{Flow-Based Features}\label{flow-AD}

Unlike features from individual packets, flow features~\cite{Bagaa2020,raghuvanshi2022intrusion,Saba2021,Manhas2021,ahmad2022supervised,zhang2022comparative,Mliki2020,el2022anomaly,Roopak2019a,Jan2019a,gaur2022analysis,Amangele2019,ren2022id,dang2022using,okey2022boostedenml,singh2022machine,Ahmad2021,Hanif2019,saheed2022machine,mohmand2022machine,basharat2022machine,tiwari2022detecting} 
are sufficient for attack detection, but they have some drawbacks. For example, in order to generate features in a flow system, the flow must end with a termination flag or timeout. In this respect, in order for a flow to be identified as an attack, it is necessary to wait for it to be terminated first. Only after the flow ends can the statistics required for attack detection be calculated. In this study, flow-based features are extracted from the raw data using the \href{https://www.unb.ca/cic/research/applications.html#CICFlowMeter}{CICFlowMeter}  tool. We chose CICFlowMeter for this, since it extracts significantly more features than alternative tools, and also specifically focuses on features that can be used in attack detection.
Note that the IoTID20 dataset also contains features extracted using CICFlowMeter from IoT-NID; however, our preliminary analysis of IoTID20 (see Appendix~\ref{IoTID20}) identified anomalies, prompting us to do our own feature extraction.

\subsubsection{Window-Based Features}\label{Window-AD}

\begin{figure}[htbp]
	\centering
	\includegraphics[width=65mm]{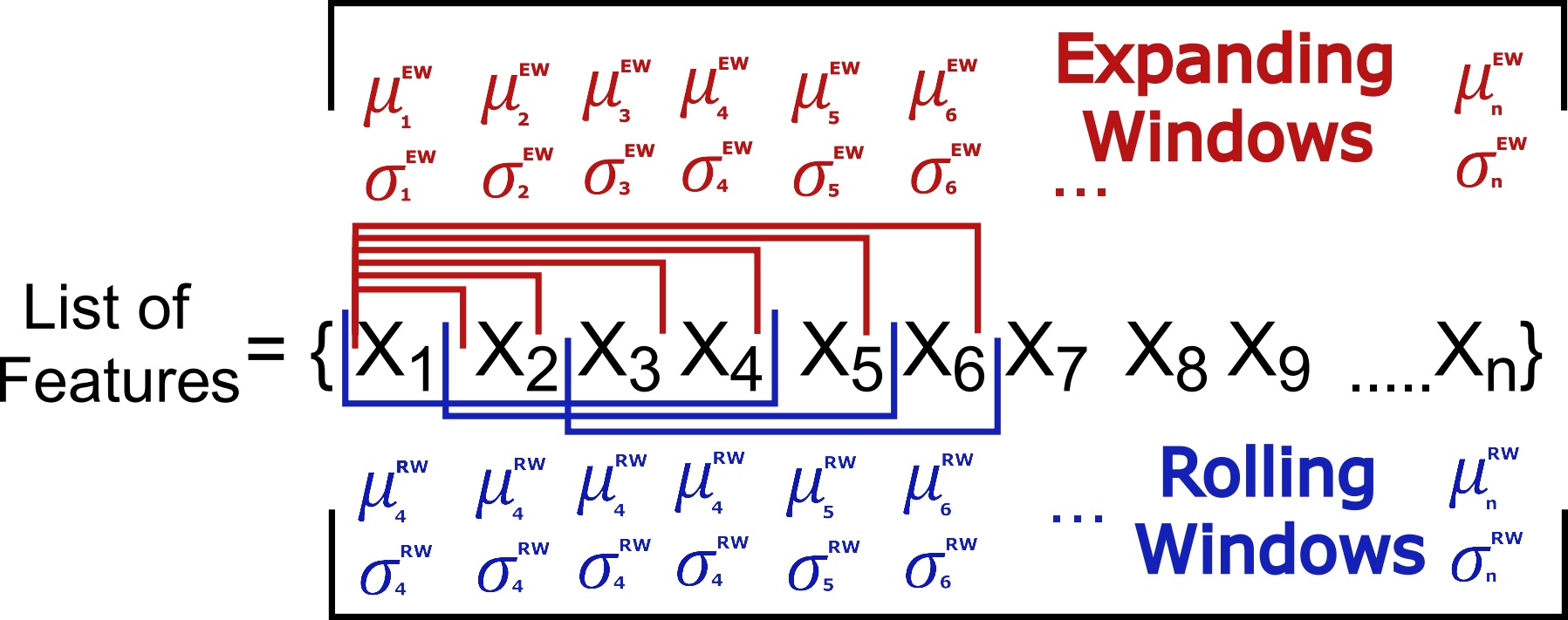}
	\caption{Window-based feature extraction. $\mu$ is mean, $\sigma$  is std.\@ deviation.}
	\label{fig:ewrw}
\end{figure}

In this approach, we focus on the changes in the data flow between the source and destination. However, in contrast to the flow-based approach, instead of generating an aggregate statistic for each flow, we utilise the changes in the network data that occur as each packet arrives. While similar approaches have been used in the literature~\cite{abu2022elba,faysal2022xgb,Desai2020}, we present a different approach in terms of both methodology and features. This uses both rolling (RW) and expanding windows (EW) that extract features from information carried by packets between the source and destination (MAC/IP address). In the rolling window approach, we observe the change between packets within a certain window size. The expanding window starts with a small window covering the initial data and gradually expands to include new data. Each time a new packet is added, the window becomes larger and statistics are recalculated. We use four different sources: size, time, destination-source and Transmission Control Protocol (TCP) flags, and also calculate the mean and standard deviation of the window values for some individual features. The process is visualised in Fig.~\ref{fig:ewrw}.

\begin{figure}[htbp]
	\centering
	\includegraphics[width=75mm]{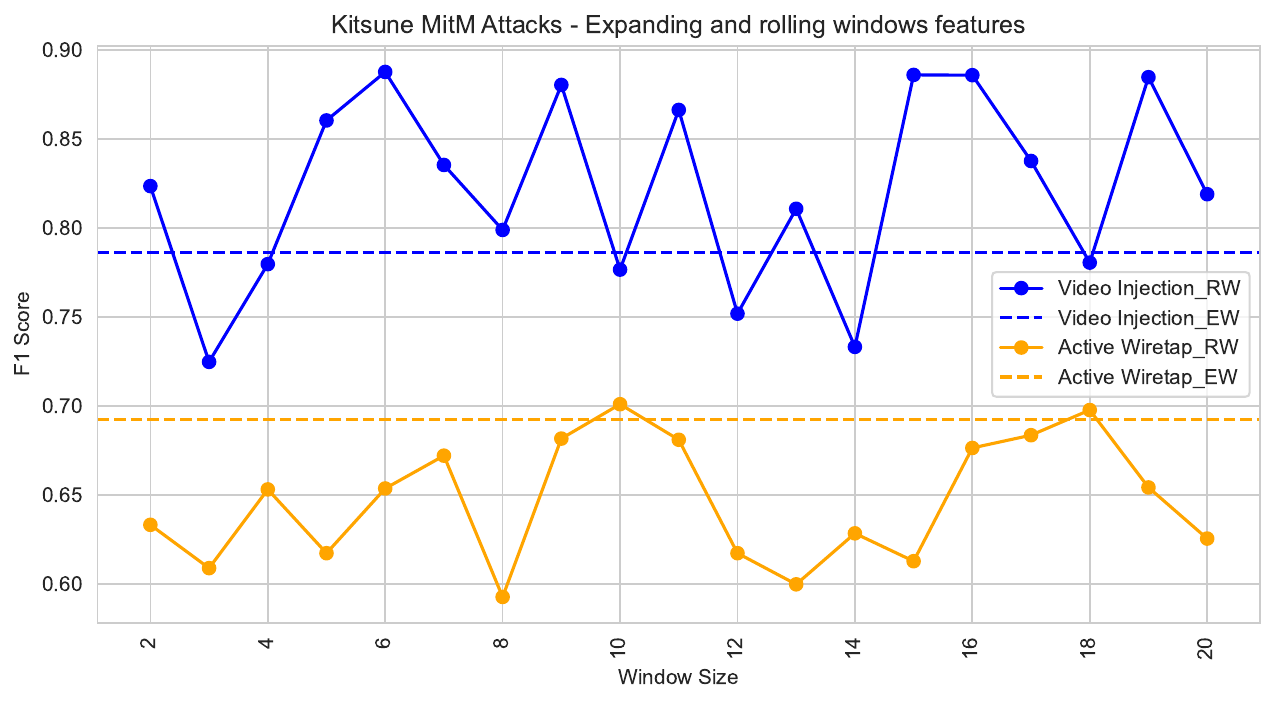}
	\caption{Utilizing diverse window sizes for Kitsune MitM attack detection and the change of performance according to the window size.}
	\label{fig:kitsune}
\end{figure}

When using the rolling window method, we also had to decide on the window size (number of packets) we would use. To avoid overfitting, we determined the window size using two MitM (Video Injection and Active Wiretap) attacks that we did not use in our other experiments. We chose these attacks in particular because MitM attacks are sensitive to packet time changes rather than flags, packet content, and destination-source features. Therefore, statistical features such as window features play an important role in detecting them. As a preliminary study, we tried to detect these attacks using only window features, with window sizes ranging from two to 20, using a combination of EW features and EW-RW features for  \textit{packet size, time, TCP window size, payload bytes, payload entropy} features. The results are shown in Fig.~\ref{fig:kitsune}. Considering the limited data generation of many IoT devices, it is impractical to employ large window sizes. Consequently, we limit the RW size to 10, and in future experiments only employ window sizes below this which were most effective at detecting the MitM attacks: specifically two, six, and nine packets. We apply the expanding window approach to TCP flags and source-destination elements. Our full feature list is given in the Appendix, Table~\ref{tab:indF}, including an explanation of each feature type mentioned in this paper.

\textcolor{black}{In order to better understand the effectiveness of the features, following model evaluation, we employ a post-hoc explainability technique to interpret the results. Specifically, we use SHapley Additive exPlanations (SHAP) to understand the output of our most successful models. This allows us to visualize the impact of each feature on the model's prediction for a given attack, ensuring that the models are learning relevant patterns rather than relying on spurious correlations. This analysis, presented in Section~\ref{FeatureEffectiveness}, is for interpretation and does not influence the training or feature selection process.}

In Section~\ref{subPerformance}, we compare our window-based approach against a flow-based implementation in terms of its ability to support attack detection and generalizability.

\subsection{Feature Selection}\label{Feature Selection-AD}

It is important to identify stable features that are capable of supporting generalizable models. We do this using a three-step method. In the first step, we remove features that are clearly inappropriate, i.e. device or session-based identifiers such as IP-checksum, ID, synchronization, and port numbers (see Table~\ref{tab:indF} in Appendix~\ref{use-ind}). In the second step, we employ a feature elimination process that considers the value of each feature in turn, and in the third step we use a genetic algorithm (GA) to further refine the remaining features, taking into account their potential interactions.

\subsubsection{Feature elimination} 

The aim of this stage is to identify and eliminate features that are a risk to generalizability. For three different validation scenarios, we consider each feature in turn, and measure its association with the target variable when used to train an ML model which is then evaluated using Cohen's kappa. We use an ML model (extremely randomized tree, or ExtraTree) which differs from those listed in Section \ref{ML-Selec} to avoid bias. A kappa value less than or equal to zero indicates an unreliable feature.

In the first scenario, cross-validation is used within a single session (Train/CV column in Fig.~\ref{fig:datasets}). In the second scenario, two independent sessions from the same dataset are used, one (Train/CV column) for training, the other (HPO column) for validation. This helps to eliminate features that do not generalize beyond a single session, a common problem when using cross-validation alone~\cite{kostas2022IoTDevID}. The third scenario uses a validation session from a different dataset (Validation column).

\begin{figure}[htbp]
	\centering
	\includegraphics[width=85mm]{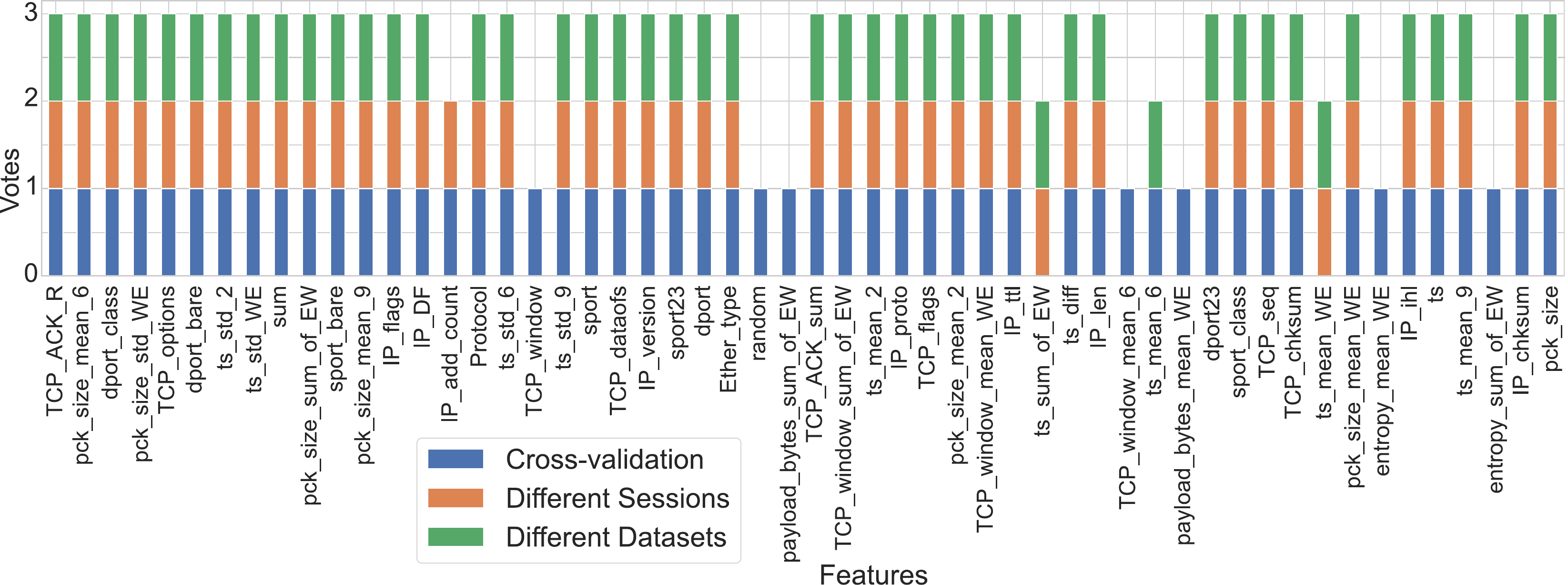}
	\caption{Voting process during the feature elimination step for the Host Discovery attack.}
	\label{fig:voting}
\end{figure}

Features are assessed in each scenario, and if a feature surpasses a kappa score of 0, it is awarded one vote for that scenario. Features with at least two votes, including one from the final scenario, are included in the feature set for the GA. This approach prioritizes features selected in the last, most stringent, scenario whilst also taking into account their utility in the other scenarios to discourage spurious correlations. Fig.~\ref{fig:voting} depicts the voting process for a particular attack.

\subsubsection{Feature selection with GA and external feedback}

We employ a GA to further optimise the feature list, taking into account the potential interactions between features. GAs are a popular approach to feature selection, since they allow the space of feature combinations to be explored in a relatively efficient manner, with a good chance of finding an optimal, or at least near-optimal, combination of features \cite{gharaee2016new}.

\begin{figure}[htbp]	
	\centering
	\includegraphics[width=75mm]{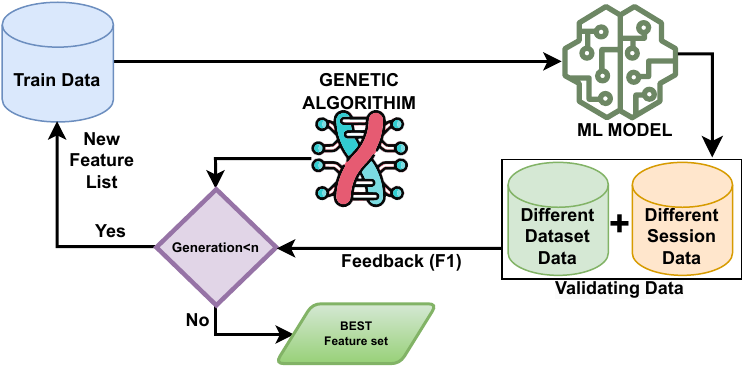}
	\caption{Visualization of feature selection with GA. The GA outputs the best feature set from n generations with the feedback it receives from validation.}
	\label{fig:ga_feature}
\end{figure}

The GA process is depicted in Fig.~\ref{fig:ga_feature}. The validation results, in terms of F1 score, are used as feedback for the GA, guiding the creation of new feature combinations. This iterative process continues for a specified number of generations (25), with the best combination of features observed during this period chosen as the final feature set.

In this procedure, the GA obtains an F1 score as external performance feedback from a distinct dataset (see Fig.~\ref{fig:datasets}). 
\textcolor{black}{Our approach presents a significant divergence from typical GA-based feature selection. Conventionally, a GA's fitness function is evaluated on the same dataset used for training (e.g., via cross-validation). In contrast, our method employs disparate data for fitness evaluation. This use of external feedback from an unseen dataset is the core of our innovation, as it fosters the selection of feature combinations that are inherently generalizable, thereby addressing overfitting concerns.} 

\subsection{Preventing Data Leakage} %
\textcolor{black}{
A central goal of IoTGeM is to produce generalizable models by rigorously preventing data leakage during the development process. Data leakage occurs when information from outside the training dataset is used to build a model, leading to overly optimistic performance estimates that fail on truly unseen data. A common pitfall is performing preprocessing steps, such as feature selection or normalization, on an entire dataset before splitting it into training and test sets. This allows the model to learn from the statistical properties of the test data, thereby contaminating the evaluation.}
Our methodology prevents this leakage through a strict, multi-stage isolation of data, as depicted in Fig.~\ref{fig:datasets}. We use entirely separate datasets for:

\textbf{Model Training and Cross-Validation (Train/CV)}: The dataset used exclusively to train the model parameters.

\textbf{Feature/Hyperparameter Tuning (HPO \& Validation)}: Separate datasets used to guide the feature selection process and for hyperparameter optimization. This ensures that the choice of features and model parameters is not influenced by the final test data.

\textbf{Final Testing (Session Test \& Dataset Test)}: Completely unseen datasets reserved exclusively for the final performance evaluation of the trained model.

By adhering to this separation, we ensure that the final test data has no influence on the training, feature selection, or tuning of the model.

\section{Performance Evaluation}\label{Performance}

\textcolor{black}{This section evaluates the performance of the attack detection models created with the IoTGeM pipeline. Our experimental setup involves training a separate binary classification model for each of the ten attack types. For each attack, we use the feature set identified by our GA-based selection process to train eight distinct ML algorithms (listed in Section~\ref{ML-Selec}). We then evaluate these models under three increasingly stringent scenarios: (a) 10-fold cross-validation, (b) testing on an independent session from the same dataset, and (c) testing on a completely different dataset the model has never seen. This final cross-dataset test is the most robust measure of generalizability. To handle imbalanced data, we primarily report the F1 score, though additional metrics are available in Appendix~\ref{MLR}. The models were developed and evaluated on a system with the following technical specifications: Intel(R) Core(TM) i7-7500U CPU @ 2.70GHz (Boost up to 2.90 GHz), 8 GB RAM (7.74 GB usable), Windows 10 Pro 64-bit operating system, and AMD Radeon(TM) 530 GPU.}

Starting with the performance of attack detection models trained with the feature sets we have created, Table~\ref{tab:mainresults} presents the results for each attack when using window-based features. In each case, F1 scores are shown for when the model is evaluated using (a) cross-validation, (b) an independent session from the same dataset it was trained on, and (c) a dataset different to the one it was trained on. This is intended to give insight into each model's generality, with each subsequent evaluation scenario being a more stringent test of the model's ability to generalize to unseen (and therefore more realistic) conditions. Additional evaluation measures, including accuracy, recall, and precision, can be found in Appendix~\ref{MLR}.

For cross-validation, near-perfect discrimination is achieved for all attacks using at least one ML model. When an isolated session is used for evaluation, high levels of discrimination are observed for all but the ARPS attack. When a different dataset is used for evaluation, high levels of discrimination are achieved for all but the ARPS, BF and OSD attacks. It is promising that the majority of attacks are still detected with good levels of discrimination even in the most realistic evaluation condition. However, the success rate clearly drops as the evaluation conditions become more stringent, and this highlights the fact that cross-validation---an approach widely used in the literature---can lead to overly-confident measures of generality, even when steps are taken to remove features that cause overfitting in the feature selection stage. 

Notably, the two decision tree ensemble approaches, XGB and RF, lead to the most successful models, with only a few exceptions. For BF and UDPF attacks, NB demonstrated the highest success, while LR performed best for OSD attacks.

\begin{table}[htbp]
  \centering
  \caption{F1 scores of models using window-based features in the cases of cross-validation, isolated sessions and isolated datasets.}
    \resizebox{0.5\textwidth}{!}{\begin{tabular}{clrrrrclrrr}

     &     & \multicolumn{1}{c}{} & \multicolumn{1}{c}{Session} & \multicolumn{1}{c}{Dataset} &       &  &     & \multicolumn{1}{c}{} & \multicolumn{1}{c}{Session} & \multicolumn{1}{c}{Dataset} \\
    Name & ML    & \multicolumn{1}{c}{CV} & \multicolumn{1}{c}{Test} & \multicolumn{1}{c}{Test} &       & Name & ML    & \multicolumn{1}{c}{CV} & \multicolumn{1}{c}{Test} & \multicolumn{1}{c}{Test} \\
\cmidrule{1-5}\cmidrule{7-11}    \multirow{8}[2]{*}{ACK} & DT    & \cellcolor[rgb]{ .388,  .745,  .482}1.000 & \cellcolor[rgb]{ .392,  .749,  .486}0.999 & \cellcolor[rgb]{ .459,  .773,  .541}0.912 &       & \multirow{8}[2]{*}{OS} & DT    & \cellcolor[rgb]{ .388,  .745,  .482}0.999 & \cellcolor[rgb]{ .478,  .78,  .561}0.890 & \cellcolor[rgb]{ .753,  .894,  .8}0.542 \\
          & KNN   & \cellcolor[rgb]{ .388,  .745,  .482}1.000 & \cellcolor[rgb]{ .388,  .745,  .482}1.000 & \cellcolor[rgb]{ .988,  .988,  1}0.212 &       &       & KNN   & \cellcolor[rgb]{ .435,  .765,  .522}0.944 & \cellcolor[rgb]{ .792,  .91,  .831}0.493 & \cellcolor[rgb]{ .824,  .922,  .859}0.457 \\
          & LR    & \cellcolor[rgb]{ .388,  .745,  .482}1.000 & \cellcolor[rgb]{ .388,  .745,  .482}1.000 & \cellcolor[rgb]{ .4,  .753,  .494}0.986 &       &       & LR    & \cellcolor[rgb]{ .392,  .749,  .486}0.998 & \cellcolor[rgb]{ .482,  .784,  .565}0.885 & \cellcolor[rgb]{ .635,  .847,  .698}0.691 \\
          & MLP   & \cellcolor[rgb]{ .388,  .745,  .482}1.000 & \cellcolor[rgb]{ .392,  .749,  .486}0.999 & \cellcolor[rgb]{ .898,  .953,  .922}0.333 &       &       & MLP   & \cellcolor[rgb]{ .8,  .914,  .839}0.485 & \cellcolor[rgb]{ .796,  .91,  .835}0.490 & \cellcolor[rgb]{ .922,  .961,  .945}0.332 \\
          & NB    & \cellcolor[rgb]{ .388,  .745,  .482}1.000 & \cellcolor[rgb]{ .388,  .745,  .482}1.000 & \cellcolor[rgb]{ .988,  .988,  1}0.212 &       &       & NB    & \cellcolor[rgb]{ .718,  .878,  .765}0.590 & \cellcolor[rgb]{ .988,  .988,  1}0.247 & \cellcolor[rgb]{ .827,  .925,  .859}0.453 \\
          & RF    & \cellcolor[rgb]{ .388,  .745,  .482}1.000 & \cellcolor[rgb]{ .392,  .749,  .486}0.999 & \cellcolor[rgb]{ .4,  .749,  .49}0.989 &       &       & RF    & \cellcolor[rgb]{ .408,  .753,  .502}0.975 & \cellcolor[rgb]{ .561,  .816,  .631}0.784 & \cellcolor[rgb]{ .694,  .871,  .745}0.618 \\
          & SVM   & \cellcolor[rgb]{ .624,  .843,  .686}0.694 & \cellcolor[rgb]{ .973,  .984,  .988}0.235 & \cellcolor[rgb]{ .898,  .953,  .922}0.333 &       &       & SVM   & \cellcolor[rgb]{ .8,  .914,  .839}0.485 & \cellcolor[rgb]{ .788,  .91,  .827}0.499 & \cellcolor[rgb]{ .922,  .961,  .941}0.333 \\
          & XGB   & \cellcolor[rgb]{ .388,  .745,  .482}1.000 & \cellcolor[rgb]{ .388,  .745,  .482}1.000 & \cellcolor[rgb]{ .4,  .749,  .49}0.989 &       &       & XGB   & \cellcolor[rgb]{ .392,  .749,  .486}0.998 & \cellcolor[rgb]{ .478,  .78,  .561}0.890 & \cellcolor[rgb]{ .749,  .89,  .792}0.551 \\
\cmidrule{1-5}\cmidrule{7-11}    \multirow{8}[2]{*}{ARP} & DT    & \cellcolor[rgb]{ .396,  .749,  .49}0.994 & \cellcolor[rgb]{ .753,  .894,  .796}0.583 & \cellcolor[rgb]{ .867,  .941,  .894}0.454 &       & \multirow{8}[2]{*}{SHD} & DT    & \cellcolor[rgb]{ .392,  .749,  .486}0.999 & \cellcolor[rgb]{ .475,  .78,  .557}0.890 & \cellcolor[rgb]{ .475,  .78,  .557}0.889 \\
          & KNN   & \cellcolor[rgb]{ .455,  .773,  .541}0.925 & \cellcolor[rgb]{ .733,  .886,  .78}0.607 & \cellcolor[rgb]{ .965,  .98,  .98}0.341 &       &       & KNN   & \cellcolor[rgb]{ .392,  .749,  .486}0.998 & \cellcolor[rgb]{ .588,  .827,  .655}0.740 & \cellcolor[rgb]{ .576,  .824,  .647}0.756 \\
          & LR    & \cellcolor[rgb]{ .557,  .816,  .627}0.808 & \cellcolor[rgb]{ .71,  .878,  .761}0.631 & \cellcolor[rgb]{ .769,  .898,  .808}0.567 &       &       & LR    & \cellcolor[rgb]{ .392,  .749,  .486}0.998 & \cellcolor[rgb]{ .69,  .867,  .741}0.611 & \cellcolor[rgb]{ .667,  .859,  .722}0.639 \\
          & MLP   & \cellcolor[rgb]{ .702,  .875,  .753}0.642 & \cellcolor[rgb]{ .769,  .902,  .812}0.563 & \cellcolor[rgb]{ .937,  .969,  .953}0.373 &       &       & MLP   & \cellcolor[rgb]{ .392,  .749,  .486}0.999 & \cellcolor[rgb]{ .729,  .882,  .776}0.560 & \cellcolor[rgb]{ .557,  .816,  .627}0.781 \\
          & NB    & \cellcolor[rgb]{ .831,  .925,  .863}0.494 & \cellcolor[rgb]{ .902,  .953,  .925}0.411 & \cellcolor[rgb]{ .973,  .984,  .988}0.329 &       &       & NB    & \cellcolor[rgb]{ .78,  .906,  .824}0.490 & \cellcolor[rgb]{ .631,  .843,  .69}0.687 & \cellcolor[rgb]{ .6,  .831,  .667}0.726 \\
          & RF    & \cellcolor[rgb]{ .392,  .749,  .486}0.997 & \cellcolor[rgb]{ .682,  .867,  .737}0.664 & \cellcolor[rgb]{ .961,  .976,  .976}0.345 &       &       & RF    & \cellcolor[rgb]{ .392,  .749,  .486}0.999 & \cellcolor[rgb]{ .533,  .804,  .608}0.816 & \cellcolor[rgb]{ .631,  .847,  .694}0.684 \\
          & SVM   & \cellcolor[rgb]{ .871,  .941,  .898}0.447 & \cellcolor[rgb]{ .882,  .949,  .91}0.432 & \cellcolor[rgb]{ .969,  .98,  .984}0.333 &       &       & SVM   & \cellcolor[rgb]{ .388,  .745,  .482}1.000 & \cellcolor[rgb]{ .776,  .902,  .816}0.497 & \cellcolor[rgb]{ .776,  .906,  .82}0.495 \\
          & XGB   & \cellcolor[rgb]{ .392,  .749,  .486}0.997 & \cellcolor[rgb]{ .624,  .839,  .686}0.733 & \cellcolor[rgb]{ .957,  .976,  .973}0.349 &       &       & XGB   & \cellcolor[rgb]{ .392,  .749,  .486}0.999 & \cellcolor[rgb]{ .424,  .761,  .514}0.954 & \cellcolor[rgb]{ .459,  .773,  .541}0.912 \\
\cmidrule{1-5}\cmidrule{7-11}    \multirow{8}[2]{*}{BF} & DT    & \cellcolor[rgb]{ .392,  .749,  .486}0.999 & \cellcolor[rgb]{ .843,  .929,  .875}0.481 & \cellcolor[rgb]{ .953,  .976,  .973}0.351 &       & \multirow{8}[2]{*}{PS} & DT    & \cellcolor[rgb]{ .388,  .745,  .482}1.000 & \cellcolor[rgb]{ .388,  .745,  .482}1.000 & \cellcolor[rgb]{ .471,  .78,  .553}0.896 \\
          & KNN   & \cellcolor[rgb]{ .408,  .753,  .502}0.979 & \cellcolor[rgb]{ .455,  .773,  .541}0.925 & \cellcolor[rgb]{ .808,  .918,  .843}0.520 &       &       & KNN   & \cellcolor[rgb]{ .388,  .745,  .482}1.000 & \cellcolor[rgb]{ .388,  .745,  .482}1.000 & \cellcolor[rgb]{ .392,  .749,  .486}0.997 \\
          & LR    & \cellcolor[rgb]{ .392,  .749,  .486}0.996 & \cellcolor[rgb]{ .812,  .918,  .847}0.516 & \cellcolor[rgb]{ .91,  .957,  .933}0.402 &       &       & LR    & \cellcolor[rgb]{ .392,  .749,  .486}0.999 & \cellcolor[rgb]{ .388,  .745,  .482}1.000 & \cellcolor[rgb]{ .392,  .749,  .486}0.997 \\
          & MLP   & \cellcolor[rgb]{ .443,  .769,  .529}0.938 & \cellcolor[rgb]{ .416,  .757,  .506}0.969 & \cellcolor[rgb]{ .835,  .929,  .871}0.486 &       &       & MLP   & \cellcolor[rgb]{ .388,  .745,  .482}1.000 & \cellcolor[rgb]{ .388,  .745,  .482}1.000 & \cellcolor[rgb]{ .392,  .749,  .486}0.997 \\
          & NB    & \cellcolor[rgb]{ .737,  .886,  .784}0.602 & \cellcolor[rgb]{ .737,  .886,  .784}0.601 & \cellcolor[rgb]{ .62,  .839,  .682}0.738 &       &       & NB    & \cellcolor[rgb]{ .412,  .757,  .502}0.972 & \cellcolor[rgb]{ .42,  .757,  .51}0.963 & \cellcolor[rgb]{ .412,  .757,  .502}0.971 \\
          & RF    & \cellcolor[rgb]{ .388,  .745,  .482}1.000 & \cellcolor[rgb]{ .412,  .757,  .502}0.974 & \cellcolor[rgb]{ .984,  .988,  .996}0.317 &       &       & RF    & \cellcolor[rgb]{ .392,  .749,  .486}0.999 & \cellcolor[rgb]{ .388,  .745,  .482}1.000 & \cellcolor[rgb]{ .494,  .788,  .576}0.862 \\
          & SVM   & \cellcolor[rgb]{ .412,  .757,  .506}0.973 & \cellcolor[rgb]{ .416,  .757,  .506}0.971 & \cellcolor[rgb]{ .988,  .988,  1}0.313 &       &       & SVM   & \cellcolor[rgb]{ .796,  .91,  .835}0.472 & \cellcolor[rgb]{ .69,  .871,  .745}0.608 & \cellcolor[rgb]{ .804,  .914,  .839}0.461 \\
          & XGB   & \cellcolor[rgb]{ .392,  .749,  .486}0.998 & \cellcolor[rgb]{ .478,  .784,  .561}0.898 & \cellcolor[rgb]{ .988,  .988,  1}0.313 &       &       & XGB   & \cellcolor[rgb]{ .392,  .749,  .486}0.999 & \cellcolor[rgb]{ .388,  .745,  .482}1.000 & \cellcolor[rgb]{ .431,  .765,  .522}0.945 \\
\cmidrule{1-5}\cmidrule{7-11}    \multirow{8}[2]{*}{HTTP} & DT    & \cellcolor[rgb]{ .388,  .745,  .482}1.000 & \cellcolor[rgb]{ .396,  .749,  .49}0.993 & \cellcolor[rgb]{ .388,  .745,  .482}1.000 &       & \multirow{8}[2]{*}{SYN} & DT    & \cellcolor[rgb]{ .388,  .745,  .482}1.000 & \cellcolor[rgb]{ .392,  .749,  .486}0.999 & \cellcolor[rgb]{ .443,  .769,  .529}0.930 \\
          & KNN   & \cellcolor[rgb]{ .392,  .749,  .486}0.998 & \cellcolor[rgb]{ .482,  .784,  .565}0.893 & \cellcolor[rgb]{ .953,  .973,  .969}0.355 &       &       & KNN   & \cellcolor[rgb]{ .392,  .749,  .486}0.999 & \cellcolor[rgb]{ .392,  .749,  .486}0.997 & \cellcolor[rgb]{ .929,  .965,  .949}0.298 \\
          & LR    & \cellcolor[rgb]{ .396,  .749,  .49}0.994 & \cellcolor[rgb]{ .396,  .749,  .49}0.993 & \cellcolor[rgb]{ .988,  .988,  1}0.310 &       &       & LR    & \cellcolor[rgb]{ .388,  .745,  .482}1.000 & \cellcolor[rgb]{ .396,  .749,  .49}0.993 & \cellcolor[rgb]{ .404,  .753,  .498}0.981 \\
          & MLP   & \cellcolor[rgb]{ .404,  .753,  .494}0.985 & \cellcolor[rgb]{ .451,  .773,  .537}0.930 & \cellcolor[rgb]{ .922,  .961,  .945}0.388 &       &       & MLP   & \cellcolor[rgb]{ .392,  .749,  .486}0.999 & \cellcolor[rgb]{ .392,  .749,  .486}0.997 & \cellcolor[rgb]{ .902,  .953,  .925}0.333 \\
          & NB    & \cellcolor[rgb]{ .71,  .878,  .761}0.632 & \cellcolor[rgb]{ .867,  .941,  .898}0.450 & \cellcolor[rgb]{ .949,  .973,  .969}0.356 &       &       & NB    & \cellcolor[rgb]{ .396,  .749,  .49}0.993 & \cellcolor[rgb]{ .396,  .749,  .49}0.993 & \cellcolor[rgb]{ .392,  .749,  .486}0.998 \\
          & RF    & \cellcolor[rgb]{ .388,  .745,  .482}1.000 & \cellcolor[rgb]{ .392,  .749,  .486}0.996 & \cellcolor[rgb]{ .388,  .745,  .482}1.000 &       &       & RF    & \cellcolor[rgb]{ .388,  .745,  .482}1.000 & \cellcolor[rgb]{ .392,  .749,  .486}0.999 & \cellcolor[rgb]{ .388,  .745,  .482}1.000 \\
          & SVM   & \cellcolor[rgb]{ .835,  .925,  .867}0.490 & \cellcolor[rgb]{ .835,  .929,  .871}0.486 & \cellcolor[rgb]{ .969,  .98,  .984}0.333 &       &       & SVM   & \cellcolor[rgb]{ .431,  .765,  .522}0.945 & \cellcolor[rgb]{ .427,  .761,  .518}0.951 & \cellcolor[rgb]{ .902,  .953,  .925}0.333 \\
          & XGB   & \cellcolor[rgb]{ .388,  .745,  .482}1.000 & \cellcolor[rgb]{ .392,  .749,  .486}0.996 & \cellcolor[rgb]{ .388,  .745,  .482}1.000 &       &       & XGB   & \cellcolor[rgb]{ .388,  .745,  .482}1.000 & \cellcolor[rgb]{ .392,  .749,  .486}0.999 & \cellcolor[rgb]{ .388,  .745,  .482}1.000 \\
\cmidrule{1-5}\cmidrule{7-11}    \multirow{8}[2]{*}{MHD} & DT    & \cellcolor[rgb]{ .396,  .749,  .49}0.994 & \cellcolor[rgb]{ .557,  .816,  .627}0.808 & \cellcolor[rgb]{ .569,  .82,  .639}0.795 &       & \multirow{8}[2]{*}{UDP} & DT    & \cellcolor[rgb]{ .388,  .745,  .482}1.000 & \cellcolor[rgb]{ .478,  .784,  .561}0.884 & \cellcolor[rgb]{ .882,  .945,  .91}0.359 \\
          & KNN   & \cellcolor[rgb]{ .408,  .753,  .498}0.981 & \cellcolor[rgb]{ .396,  .749,  .49}0.992 & \cellcolor[rgb]{ .396,  .749,  .49}0.993 &       &       & KNN   & \cellcolor[rgb]{ .388,  .745,  .482}1.000 & \cellcolor[rgb]{ .388,  .745,  .482}1.000 & \cellcolor[rgb]{ .988,  .988,  1}0.218 \\
          & LR    & \cellcolor[rgb]{ .404,  .753,  .498}0.984 & \cellcolor[rgb]{ .396,  .749,  .49}0.994 & \cellcolor[rgb]{ .392,  .749,  .486}0.998 &       &       & LR    & \cellcolor[rgb]{ .388,  .745,  .482}1.000 & \cellcolor[rgb]{ .388,  .745,  .482}1.000 & \cellcolor[rgb]{ .988,  .988,  1}0.218 \\
          & MLP   & \cellcolor[rgb]{ .396,  .749,  .486}0.995 & \cellcolor[rgb]{ .396,  .749,  .49}0.994 & \cellcolor[rgb]{ .396,  .749,  .486}0.995 &       &       & MLP   & \cellcolor[rgb]{ .392,  .749,  .486}0.999 & \cellcolor[rgb]{ .388,  .745,  .482}1.000 & \cellcolor[rgb]{ .533,  .804,  .608}0.812 \\
          & NB    & \cellcolor[rgb]{ .773,  .902,  .812}0.562 & \cellcolor[rgb]{ .569,  .82,  .639}0.795 & \cellcolor[rgb]{ .537,  .808,  .608}0.833 &       &       & NB    & \cellcolor[rgb]{ .643,  .851,  .702}0.669 & \cellcolor[rgb]{ .404,  .753,  .494}0.983 & \cellcolor[rgb]{ .435,  .765,  .522}0.941 \\
          & RF    & \cellcolor[rgb]{ .392,  .749,  .486}0.996 & \cellcolor[rgb]{ .408,  .753,  .498}0.980 & \cellcolor[rgb]{ .404,  .753,  .494}0.985 &       &       & RF    & \cellcolor[rgb]{ .388,  .745,  .482}1.000 & \cellcolor[rgb]{ .388,  .745,  .482}1.000 & \cellcolor[rgb]{ .867,  .941,  .898}0.377 \\
          & SVM   & \cellcolor[rgb]{ .776,  .902,  .816}0.557 & \cellcolor[rgb]{ .753,  .894,  .8}0.581 & \cellcolor[rgb]{ .749,  .894,  .796}0.587 &       &       & SVM   & \cellcolor[rgb]{ .388,  .745,  .482}1.000 & \cellcolor[rgb]{ .388,  .745,  .482}1.000 & \cellcolor[rgb]{ .729,  .886,  .776}0.556 \\
          & XGB   & \cellcolor[rgb]{ .392,  .749,  .486}0.996 & \cellcolor[rgb]{ .569,  .82,  .639}0.793 & \cellcolor[rgb]{ .553,  .812,  .627}0.811 &       &       & XGB   & \cellcolor[rgb]{ .388,  .745,  .482}1.000 & \cellcolor[rgb]{ .388,  .745,  .482}1.000 & \cellcolor[rgb]{ .988,  .988,  1}0.218 \\

\cmidrule{1-5}\cmidrule{7-11}    \end{tabular}}%
  \label{tab:mainresults}%
\end{table}%

\begin{table}[htbp]
  \centering
  \caption{F1 score comparison of window and flow-based approaches.}
    \resizebox{0.46\textwidth}{!}{\begin{tabular}{llrlrlrl}
          &       & \multicolumn{2}{c}{CV} & \multicolumn{2}{c}{Session Test} & \multicolumn{2}{c}{Dataset Test} \\
    Name & ML    & \multicolumn{1}{c}{Window} & \multicolumn{1}{c}{Flow} & \multicolumn{1}{c}{Window} & \multicolumn{1}{c}{Flow} & \multicolumn{1}{c}{Window} & \multicolumn{1}{c}{Flow} \\
    \midrule
    \multirow{8}[2]{*}{ACK} & DT    & \cellcolor[rgb]{ .388,  .745,  .482}1.000 & \multicolumn{1}{r}{\cellcolor[rgb]{ .455,  .773,  .541}0.893} & \cellcolor[rgb]{ .392,  .749,  .486}0.999 & \multicolumn{1}{r}{\cellcolor[rgb]{ .412,  .757,  .502}0.966} & \cellcolor[rgb]{ .443,  .769,  .529}0.912 & \multicolumn{1}{r}{\cellcolor[rgb]{ .933,  .969,  .953}0.113} \\
          & KNN   & \cellcolor[rgb]{ .388,  .745,  .482}1.000 & \multicolumn{1}{r}{\cellcolor[rgb]{ .776,  .902,  .82}0.370} & \cellcolor[rgb]{ .388,  .745,  .482}1.000 & \multicolumn{1}{r}{\cellcolor[rgb]{ .988,  .988,  1}0.022} & \cellcolor[rgb]{ .875,  .941,  .902}0.212 & \multicolumn{1}{r}{\cellcolor[rgb]{ .945,  .973,  .965}0.093} \\
          & LR    & \cellcolor[rgb]{ .388,  .745,  .482}1.000 & \multicolumn{1}{r}{\cellcolor[rgb]{ .455,  .773,  .541}0.894} & \cellcolor[rgb]{ .388,  .745,  .482}1.000 & \multicolumn{1}{r}{\cellcolor[rgb]{ .412,  .757,  .502}0.966} & \cellcolor[rgb]{ .4,  .749,  .49}0.986 & \multicolumn{1}{r}{\cellcolor[rgb]{ .933,  .969,  .953}0.113} \\
          & MLP   & \cellcolor[rgb]{ .388,  .745,  .482}1.000 & \multicolumn{1}{r}{\cellcolor[rgb]{ .455,  .773,  .541}0.895} & \cellcolor[rgb]{ .392,  .749,  .486}0.999 & \multicolumn{1}{r}{\cellcolor[rgb]{ .412,  .757,  .502}0.966} & \cellcolor[rgb]{ .8,  .914,  .839}0.333 & \multicolumn{1}{r}{\cellcolor[rgb]{ .933,  .969,  .953}0.113} \\
          & NB    & \cellcolor[rgb]{ .388,  .745,  .482}1.000 & \multicolumn{1}{r}{\cellcolor[rgb]{ .514,  .796,  .592}0.799} & \cellcolor[rgb]{ .388,  .745,  .482}1.000 & \multicolumn{1}{r}{\cellcolor[rgb]{ .459,  .776,  .545}0.886} & \cellcolor[rgb]{ .875,  .941,  .902}0.212 & \multicolumn{1}{r}{\cellcolor[rgb]{ .933,  .969,  .953}0.115} \\
          & RF    & \cellcolor[rgb]{ .388,  .745,  .482}1.000 & \multicolumn{1}{r}{\cellcolor[rgb]{ .455,  .773,  .541}0.893} & \cellcolor[rgb]{ .392,  .749,  .486}0.999 & \multicolumn{1}{r}{\cellcolor[rgb]{ .412,  .757,  .502}0.966} & \cellcolor[rgb]{ .396,  .749,  .49}0.989 & \multicolumn{1}{r}{\cellcolor[rgb]{ .933,  .969,  .953}0.113} \\
          & SVM   & \cellcolor[rgb]{ .576,  .824,  .647}0.694 & \multicolumn{1}{r}{\cellcolor[rgb]{ .455,  .773,  .541}0.894} & \cellcolor[rgb]{ .859,  .937,  .89}0.235 & \multicolumn{1}{r}{\cellcolor[rgb]{ .412,  .757,  .502}0.966} & \cellcolor[rgb]{ .8,  .914,  .839}0.333 & \multicolumn{1}{r}{\cellcolor[rgb]{ .933,  .969,  .953}0.113} \\
          & XGB   & \cellcolor[rgb]{ .388,  .745,  .482}1.000 & \multicolumn{1}{r}{\cellcolor[rgb]{ .455,  .773,  .541}0.893} & \cellcolor[rgb]{ .388,  .745,  .482}1.000 & \multicolumn{1}{r}{\cellcolor[rgb]{ .412,  .757,  .502}0.967} & \cellcolor[rgb]{ .396,  .749,  .49}0.989 & \multicolumn{1}{r}{\cellcolor[rgb]{ .933,  .969,  .953}0.113} \\
    \midrule
    \multirow{8}[2]{*}{ARP} & DT    & \cellcolor[rgb]{ .396,  .749,  .49}0.994 & \multicolumn{1}{r}{\cellcolor[rgb]{ .651,  .851,  .71}0.659} & \cellcolor[rgb]{ .706,  .875,  .757}0.583 & \multicolumn{1}{r}{\cellcolor[rgb]{ .62,  .839,  .682}0.698} & \cellcolor[rgb]{ .804,  .914,  .843}0.454 & \multicolumn{1}{r}{\cellcolor[rgb]{ .847,  .933,  .878}0.401} \\
          & KNN   & \cellcolor[rgb]{ .447,  .769,  .533}0.925 & \multicolumn{1}{r}{\cellcolor[rgb]{ .667,  .859,  .722}0.636} & \cellcolor[rgb]{ .69,  .867,  .741}0.607 & \multicolumn{1}{r}{\cellcolor[rgb]{ .612,  .835,  .675}0.710} & \cellcolor[rgb]{ .89,  .949,  .918}0.341 & \multicolumn{1}{r}{\cellcolor[rgb]{ .694,  .871,  .749}0.599} \\
          & LR    & \cellcolor[rgb]{ .537,  .808,  .612}0.808 & \multicolumn{1}{r}{\cellcolor[rgb]{ .647,  .851,  .706}0.661} & \cellcolor[rgb]{ .671,  .859,  .725}0.631 & \multicolumn{1}{r}{\cellcolor[rgb]{ .639,  .847,  .698}0.674} & \cellcolor[rgb]{ .718,  .882,  .769}0.567 & \multicolumn{1}{r}{\cellcolor[rgb]{ .894,  .953,  .922}0.336} \\
          & MLP   & \cellcolor[rgb]{ .663,  .859,  .718}0.642 & \multicolumn{1}{r}{\cellcolor[rgb]{ .663,  .859,  .722}0.641} & \cellcolor[rgb]{ .722,  .882,  .773}0.563 & \multicolumn{1}{r}{\cellcolor[rgb]{ .627,  .843,  .69}0.686} & \cellcolor[rgb]{ .867,  .941,  .894}0.373 & \multicolumn{1}{r}{\cellcolor[rgb]{ .573,  .82,  .643}0.759} \\
          & NB    & \cellcolor[rgb]{ .776,  .902,  .816}0.494 & \multicolumn{1}{r}{\cellcolor[rgb]{ .8,  .914,  .835}0.463} & \cellcolor[rgb]{ .839,  .929,  .871}0.411 & \multicolumn{1}{r}{\cellcolor[rgb]{ .824,  .922,  .859}0.428} & \cellcolor[rgb]{ .902,  .953,  .925}0.329 & \multicolumn{1}{r}{\cellcolor[rgb]{ .596,  .831,  .663}0.730} \\
          & RF    & \cellcolor[rgb]{ .392,  .749,  .486}0.997 & \multicolumn{1}{r}{\cellcolor[rgb]{ .631,  .843,  .694}0.682} & \cellcolor[rgb]{ .647,  .851,  .706}0.664 & \multicolumn{1}{r}{\cellcolor[rgb]{ .596,  .831,  .663}0.727} & \cellcolor[rgb]{ .89,  .949,  .914}0.345 & \multicolumn{1}{r}{\cellcolor[rgb]{ .831,  .925,  .863}0.421} \\
          & SVM   & \cellcolor[rgb]{ .812,  .918,  .847}0.447 & \multicolumn{1}{r}{\cellcolor[rgb]{ .867,  .941,  .894}0.375} & \cellcolor[rgb]{ .824,  .922,  .859}0.432 & \multicolumn{1}{r}{\cellcolor[rgb]{ .827,  .925,  .863}0.424} & \cellcolor[rgb]{ .898,  .953,  .922}0.333 & \multicolumn{1}{r}{\cellcolor[rgb]{ .988,  .988,  1}0.211} \\
          & XGB   & \cellcolor[rgb]{ .392,  .749,  .486}0.997 & \multicolumn{1}{r}{\cellcolor[rgb]{ .631,  .847,  .694}0.681} & \cellcolor[rgb]{ .592,  .827,  .659}0.733 & \multicolumn{1}{r}{\cellcolor[rgb]{ .596,  .831,  .663}0.731} & \cellcolor[rgb]{ .886,  .949,  .91}0.349 & \multicolumn{1}{r}{\cellcolor[rgb]{ .855,  .937,  .886}0.389} \\
    \midrule
    \multirow{8}[2]{*}{BF} & DT    & \cellcolor[rgb]{ .392,  .749,  .486}0.999 & \multicolumn{1}{r}{\cellcolor[rgb]{ .502,  .792,  .58}0.851} & \cellcolor[rgb]{ .784,  .906,  .824}0.481 & \multicolumn{1}{r}{\cellcolor[rgb]{ .569,  .82,  .635}0.767} & \cellcolor[rgb]{ .882,  .945,  .91}0.351 & \multicolumn{1}{r}{\cellcolor[rgb]{ .82,  .922,  .855}0.433} \\
          & KNN   & \cellcolor[rgb]{ .408,  .753,  .498}0.979 & \multicolumn{1}{r}{\cellcolor[rgb]{ .557,  .816,  .627}0.781} & \cellcolor[rgb]{ .447,  .769,  .533}0.925 & \multicolumn{1}{r}{\cellcolor[rgb]{ .6,  .831,  .663}0.726} & \cellcolor[rgb]{ .757,  .894,  .8}0.520 & \multicolumn{1}{r}{\cellcolor[rgb]{ .878,  .945,  .906}0.359} \\
          & LR    & \cellcolor[rgb]{ .392,  .749,  .486}0.996 & \multicolumn{1}{r}{\cellcolor[rgb]{ .804,  .914,  .843}0.456} & \cellcolor[rgb]{ .757,  .898,  .8}0.516 & \multicolumn{1}{r}{\cellcolor[rgb]{ .804,  .914,  .839}0.458} & \cellcolor[rgb]{ .843,  .929,  .878}0.402 & \multicolumn{1}{r}{\cellcolor[rgb]{ .91,  .957,  .933}0.316} \\
          & MLP   & \cellcolor[rgb]{ .439,  .765,  .525}0.938 & \multicolumn{1}{r}{\cellcolor[rgb]{ .663,  .859,  .722}0.640} & \cellcolor[rgb]{ .416,  .757,  .506}0.969 & \multicolumn{1}{r}{\cellcolor[rgb]{ .722,  .882,  .769}0.564} & \cellcolor[rgb]{ .78,  .906,  .82}0.486 & \multicolumn{1}{r}{\cellcolor[rgb]{ .816,  .918,  .851}0.443} \\
          & NB    & \cellcolor[rgb]{ .694,  .871,  .745}0.602 & \multicolumn{1}{r}{\cellcolor[rgb]{ .78,  .906,  .82}0.486} & \cellcolor[rgb]{ .694,  .871,  .745}0.601 & \multicolumn{1}{r}{\cellcolor[rgb]{ .765,  .898,  .808}0.506} & \cellcolor[rgb]{ .588,  .827,  .655}0.738 & \multicolumn{1}{r}{\cellcolor[rgb]{ .627,  .843,  .69}0.686} \\
          & RF    & \cellcolor[rgb]{ .388,  .745,  .482}1.000 & \multicolumn{1}{r}{\cellcolor[rgb]{ .478,  .784,  .561}0.883} & \cellcolor[rgb]{ .412,  .757,  .502}0.974 & \multicolumn{1}{r}{\cellcolor[rgb]{ .533,  .804,  .608}0.810} & \cellcolor[rgb]{ .91,  .957,  .933}0.317 & \multicolumn{1}{r}{\cellcolor[rgb]{ .89,  .949,  .914}0.343} \\
          & SVM   & \cellcolor[rgb]{ .412,  .757,  .502}0.973 & \multicolumn{1}{r}{\cellcolor[rgb]{ .788,  .91,  .827}0.474} & \cellcolor[rgb]{ .412,  .757,  .502}0.971 & \multicolumn{1}{r}{\cellcolor[rgb]{ .784,  .906,  .824}0.480} & \cellcolor[rgb]{ .914,  .957,  .933}0.313 & \multicolumn{1}{r}{\cellcolor[rgb]{ .91,  .957,  .933}0.316} \\
          & XGB   & \cellcolor[rgb]{ .392,  .749,  .486}0.998 & \multicolumn{1}{r}{\cellcolor[rgb]{ .471,  .78,  .557}0.892} & \cellcolor[rgb]{ .467,  .78,  .553}0.898 & \multicolumn{1}{r}{\cellcolor[rgb]{ .541,  .808,  .612}0.803} & \cellcolor[rgb]{ .914,  .957,  .933}0.313 & \multicolumn{1}{r}{\cellcolor[rgb]{ .769,  .902,  .812}0.503} \\
    \midrule
    \multirow{8}[2]{*}{HTTP} & DT    & \cellcolor[rgb]{ .388,  .745,  .482}1.000 & \multicolumn{1}{r}{\cellcolor[rgb]{ .42,  .757,  .51}0.962} & \cellcolor[rgb]{ .396,  .749,  .49}0.993 & \multicolumn{1}{r}{\cellcolor[rgb]{ .439,  .769,  .529}0.933} & \cellcolor[rgb]{ .388,  .745,  .482}1.000 & \multicolumn{1}{r}{\cellcolor[rgb]{ .631,  .847,  .694}0.681} \\
          & KNN   & \cellcolor[rgb]{ .392,  .749,  .486}0.998 & \multicolumn{1}{r}{\cellcolor[rgb]{ .475,  .78,  .557}0.890} & \cellcolor[rgb]{ .471,  .78,  .553}0.893 & \multicolumn{1}{r}{\cellcolor[rgb]{ .514,  .796,  .588}0.840} & \cellcolor[rgb]{ .882,  .945,  .906}0.355 & \multicolumn{1}{r}{\cellcolor[rgb]{ .722,  .882,  .769}0.566} \\
          & LR    & \cellcolor[rgb]{ .396,  .749,  .49}0.994 & \multicolumn{1}{r}{\cellcolor[rgb]{ .439,  .769,  .529}0.934} & \cellcolor[rgb]{ .396,  .749,  .49}0.993 & \multicolumn{1}{r}{\cellcolor[rgb]{ .451,  .773,  .537}0.919} & \cellcolor[rgb]{ .914,  .961,  .937}0.310 & \multicolumn{1}{r}{\cellcolor[rgb]{ .451,  .773,  .537}0.920} \\
          & MLP   & \cellcolor[rgb]{ .4,  .753,  .494}0.985 & \multicolumn{1}{r}{\cellcolor[rgb]{ .573,  .82,  .639}0.761} & \cellcolor[rgb]{ .443,  .769,  .529}0.930 & \multicolumn{1}{r}{\cellcolor[rgb]{ .549,  .812,  .62}0.791} & \cellcolor[rgb]{ .855,  .937,  .886}0.388 & \multicolumn{1}{r}{\cellcolor[rgb]{ .878,  .945,  .906}0.359} \\
          & NB    & \cellcolor[rgb]{ .671,  .859,  .725}0.632 & \multicolumn{1}{r}{\cellcolor[rgb]{ .541,  .808,  .612}0.804} & \cellcolor[rgb]{ .808,  .918,  .847}0.450 & \multicolumn{1}{r}{\cellcolor[rgb]{ .584,  .824,  .651}0.747} & \cellcolor[rgb]{ .878,  .945,  .906}0.356 & \multicolumn{1}{r}{\cellcolor[rgb]{ .412,  .757,  .502}0.971} \\
          & RF    & \cellcolor[rgb]{ .388,  .745,  .482}1.000 & \multicolumn{1}{r}{\cellcolor[rgb]{ .412,  .757,  .502}0.971} & \cellcolor[rgb]{ .392,  .749,  .486}0.996 & \multicolumn{1}{r}{\cellcolor[rgb]{ .42,  .757,  .51}0.962} & \cellcolor[rgb]{ .388,  .745,  .482}1.000 & \multicolumn{1}{r}{\cellcolor[rgb]{ .698,  .871,  .749}0.596} \\
          & SVM   & \cellcolor[rgb]{ .776,  .906,  .82}0.490 & \multicolumn{1}{r}{\cellcolor[rgb]{ .851,  .933,  .882}0.392} & \cellcolor[rgb]{ .78,  .906,  .82}0.486 & \multicolumn{1}{r}{\cellcolor[rgb]{ .871,  .941,  .898}0.369} & \cellcolor[rgb]{ .898,  .953,  .922}0.333 & \multicolumn{1}{r}{\cellcolor[rgb]{ .8,  .914,  .839}0.462} \\
          & XGB   & \cellcolor[rgb]{ .388,  .745,  .482}1.000 & \multicolumn{1}{r}{\cellcolor[rgb]{ .416,  .757,  .506}0.968} & \cellcolor[rgb]{ .392,  .749,  .486}0.996 & \multicolumn{1}{r}{\cellcolor[rgb]{ .435,  .765,  .522}0.941} & \cellcolor[rgb]{ .388,  .745,  .482}1.000 & \multicolumn{1}{r}{\cellcolor[rgb]{ .655,  .855,  .714}0.652} \\
    \midrule
    \multirow{8}[2]{*}{PS} & DT    & \cellcolor[rgb]{ .388,  .745,  .482}1.000 & \cellcolor[rgb]{ .396,  .749,  .49}0.995 & \cellcolor[rgb]{ .388,  .745,  .482}1.000 & \cellcolor[rgb]{ .396,  .749,  .49}0.994 & \cellcolor[rgb]{ .506,  .792,  .584}0.896 & \cellcolor[rgb]{ .447,  .769,  .533}0.949 \\
          & KNN   & \cellcolor[rgb]{ .388,  .745,  .482}1.000 & \cellcolor[rgb]{ .427,  .761,  .518}0.966 & \cellcolor[rgb]{ .388,  .745,  .482}1.000 & \cellcolor[rgb]{ .486,  .788,  .569}0.912 & \cellcolor[rgb]{ .392,  .749,  .486}0.997 & \cellcolor[rgb]{ .729,  .882,  .776}0.696 \\
          & LR    & \cellcolor[rgb]{ .392,  .749,  .486}0.999 & \cellcolor[rgb]{ .4,  .753,  .494}0.990 & \cellcolor[rgb]{ .388,  .745,  .482}1.000 & \cellcolor[rgb]{ .392,  .749,  .486}0.999 & \cellcolor[rgb]{ .392,  .749,  .486}0.997 & \cellcolor[rgb]{ .396,  .749,  .49}0.994 \\
          & MLP   & \cellcolor[rgb]{ .388,  .745,  .482}1.000 & \cellcolor[rgb]{ .475,  .78,  .557}0.924 & \cellcolor[rgb]{ .388,  .745,  .482}1.000 & \cellcolor[rgb]{ .51,  .796,  .588}0.891 & \cellcolor[rgb]{ .392,  .749,  .486}0.997 & \cellcolor[rgb]{ .741,  .89,  .788}0.685 \\
          & NB    & \cellcolor[rgb]{ .42,  .761,  .51}0.972 & \cellcolor[rgb]{ .435,  .765,  .525}0.959 & \cellcolor[rgb]{ .431,  .765,  .522}0.963 & \cellcolor[rgb]{ .427,  .761,  .518}0.966 & \cellcolor[rgb]{ .424,  .761,  .514}0.971 & \cellcolor[rgb]{ .682,  .867,  .733}0.739 \\
          & RF    & \cellcolor[rgb]{ .392,  .749,  .486}0.999 & \cellcolor[rgb]{ .388,  .745,  .482}1.000 & \cellcolor[rgb]{ .388,  .745,  .482}1.000 & \cellcolor[rgb]{ .392,  .749,  .486}0.998 & \cellcolor[rgb]{ .545,  .808,  .616}0.862 & \cellcolor[rgb]{ .412,  .757,  .502}0.980 \\
          & SVM   & \cellcolor[rgb]{ .976,  .984,  .992}0.472 & \cellcolor[rgb]{ .929,  .965,  .949}0.516 & \cellcolor[rgb]{ .827,  .925,  .859}0.608 & \cellcolor[rgb]{ .976,  .984,  .988}0.475 & \cellcolor[rgb]{ .988,  .988,  1}0.461 & \cellcolor[rgb]{ .894,  .953,  .922}0.546 \\
          & XGB   & \cellcolor[rgb]{ .392,  .749,  .486}0.999 & \cellcolor[rgb]{ .388,  .745,  .482}1.000 & \cellcolor[rgb]{ .388,  .745,  .482}1.000 & \cellcolor[rgb]{ .392,  .749,  .486}0.997 & \cellcolor[rgb]{ .451,  .773,  .537}0.945 & \cellcolor[rgb]{ .412,  .757,  .502}0.980 \\
    \midrule
    \multirow{8}[2]{*}{SYN} & DT    & \cellcolor[rgb]{ .388,  .745,  .482}1.000 & \cellcolor[rgb]{ .404,  .753,  .498}0.977 & \cellcolor[rgb]{ .392,  .749,  .486}0.999 & \cellcolor[rgb]{ .396,  .749,  .49}0.988 & \cellcolor[rgb]{ .431,  .765,  .522}0.930 & \cellcolor[rgb]{ .914,  .961,  .937}0.127 \\
          & KNN   & \cellcolor[rgb]{ .392,  .749,  .486}0.999 & \cellcolor[rgb]{ .431,  .765,  .518}0.932 & \cellcolor[rgb]{ .392,  .749,  .486}0.997 & \cellcolor[rgb]{ .443,  .769,  .529}0.913 & \cellcolor[rgb]{ .812,  .918,  .847}0.298 & \cellcolor[rgb]{ .769,  .902,  .812}0.367 \\
          & LR    & \cellcolor[rgb]{ .388,  .745,  .482}1.000 & \cellcolor[rgb]{ .416,  .757,  .506}0.960 & \cellcolor[rgb]{ .396,  .749,  .486}0.993 & \cellcolor[rgb]{ .451,  .773,  .537}0.901 & \cellcolor[rgb]{ .4,  .753,  .494}0.981 & \cellcolor[rgb]{ .737,  .886,  .784}0.422 \\
          & MLP   & \cellcolor[rgb]{ .392,  .749,  .486}0.999 & \cellcolor[rgb]{ .404,  .753,  .498}0.975 & \cellcolor[rgb]{ .392,  .749,  .486}0.997 & \cellcolor[rgb]{ .443,  .769,  .529}0.913 & \cellcolor[rgb]{ .792,  .91,  .831}0.333 & \cellcolor[rgb]{ .867,  .941,  .894}0.208 \\
          & NB    & \cellcolor[rgb]{ .396,  .749,  .486}0.993 & \cellcolor[rgb]{ .541,  .808,  .616}0.746 & \cellcolor[rgb]{ .396,  .749,  .486}0.993 & \cellcolor[rgb]{ .49,  .788,  .573}0.832 & \cellcolor[rgb]{ .392,  .749,  .486}0.998 & \cellcolor[rgb]{ .769,  .898,  .808}0.372 \\
          & RF    & \cellcolor[rgb]{ .388,  .745,  .482}1.000 & \cellcolor[rgb]{ .396,  .749,  .49}0.992 & \cellcolor[rgb]{ .392,  .749,  .486}0.999 & \cellcolor[rgb]{ .396,  .749,  .49}0.992 & \cellcolor[rgb]{ .388,  .745,  .482}1.000 & \cellcolor[rgb]{ .988,  .988,  1}0.000 \\
          & SVM   & \cellcolor[rgb]{ .424,  .761,  .514}0.945 & \cellcolor[rgb]{ .69,  .871,  .745}0.499 & \cellcolor[rgb]{ .42,  .761,  .51}0.951 & \cellcolor[rgb]{ .69,  .871,  .745}0.498 & \cellcolor[rgb]{ .792,  .91,  .831}0.333 & \cellcolor[rgb]{ .69,  .867,  .741}0.500 \\
          & XGB   & \cellcolor[rgb]{ .388,  .745,  .482}1.000 & \cellcolor[rgb]{ .404,  .753,  .498}0.977 & \cellcolor[rgb]{ .392,  .749,  .486}0.999 & \cellcolor[rgb]{ .4,  .749,  .49}0.986 & \cellcolor[rgb]{ .388,  .745,  .482}1.000 & \cellcolor[rgb]{ .925,  .965,  .945}0.110 \\
    \midrule
    \multirow{8}[2]{*}{UDP} & DT    & \cellcolor[rgb]{ .388,  .745,  .482}1.000 & \cellcolor[rgb]{ .388,  .745,  .482}1.000 & \cellcolor[rgb]{ .459,  .776,  .545}0.884 & \cellcolor[rgb]{ .388,  .745,  .482}1.000 & \cellcolor[rgb]{ .776,  .902,  .816}0.359 & \cellcolor[rgb]{ .988,  .988,  1}0.001 \\
          & KNN   & \cellcolor[rgb]{ .388,  .745,  .482}1.000 & \cellcolor[rgb]{ .435,  .765,  .522}0.926 & \cellcolor[rgb]{ .388,  .745,  .482}1.000 & \cellcolor[rgb]{ .725,  .882,  .773}0.441 & \cellcolor[rgb]{ .859,  .937,  .89}0.218 & \cellcolor[rgb]{ .988,  .988,  1}0.001 \\
          & LR    & \cellcolor[rgb]{ .388,  .745,  .482}1.000 & \cellcolor[rgb]{ .404,  .753,  .498}0.975 & \cellcolor[rgb]{ .388,  .745,  .482}1.000 & \cellcolor[rgb]{ .388,  .745,  .482}1.000 & \cellcolor[rgb]{ .859,  .937,  .89}0.218 & \cellcolor[rgb]{ .988,  .988,  1}0.001 \\
          & MLP   & \cellcolor[rgb]{ .392,  .749,  .486}0.999 & \cellcolor[rgb]{ .424,  .761,  .514}0.945 & \cellcolor[rgb]{ .388,  .745,  .482}1.000 & \cellcolor[rgb]{ .729,  .882,  .776}0.437 & \cellcolor[rgb]{ .502,  .792,  .58}0.812 & \cellcolor[rgb]{ .988,  .988,  1}0.001 \\
          & NB    & \cellcolor[rgb]{ .588,  .827,  .655}0.669 & \cellcolor[rgb]{ .612,  .839,  .678}0.628 & \cellcolor[rgb]{ .4,  .753,  .494}0.983 & \cellcolor[rgb]{ .576,  .824,  .647}0.687 & \cellcolor[rgb]{ .427,  .761,  .514}0.941 & \cellcolor[rgb]{ .988,  .988,  1}0.001 \\
          & RF    & \cellcolor[rgb]{ .388,  .745,  .482}1.000 & \cellcolor[rgb]{ .388,  .745,  .482}1.000 & \cellcolor[rgb]{ .388,  .745,  .482}1.000 & \cellcolor[rgb]{ .388,  .745,  .482}1.000 & \cellcolor[rgb]{ .765,  .898,  .808}0.377 & \cellcolor[rgb]{ .988,  .988,  1}0.001 \\
          & SVM   & \cellcolor[rgb]{ .388,  .745,  .482}1.000 & \cellcolor[rgb]{ .694,  .871,  .745}0.493 & \cellcolor[rgb]{ .388,  .745,  .482}1.000 & \cellcolor[rgb]{ .816,  .922,  .851}0.289 & \cellcolor[rgb]{ .655,  .855,  .714}0.556 & \cellcolor[rgb]{ .988,  .988,  1}0.001 \\
          & XGB   & \cellcolor[rgb]{ .388,  .745,  .482}1.000 & \cellcolor[rgb]{ .396,  .749,  .49}0.990 & \cellcolor[rgb]{ .388,  .745,  .482}1.000 & \cellcolor[rgb]{ .388,  .745,  .482}1.000 & \cellcolor[rgb]{ .859,  .937,  .89}0.218 & \cellcolor[rgb]{ .988,  .988,  1}0.001 \\
    \bottomrule
    \end{tabular}}%
  \label{tab:compare2}%
\end{table}%

\subsection{Comparison with Flow-Based Features}\label{subPerformance}

Next, we compare models trained using our window-based features against those trained using flow-based features, the latter being more typical 
in the existing attack detection literature. In order to ensure a fair comparison, we use the same raw data and apply the same feature selection steps. 

Table~\ref{tab:compare2} shows the results. For cross-validation, both approaches achieve near-perfect scores in at least one ML model for UDP, PS, and SYN attacks. However, the window-based approach outperforms the flow-based approach, 
especially for the ARPS, ACK and BF attacks.

When evaluated using an independent session, both methods achieve near-perfect scores in at least one model for UDPF, PS, and SYNF attacks. The window-based approach exhibits superior performance for ACKF, BF, and HTTPF attacks, with approximately 3, 16, and 6 percentage point differences, respectively. Scores for the ARPS attack were almost identical between the two methods.

When evaluated using an independent dataset, the flow-based approach fails to detect ACKF, SYNF, and UDPF attacks, while the window-based approach achieves scores above 0.94 for at least one ML model for all three attacks. Both approaches demonstrate similar performance in PS attacks, while the window-based approach outperforms the flow-based approach for ACKF, SYNF, and UDPF attacks.

\subsection{Feature Effectiveness} \label{FeatureEffectiveness}

These results strongly suggest that a window-based approach is beneficial in terms of both model performance and generality. To further investigate this, we analyze the behavior of the trained models using an explainable AI technique, SHapley Additive exPlanations (SHAP)~\cite{lundberg2017unified,molnar2020interpretable}, which enables us to understand how input features contribute to individual predictions.

In the following analysis, we apply SHAP to interpret the decisions of the single best-performing machine learning model identified for each attack type. This allows us to identify which features are most critical for achieving high detection performance, offering deeper insight into how different types of attacks are distinguished by the model.

\subsubsection{Flood attacks}

First we consider the four flood attacks: 
ACK, HTTP, SYN, and UDP floods.

\begin{figure}[htbp]
	
	\subfloat[\label{fig:CACK1}\centering Window-Based XGB.]{{\includegraphics[width=37mm]{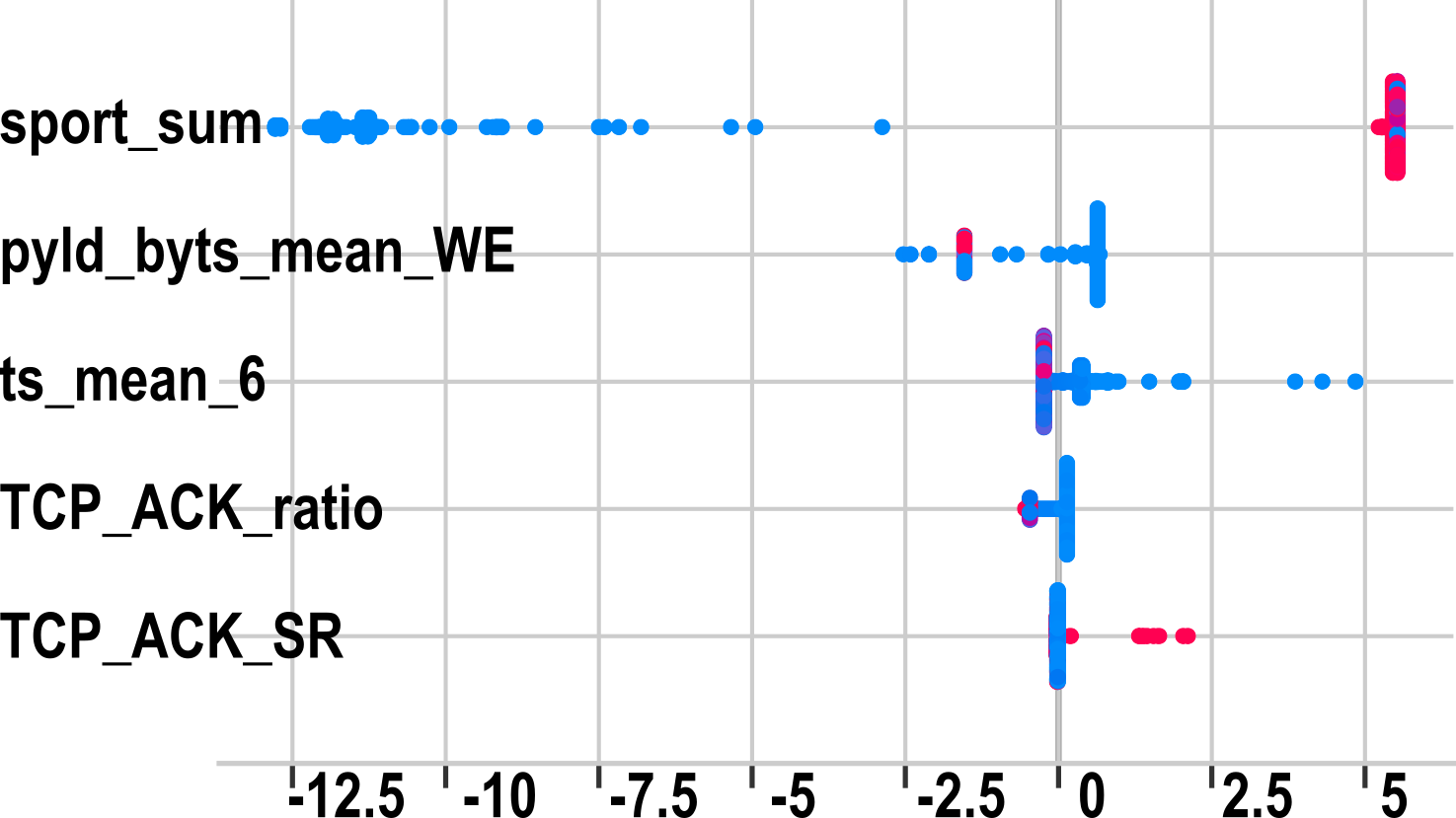}}} \hfill 
	\subfloat[\label{fig:CACK2}\centering  Flow-Based XGB.]{{\includegraphics[width=45mm]{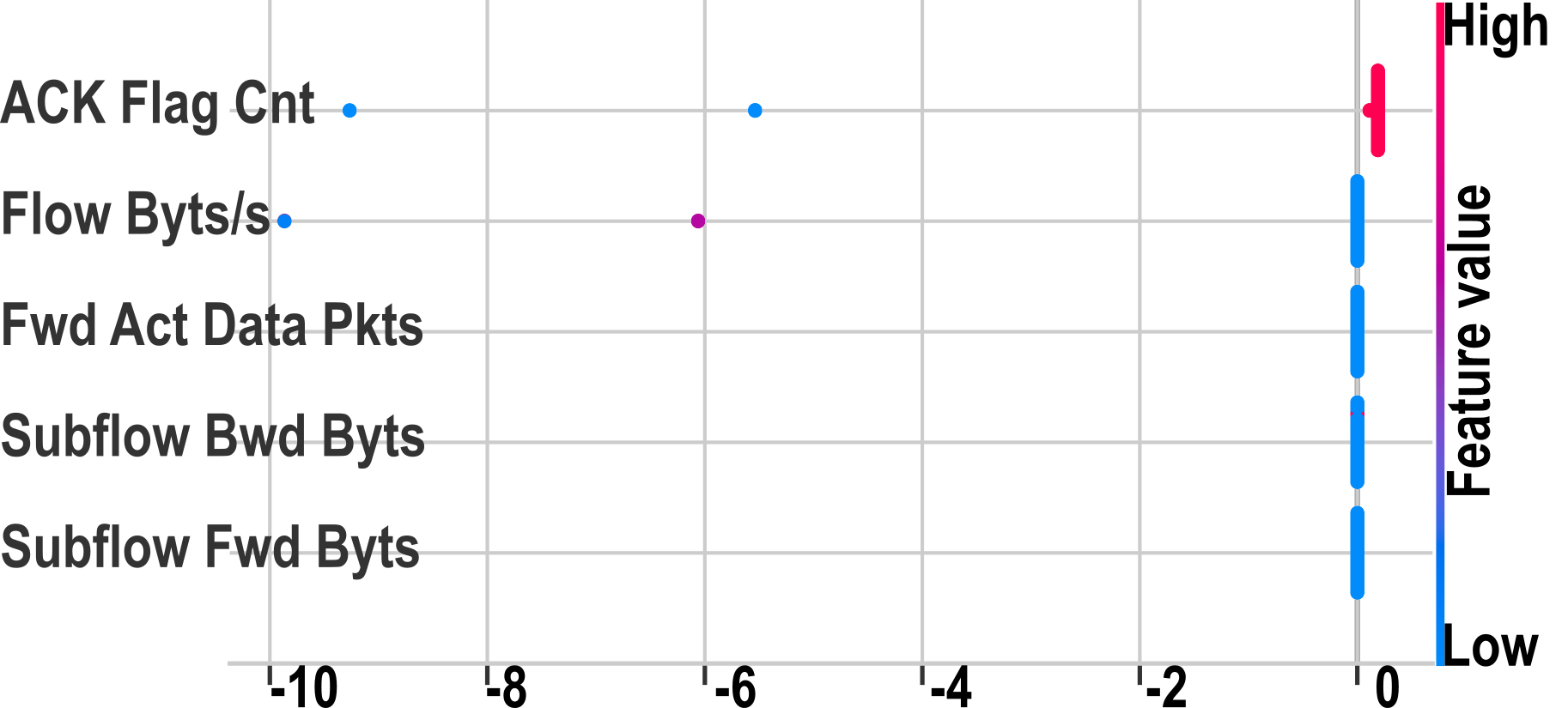}}}\\\caption{Comparison of window and flow-based features for the ACK attack, showing the effect of different features with their importance. Each point in this SHAP plot represents a sample in the dataset. On the y-axis, the attributes are listed in descending order of importance from top to bottom (with a maximum of 20 features shown per plot, otherwise all features used are shown). The x-axis shows the SHAP value: a high value indicates that it contributes to the attack class, while a low value contributes to the benign class. The colour transition indicates whether the value of the feature is high (red) or low (blue).}
	\label{fig:CACK0}	
\end{figure}

\paragraph{ACK flood (ACKF)}
For this attack, we see that both types of feature support useful models for the first two evaluation scenarios, but only window-based features support models that generalize to an independent dataset.
Fig.~\ref{fig:CACK1} shows that the most important feature in the window approach is the \textit{sport\_sum}. While this value is low in benign samples, it is usually high in attack data. In the case of ACKF attack, this value gives the idea that the source port diversity is high. We see that \textit{payload\_bytes\_mean\_WE}  and \textit{ts\_mean\_6}  features also play an active role due to the relatively small size of the packet flow in a short time in the attack situation. \textit{TCP\_ACK\_SR} and \textit{TCP\_ACK\_ratio} are also important, because they give statistics about the ACK flag. In the flow approach, two prominent features stand out: \textit{Flow Byts/s}\footnote{Feature list:~\href{https://github.com/ahlashkari/CICFlowMeter/blob/master/ReadMe.txt}{github.com/ahlashkari/CICFlowMeter/blob/master/ReadMe.txt} } and \textit{ACK Flag Cnt}. This is expected as an ACKF attack involves an abundance of ACK flags and relatively smaller data flow in terms of size. However, relying on ACK flag packets or flow size alone may not adequately indicate an attack's presence. This limitation could be why using these features restricts the model's applicability when tested on diverse datasets.

\begin{figure}[htbp]
	
	\begin{tabular}{@{}cc@{}}
		\subfloat[\label{fig:HTTPC1}\centering WB XGB.]{
			\raisebox{-\height}{\includegraphics[width=0.21\textwidth]{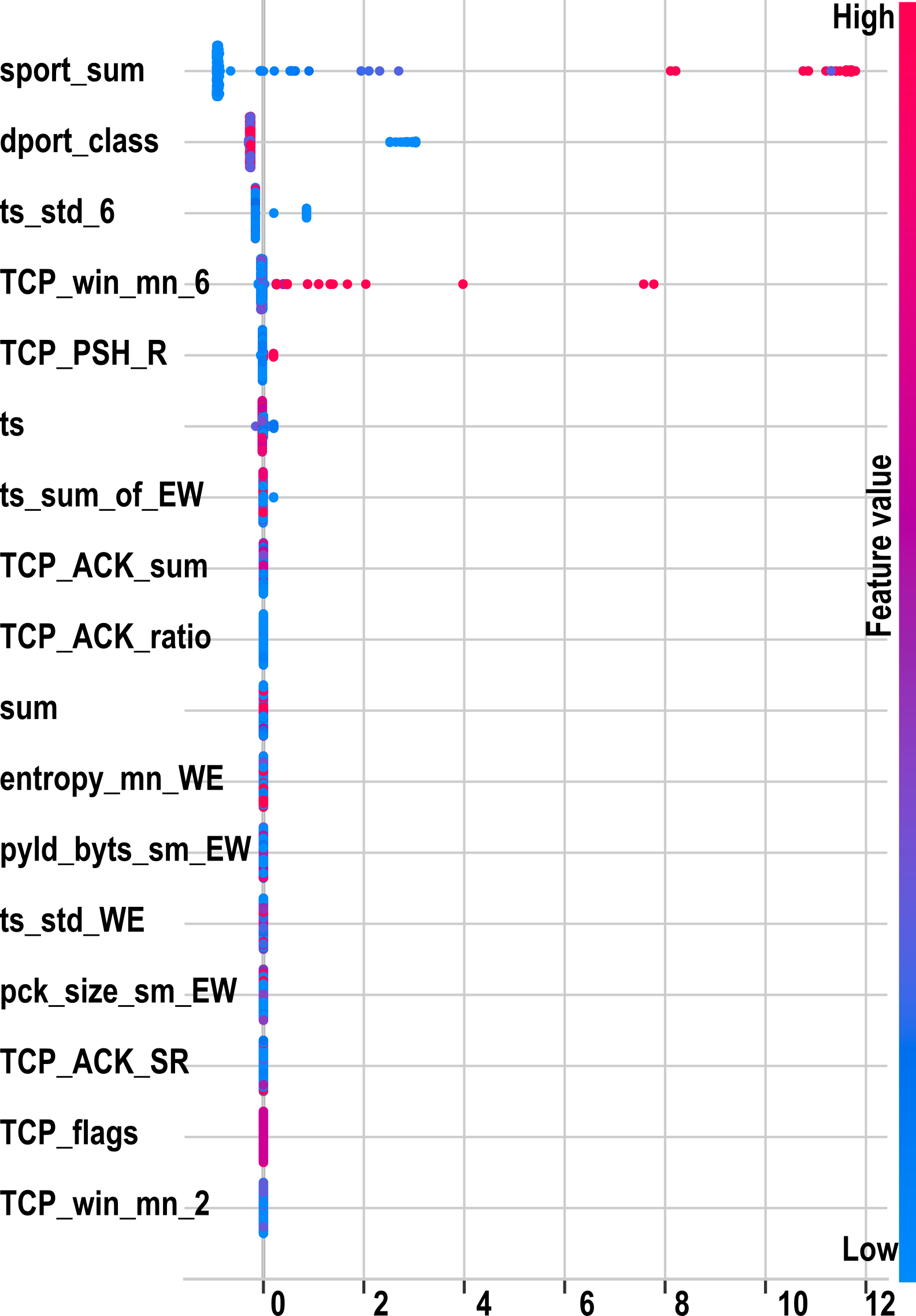}}} & 
		\begin{tabular}[t]{@{}cc@{}}
			\subfloat[\label{fig:HTTPC3}\centering FB NB.]{\raisebox{-\height}{\includegraphics[width=0.24\textwidth]{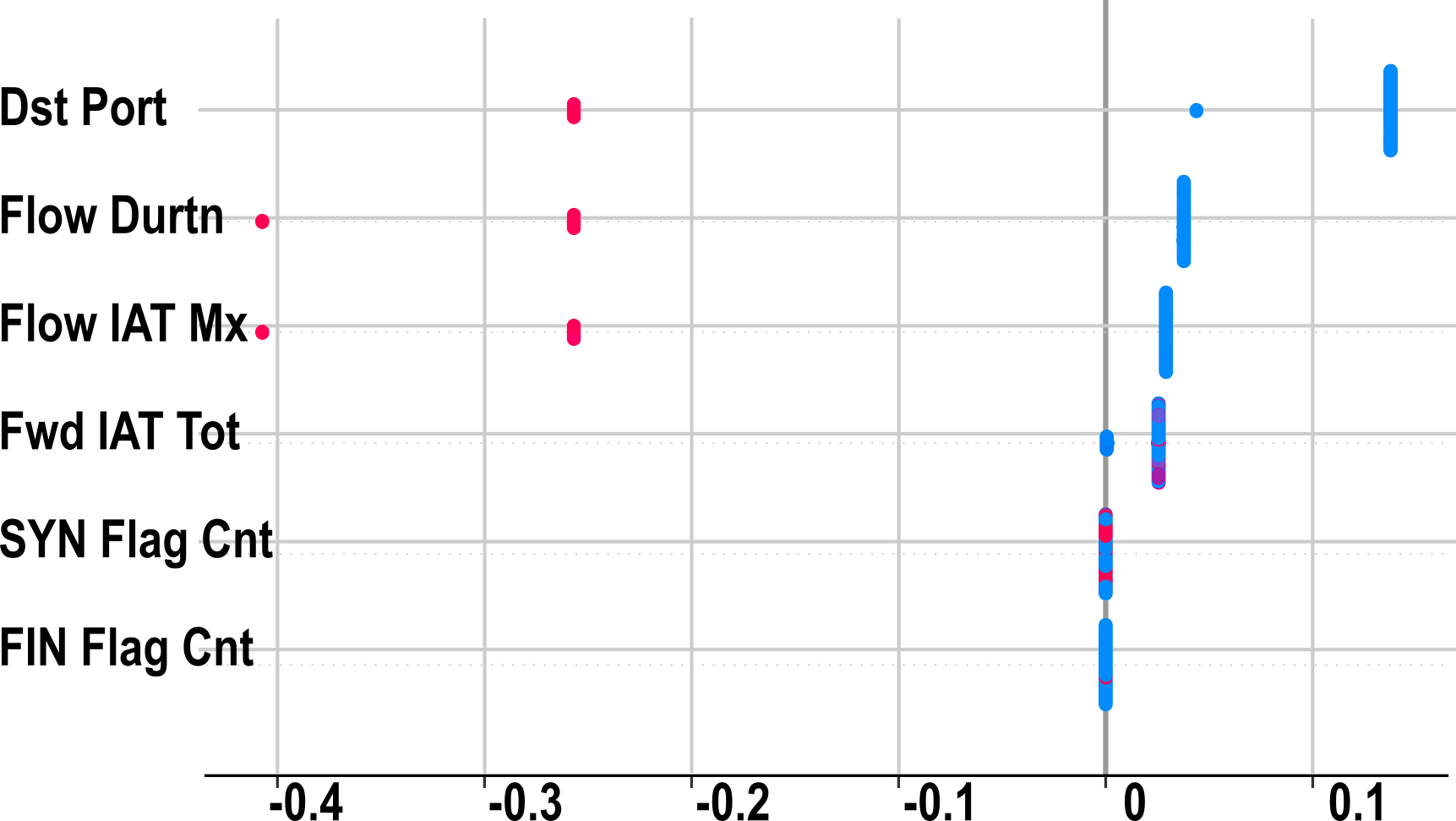}}} & \\[-0.30cm]
			\subfloat[\label{fig:HTTPC2}\centering FB XGB.]{\raisebox{-\height}{\includegraphics[width=0.24\textwidth]{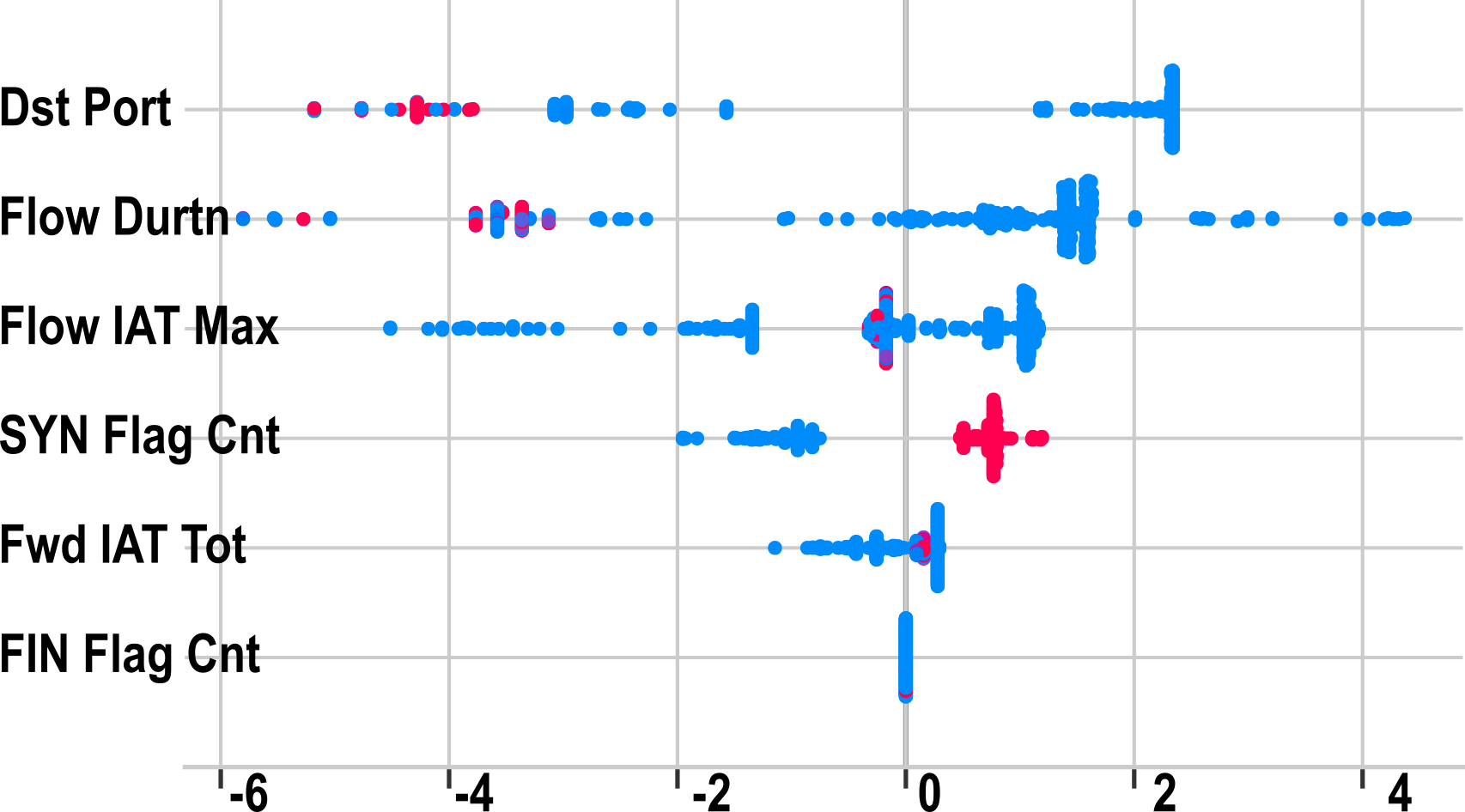}}} & 
		\end{tabular}
	\end{tabular}
	\caption{Comparison of window (WB) and flow (FB) models for HTTP attack.}
	\label{fig:HTTPC0}
\end{figure}

\paragraph{HTTP flood (HTTPF)}
Both feature approaches support the training of discriminative models, though the models build from window-based features are slightly better. Interestingly, ML model type is important here: for window features, decision tree models generalize well to an independent dataset, whereas for flow features, it is the the NB and LR models that generalize. Taking this into account, Fig.~\ref{fig:HTTPC0} shows the SHAP analysis of the XGB and NB models. 
For window features (Fig.~\ref{fig:HTTPC1}), the two most important features are related to port numbers. The \textit{sport\_sum} feature shows a high port diversity in the attack data, while the \textit{dest\_port\_class} is varied for benign and uniform for attacks. Apart from this, another distinguishing feature is related to TCP-window size. This value appears to be higher in the attack data. For flow features (Fig.~\ref{fig:HTTPC2}), it is similarly observed that the most important characteristic is the destination port. While the normal data shows variation, the port number is stable in the attack data. Another important feature is the flow time, which can be explained by HTTPF attack-generating flows tending to be shorter. When we compare with Fig.~\ref{fig:HTTPC3}, we see that similar features are also prominent. However, while the flag-related feature creates a clear distinction in XGB, it is not noticeable in the NB model. When evaluated on an independent dataset, we believe the main reason for the failure of the flow-based XGB model is that this flag feature causes over-fitting.

\begin{figure}[htbp]
	\centering
	
	\subfloat[\label{fig:SYNC1}\centering Window-Based XGB.]{\includegraphics[width=36mm]{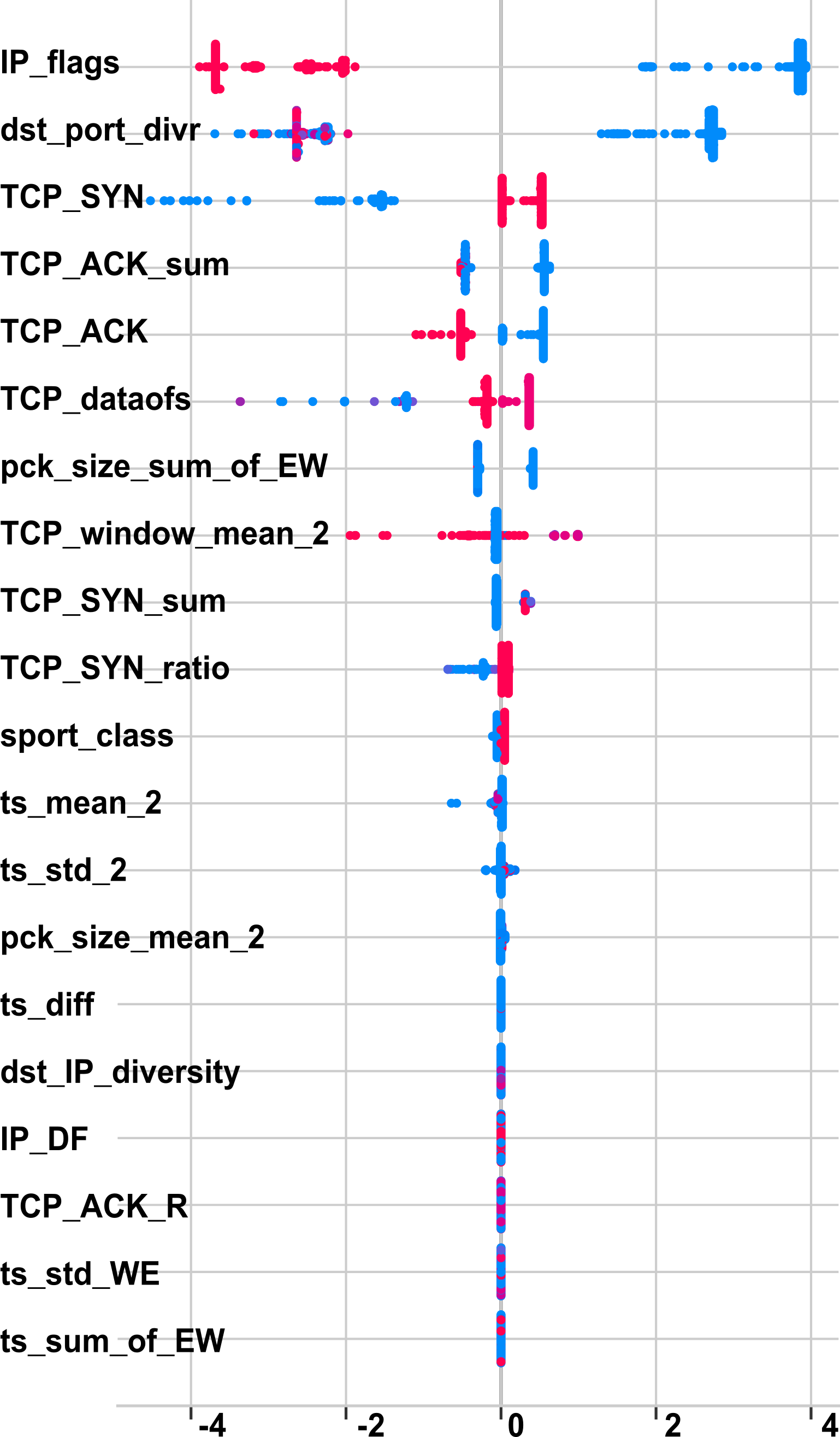}\hfill}\hspace{3mm}
	\subfloat[\label{fig:SYNC2}\centering Flow-Based XGB.]{\includegraphics[width=43mm]{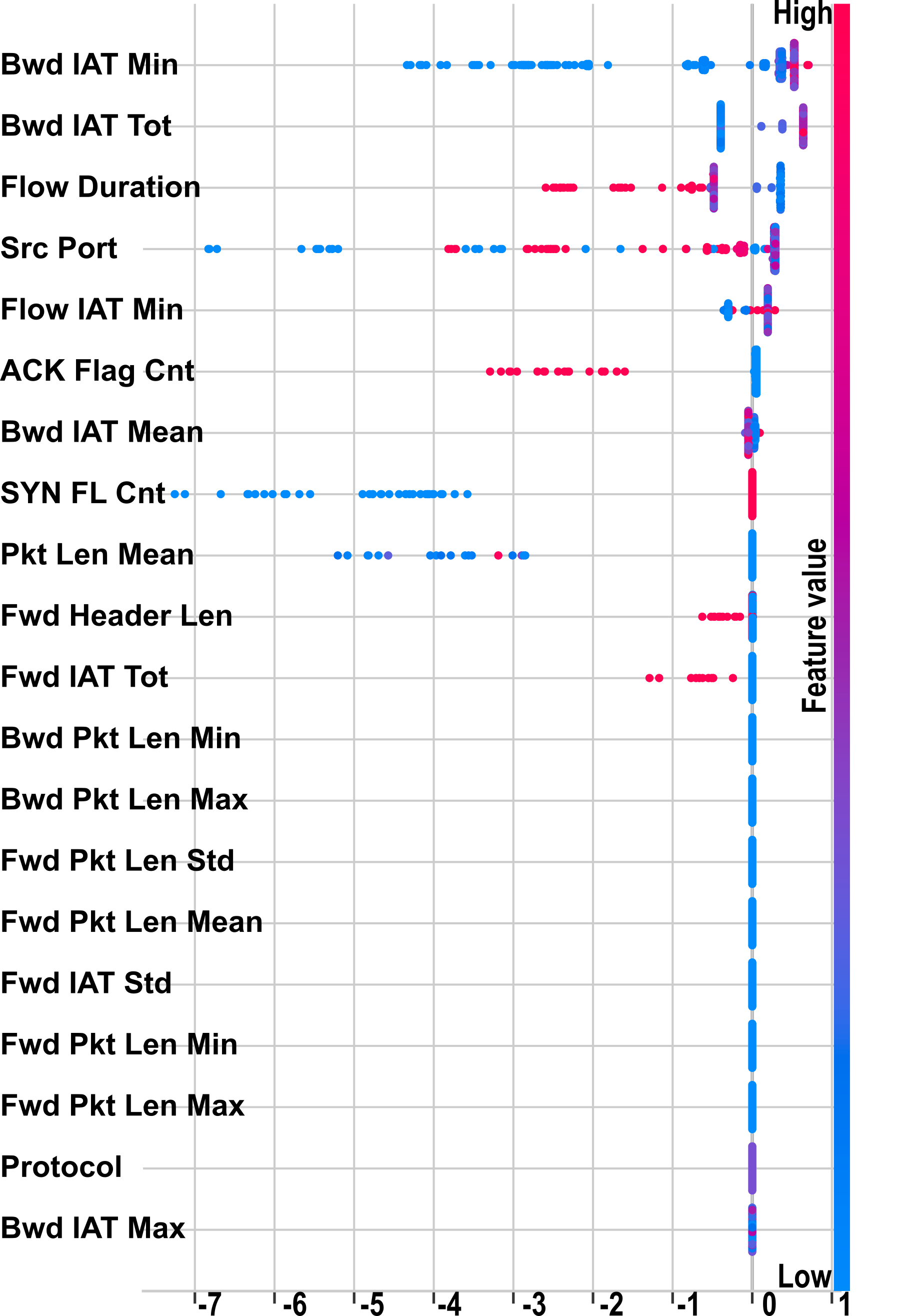}}\hfill
	\caption{Comparison of window (WB) and flow (FB) models for SYN attack.}
	\label{fig:SYNC0}
\end{figure}

\paragraph{SYN flood (SYNF)}
For the SYNF attack 
both approaches are 
successful in the first two evaluation conditions, but only the window approach generalizes to an independent dataset. In Fig.~\ref{fig:SYNC0}  the XGB model for both approaches is analysed with the SHAP plot. In Fig.~\ref{fig:SYNC1}, we can see that the variety of ports 
provides good discrimination. This is likely due to the fact that this attack targets a specific port. On the other hand, it can be seen that flag statistics features are concentrated. One of the biggest effects of SYN attacks on the network is the anomaly in flags. In this context, it is quite reasonable that features related to SYN and ACK flags are particularly important. 
While the \textit{IP\_flag} feature is not expected to make a significant contribution to the classification between the two groups, it was observed as a distinguishing feature in the model. When Fig.~\ref{fig:SYNC2} is analysed, we can see that the inter-arrival time (IAT) features stand out and make a significant distinction. Although they have a lower importance score, flow time, port number, flag and size statistics (such as \textit{Flow Duration, Src Port, SYN Flag Cnt, ACK Flag Cnt, Fwd Pkt Len Mean, Fwd Pkt Len Std, Bwd Pkt Len Max} etc.) are important discriminating features. Considering the attack characteristics, we expect significant success by utilizing these features alone. However, the overfitting of the model due to the prominence of IAT features overshadows the contributions of these features. The potential drawbacks associated with IAT features are outlined below. These features can lead to high-dimensional data sets, increasing model complexity and allowing patterns in the noise to cause overfitting. In addition, not all IAT-based features contain the same information, potentially introducing redundancy and irrelevance, further confounding the model. Sparse data distributions can result, with more features than meaningful data instances, leading to poor performance on unseen data. Complex temporal patterns captured by IAT-based features may struggle to generalize effectively, and insufficient data can exacerbate the risk of overfitting by preventing accurate pattern learning and signal/noise discrimination.

\paragraph{UDP flood (UDPF)}

\begin{figure}[htbp]
	\centering
	
	\subfloat[\label{fig:UDPC1}\centering WB MLP.]{\includegraphics[width=24mm]{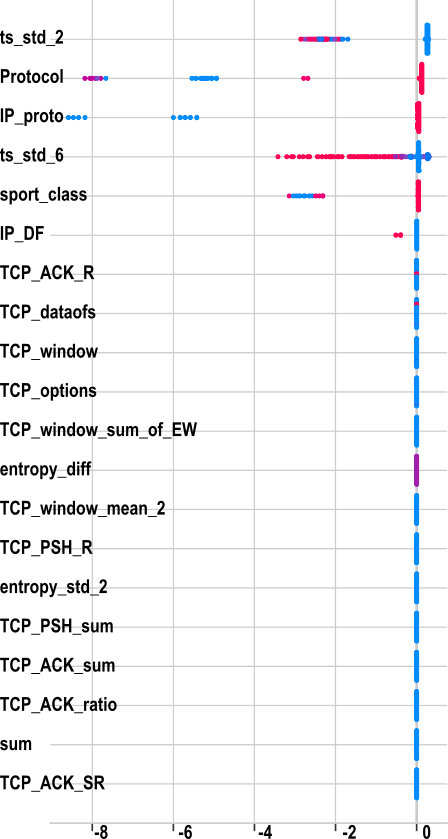}\hfill}\hspace{0.1mm}
	\subfloat[\label{fig:UDPC2}\centering FB XGB.]{\includegraphics[width=25mm]{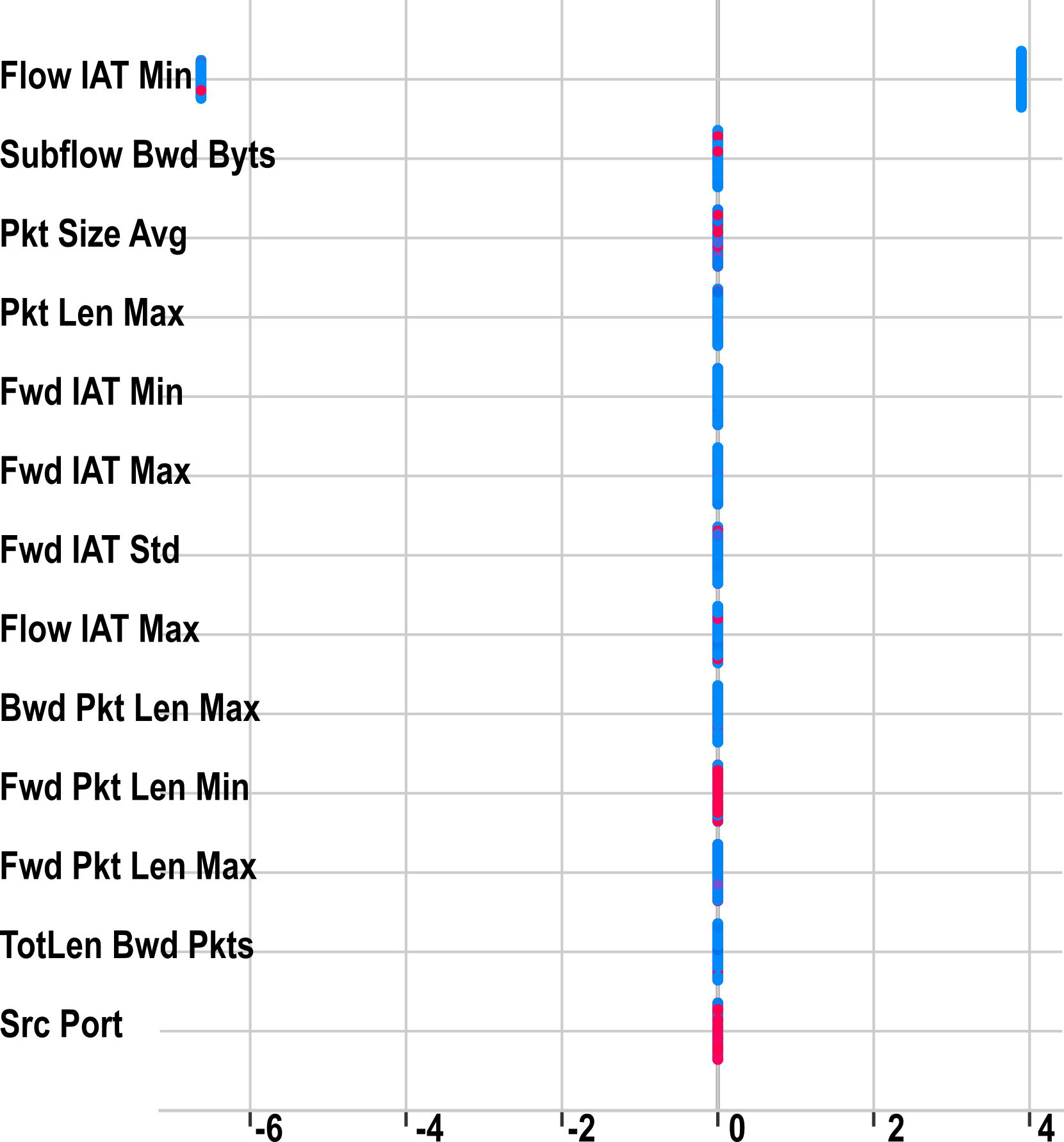}}\hspace{0.1mm}
	\subfloat[\label{fig:UDPC3}\centering WB NB.]{\includegraphics[width=26mm]{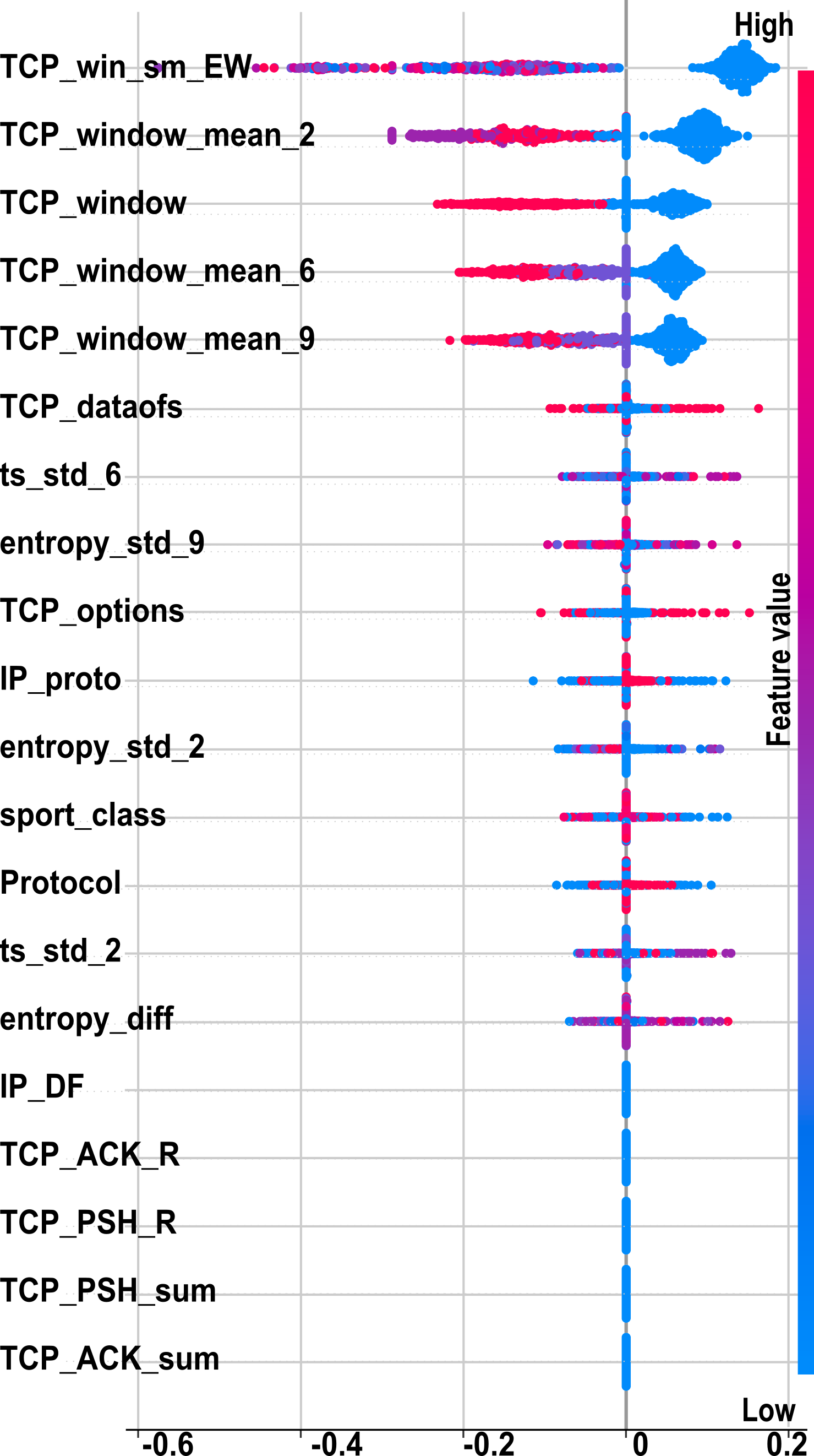}}
	\caption{Comparison of window (WB) and flow (FB) models for UDP attack.}
	
	\label{fig:UDPC0}
\end{figure}

Only the NB and MLP models in the window approach achieved significant success when evaluated on an independent dataset. The SHAP plots for these models are presented in Fig.~\ref{fig:UDPC0}. 
For the flow models, Fig.~\ref{fig:UDPC2} illustrates the prominence of a solitary feature, denoted as \textit{Flow IAT Min} which pertains to the minimum time interval between packets within a flow. The clear dominance of this particular feature triggers concerns about overfitting which implies an excessive dependence on this attribute, hindering the model's ability to generalize effectively to other datasets. Examining Fig.~\ref{fig:UDPC1} and Fig.~\ref{fig:UDPC3}, it can be seen that features that focus on the time between packets cause separation in both models (\textit{ts\_std\_6, ts\_std\_2}). However, protocol- and port-centric features (\textit{IP\_proto, Protocol, sport\_class}) are also prominent. These features are quite reasonable for this attack's detection.
Apart from these, the use of TCP-based features in the model is  significant. The TCP header has a much more complex structure than the UDP header and therefore contributed more to the feature pool. However, since all attacks in this dataset use the UDP protocol, even the fact that a packet is TCP is a serious discriminator (if it is TCP, it is benign). In an experiment using a dataset with a predominance of UDP-based benign packets, TCP-based characteristics would not be expected to be so dominant.

\subsubsection{ARPS, BF, and PS Attacks}

\begin{figure}[htbp]
	\centering
	
	\subfloat[\label{fig:ARPC1}\centering WB XGB.]{\includegraphics[width=25.75mm]{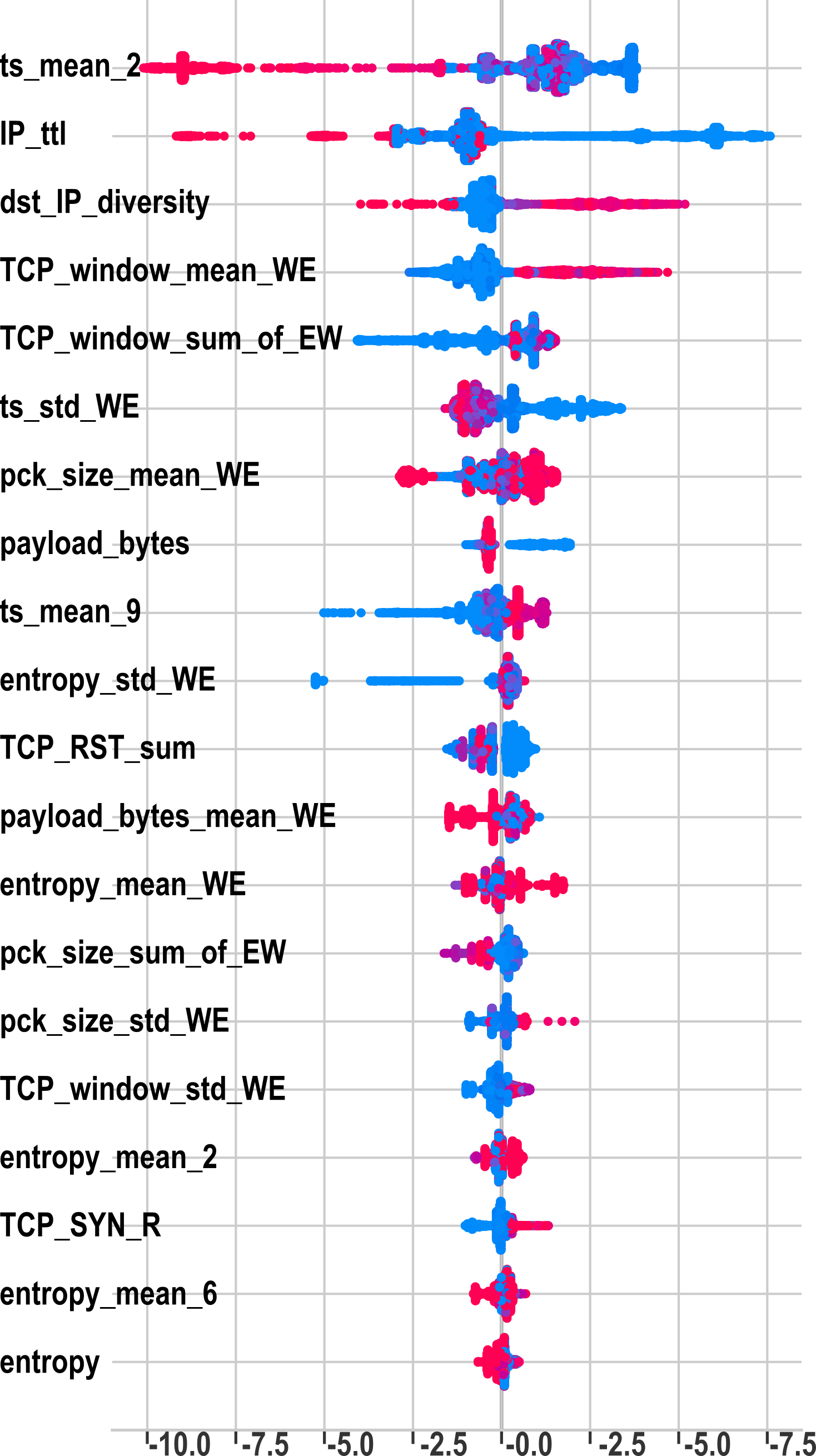}}\hspace{0.1mm}
	\subfloat[\label{fig:ARPC2}\centering FB XGB.]{\includegraphics[width=27mm]{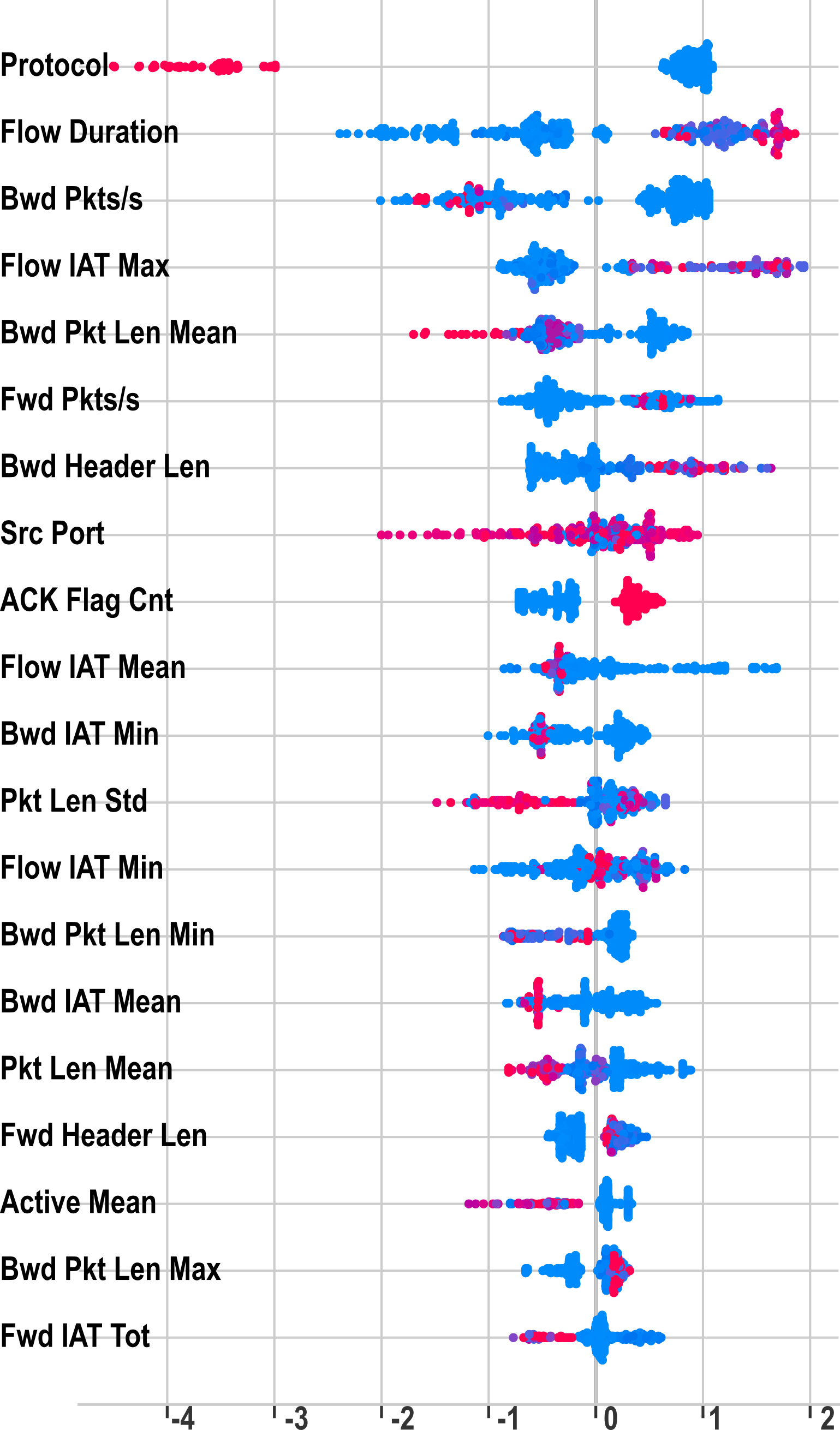}}\hspace{0.1mm}
	\subfloat[\label{fig:ARPC3}\centering FB MLP.]{\includegraphics[width=26mm]{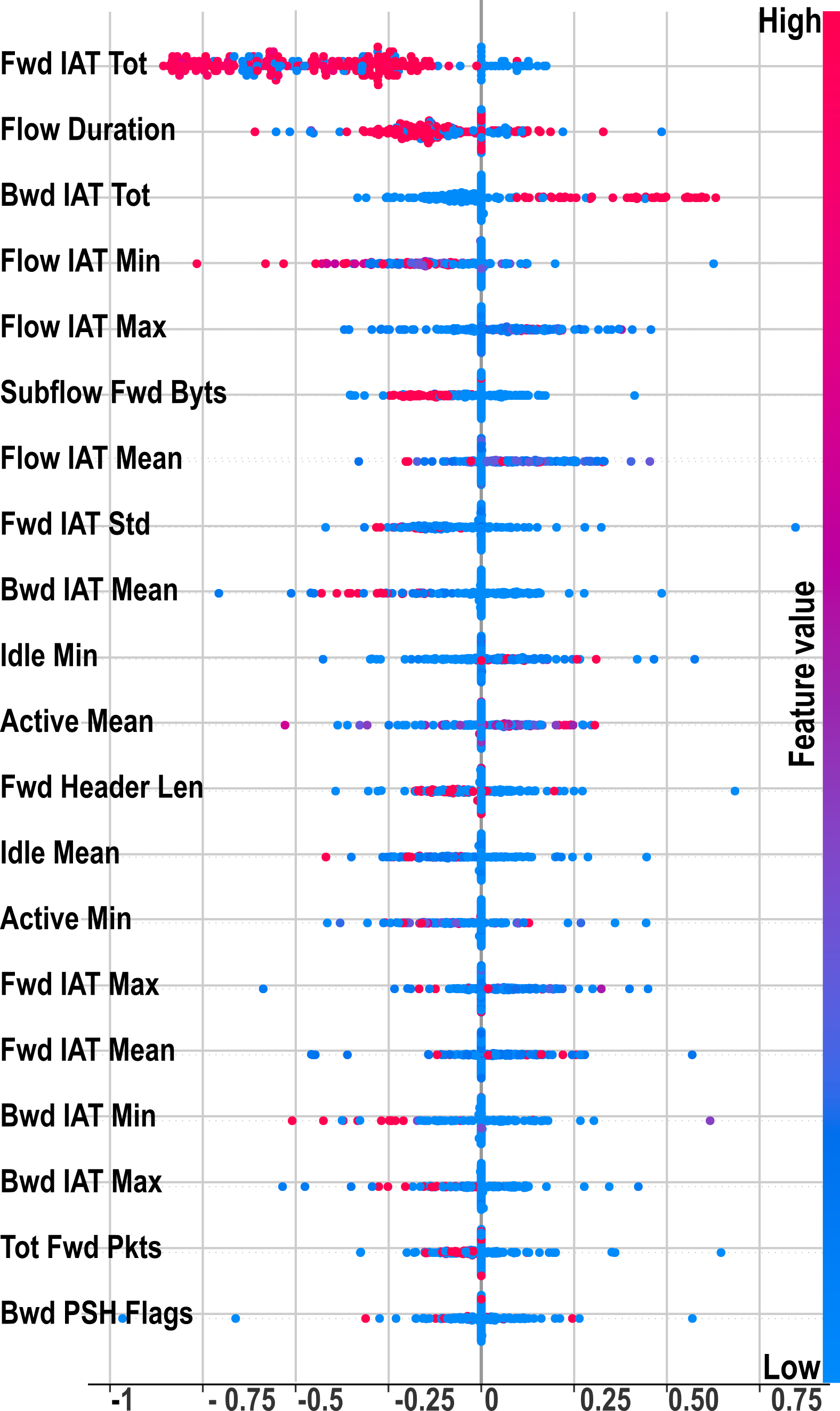}}
	\caption{Comparison of window (WB) and flow (FB) models for ARPS attack.}
	\label{fig:ARPC0}
\end{figure}

This section examines the feature effectiveness of three  other attacks.
\paragraph{ARP Spoofing (ARPS)}
This attack is notable for being the only one where the flow based approach outperforms the window based approach when evaluated on an independent dataset. It is also notable in that the attack packets are very similar to benign packets. The only difference is the time fluctuations in packet transport due to the involvement of the attacker. Therefore, it is  very difficult to detect. As a consequence, the detection rate of this attack is generally low. Fig.~\ref{fig:ARPC0} shows the SHAP analysis of the XGB models, which are the most successful for the first two evaluation conditions, and the flow based MLP models, which generalize best to the independent dataset. When we examine Fig.~\ref{fig:ARPC1}, we can see that only 3 of the 10 most important features are able to give information about the time between packets. \textit{ts\_mean\_2} takes the first place, but \textit{ts\_std\_WE, ts\_mean\_9} are far behind. \textit{IP\_ttl} seems to be a useful feature for distinguishing IP addresses that look the same, rather than being an individual feature. The prominence of the \textit{dst\_IP\_diversity} feature, which provides information about the IP-MAC relationship, is also quite understandable. Even if size, TCP-window or flags related features were highlighted, we do not think they would be useful in detecting this attack. Examination of Fig.~\ref{fig:ARPC2} shows that protocol, size, and flag based features stand out. We do not consider these features to be of any importance for an ARPS attack. In particular, although the \textit{Protocol} and \textit{ACK Flag Cnt} features provide a clear distinction, we do not consider this to be specific to the dataset and therefore not generalizable. On the other hand, the inter-arrival times (IAT) features (\textit{Fwd IAT Tot, Bwd IAT Mean, Flow IAT Min, Bwd IAT Min, Flow IAT Mean, Flow IAT Max}), which express the time statistics between packets, have the potential to provide discrimination, although they are suppressed by other features. When Fig.~\ref{fig:ARPC3} is examined, it is seen that with a few exceptions (\textit{Fwd Header Len, Subflow Fwd Byts, Tot Fwd Pkts}), all features are time-dependent, especially IAT based. In this respect, the overfitting problem caused by irrelevant features in the first two models is not observed in this model. Thanks to this characteristic, this model is relatively successful in comparison to the others.  The distinguishing characteristic of this attack—setting it apart from benign scenarios—is the temporal fluctuations between packets. Due to the comprehensive integration of inter-packet time features in the flow method, it's clear why this approach is more effective in detecting the attack compared to our method.

\begin{figure}[htbp]
	\centering

	\subfloat[\label{fig:BF1}\centering Window-Based XGB.]{\includegraphics[width=37mm]{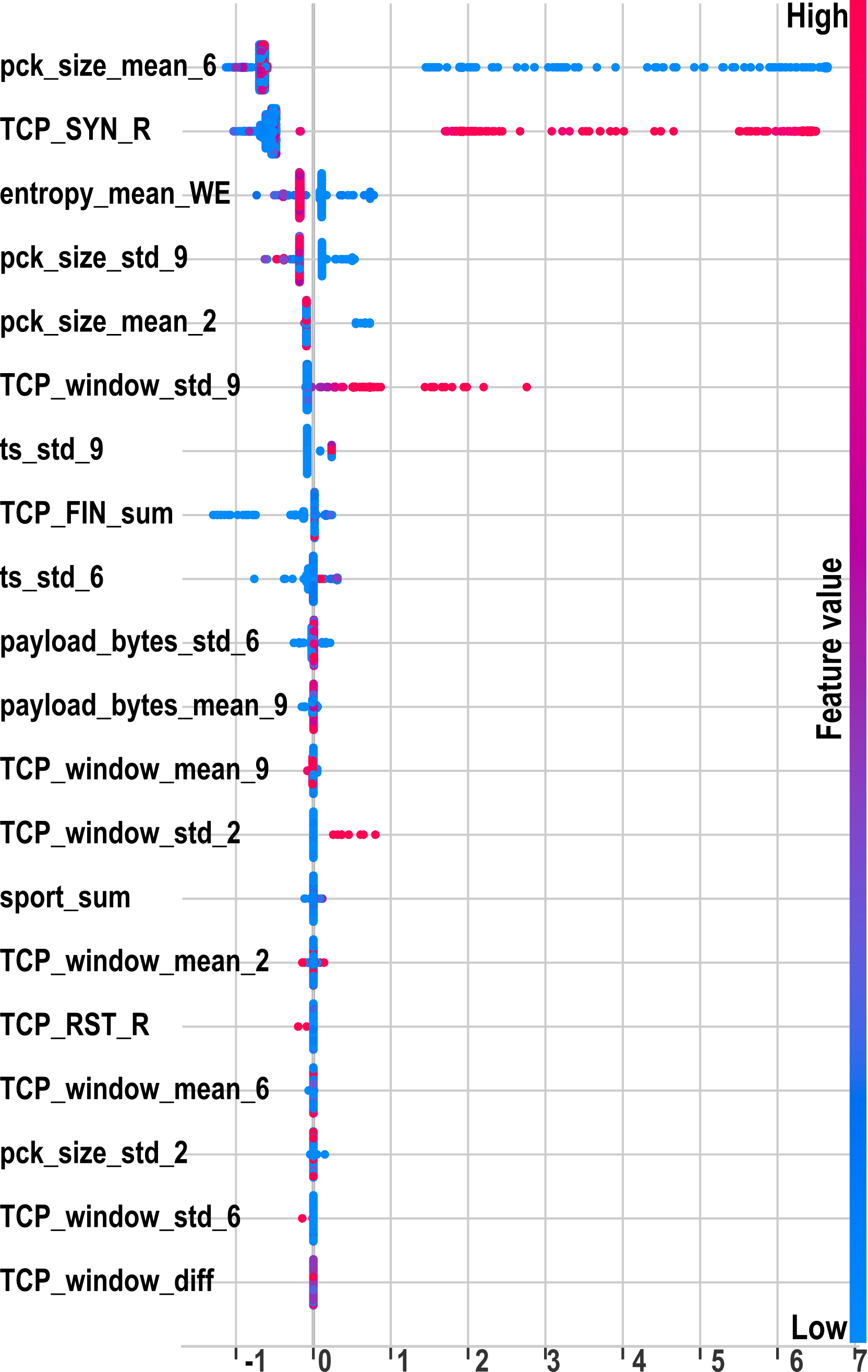}} \hfill\subfloat[\label{fig:BF2}\centering Flow-Based XGB.]{\includegraphics[width=37mm]{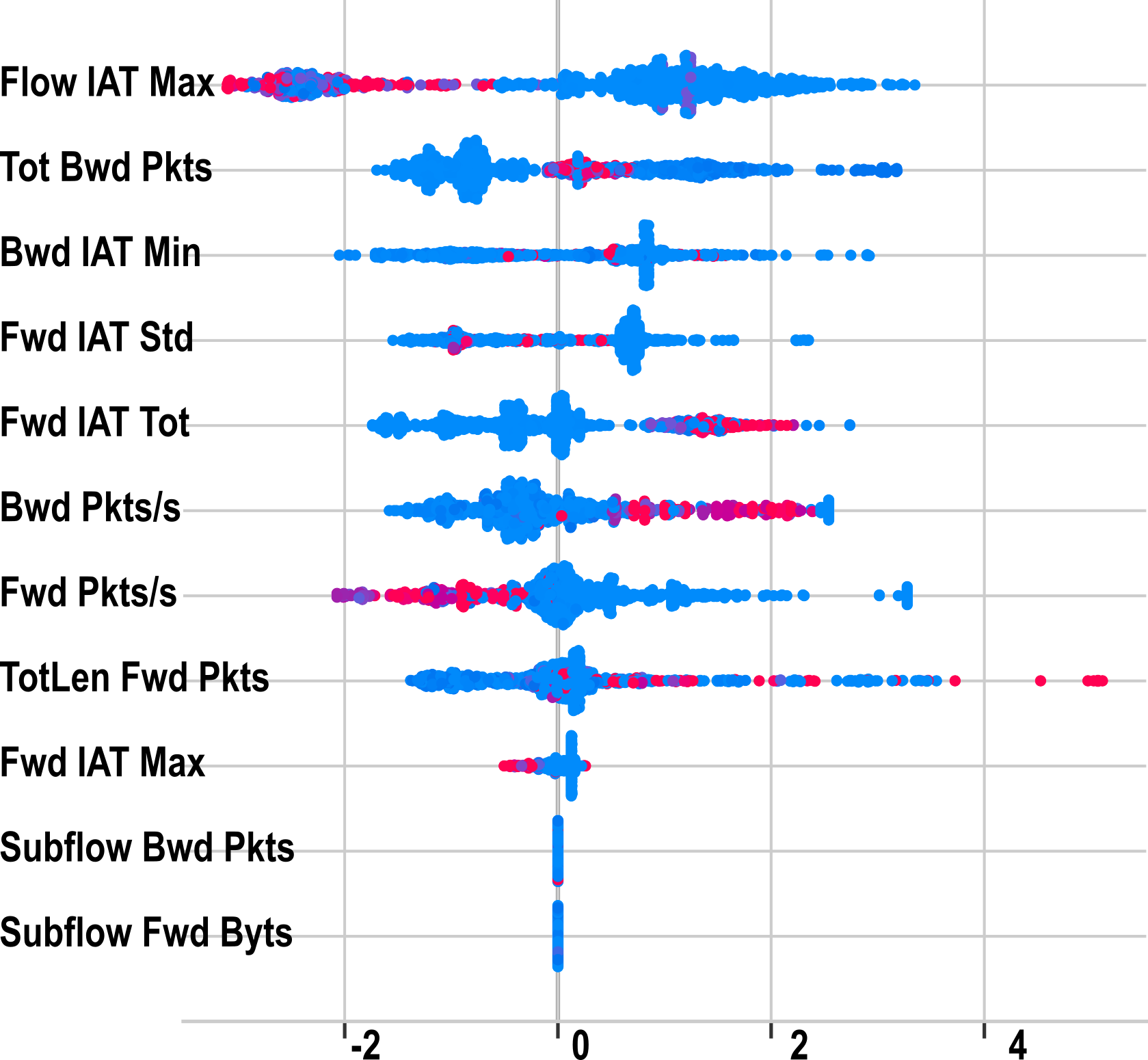}}
	\caption{Comparison of window and flow-based models for BF attack.}
	\label{fig:BF0}
\end{figure}

\paragraph{Brute-Force (BF)}
The window based approach performs best across the three evaluation scenarios. Model type is a significant factor: RF and XGB perform well in the first two scenarios, but only NB achieves reasonable generality on the independent dataset. The telnet BF attack uses the TCP protocol to target specific ports and aims to obtain the username/password by trial and error. In this context, this attack is characterised by a large number of TCP packets targeting specific ports (23,2323 - telnet ports) in a short period of time. The fact that these packets contain passwords and usernames makes the payload-based characteristics important, and the size is expected to be in a certain range. Fig.~\ref{fig:BF0}  shows the SHAP analysis of the XGB models. When Fig.~\ref{fig:BF1} is examined, it can be seen that size dependent properties such as \textit{pck\_size\_mean6,  pck\_size\_mean2, pck\_size\_std9 and entropy\_mean\_WE, payload\_bytes\_std6,
	payload\_bytes\_mean9} are used, and these provide information about the payload. In addition, TCP window size-related properties also provide good discrimination; this may be due to the fact that BF tools use a predefined window size~\cite{BF2}. So, it appears that the window-based models are using suitable features, suggesting that the relatively poor performance on the third evaluation scenario is due to the use of RTSP BF attack data for this. Despite both being brute force attacks, there exist notable differences in the tools and approaches employed for each~\cite{BF1}, and this may limit the ability of telnet BF models to generalize to RTSP BF attacks. For flow-based models, although features related to the number of outgoing and incoming packets (\textit{Tot Bwd Pkts, Fwd Pkts/s, Bwd Pkts/s}) and packet size features (\textit{TotLen Fwd Pkts}) are used, Fig.~\ref{fig:BF2} indicates that the IAT features (\textit{Fwd IAT std, Bwd IAT Min, Tot Bwd Pkts} etc.\@) are more prominent, and this may explain their poorer performance.

\begin{figure}[htbp]
	\centering
	
	\subfloat[\label{fig:PSC1}\centering Window-Based XGB.]{\includegraphics[width=37mm]{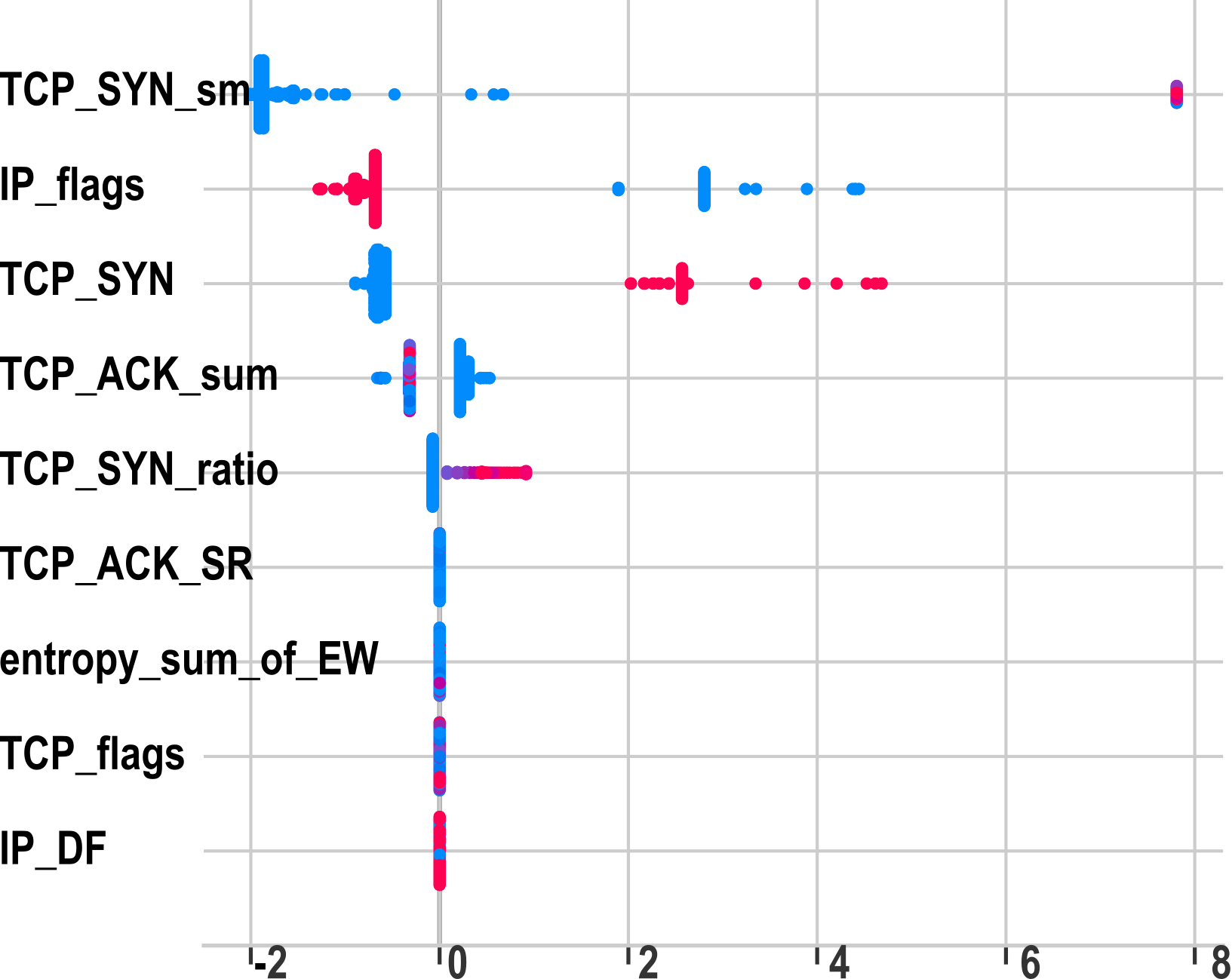}\hfill}
	\subfloat[\label{fig:PSC2}\centering Flow-Based XGB.]{\includegraphics[width=37mm]{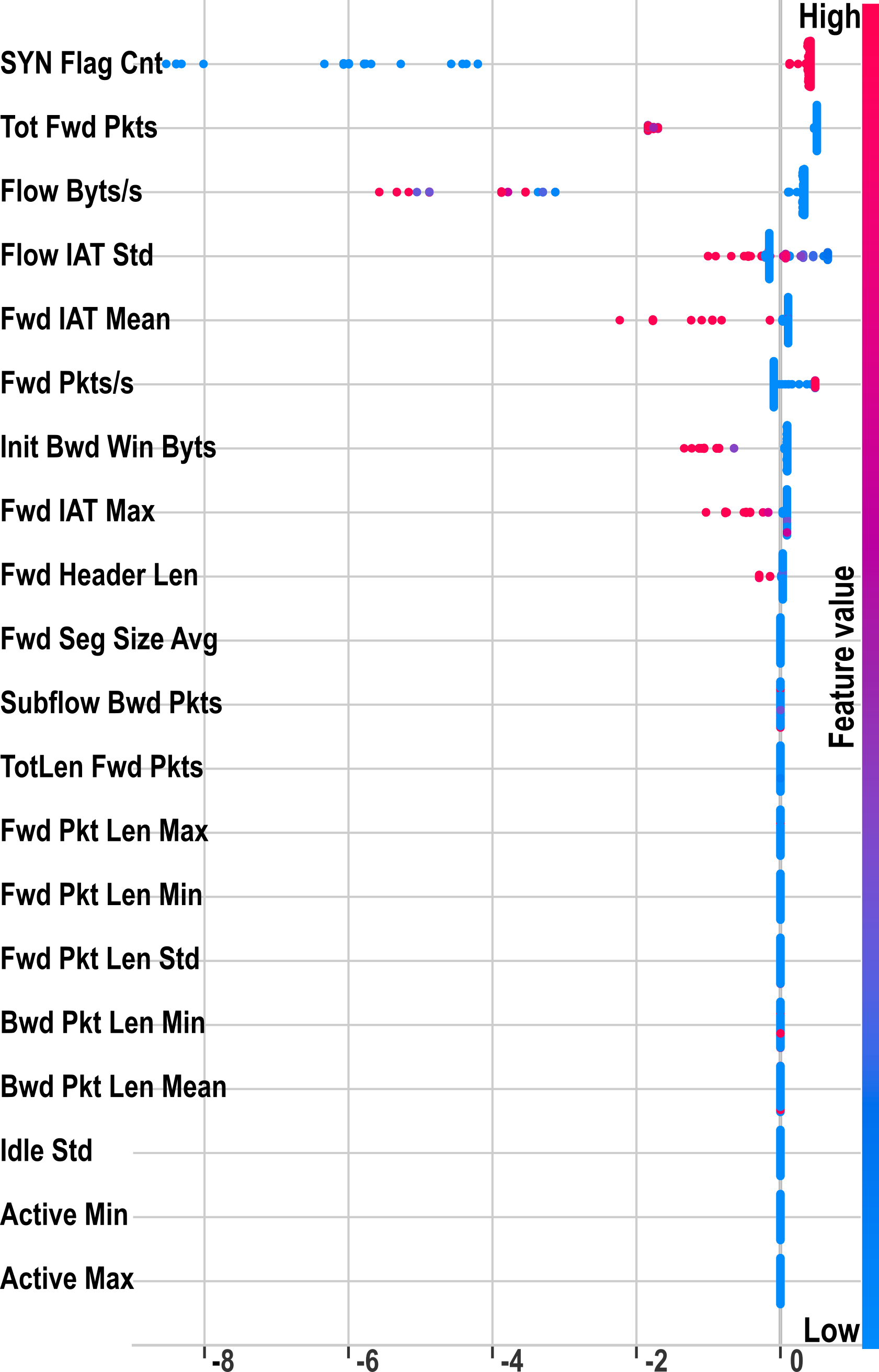}}\hfill
	\caption{Comparison of window and flow-based models for the PS attack.}
	\label{fig:PSC0}
\end{figure}

\paragraph{Port Scanning (PS)}
For this attack, generalizable models could be trained using both window and flow-based features. 
In the port scanning attack scenario, the attacker sends many TCP packets to the ports with SYN flags set. In this context, flag-based features are important. However, features that provide information about the ports can also be distinctive.
When Fig.~\ref{fig:PSC1} is analysed, it can be seen that the features related to SYN flag statistics (\textit{TCP\_ACK\_SR, TCP\_SYN\_ratio, TCP\_ACK\_sum, TCP\_SYN, TCP\_SYN\_sum}) stand out in the window approach. In addition, the effect of payload-based properties (\textit{entropy\_sum\_of\_EW}) is also used, presumably since attack packets typically lack payload. Unexpected, however, is the use of an IP flag-based feature. This appears to be due to a spurious correlation in the training data, showing that even a robust approach to feature selection can not always prevent models from picking up on irrelevant features.
For the flow models, Fig.~\ref{fig:PSC2} shows that the most important feature is the SYN packet count. Features related to packet count and packet size (\textit{Init Bwd Win Byts,Fwd Pkts/s,Flow Byts/s,Tot Fwd Pkts} etc.\@) are  prominent, and IAT features are also used, which seems reasonable for the analysis of the time between packets, due to heavy packet flow during attacks. However, it is quite interesting that the port-related features are not ranked highly in either SHAP analysis.

\subsubsection{Flow-less Attacks}

Table~\ref{tab:compare2} does not include SHD, MHD, and OSD attacks because it is not possible to extract features using CICFlowMeter for these attacks. The destination and source addresses of HD attacks are MAC addresses, and IP-based approaches like the CICFlowMeter cannot generate features for this attack. The OSD attack requires access to individual packet properties such as TCP window size and TCP flags, which is not possible in CICFlowMeter. In this context, one of the advantages of our approach is that it makes it possible to obtain a wider range of layer features and thus potentially detect many more types of attacks. 
Although it is not possible to compare with the flow-based approach, we have performed feature analysis for these attacks using SHAP. Below, we analyse the most successful models for OSD, MHD, and SHD attacks: LR, LR, and XGB, respectively.

\begin{figure}[htbp]
	\centering
	
	\subfloat[\label{fig:os}\centering OSD LR.]{\includegraphics[width=25mm]{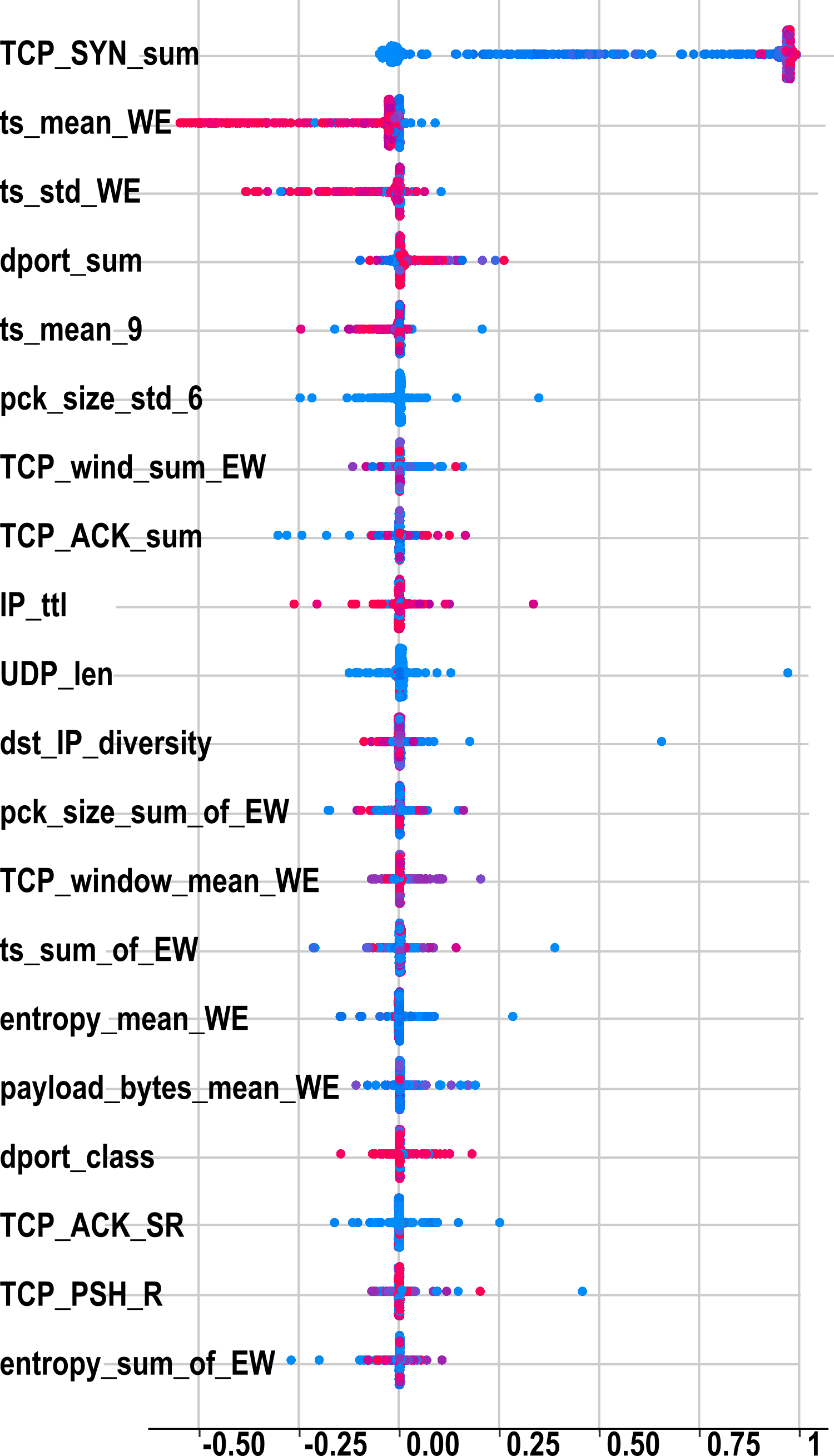}}\hspace{0.1mm}
	\subfloat[\label{fig:mhd}\centering MHD LR.]{\includegraphics[width=26mm]{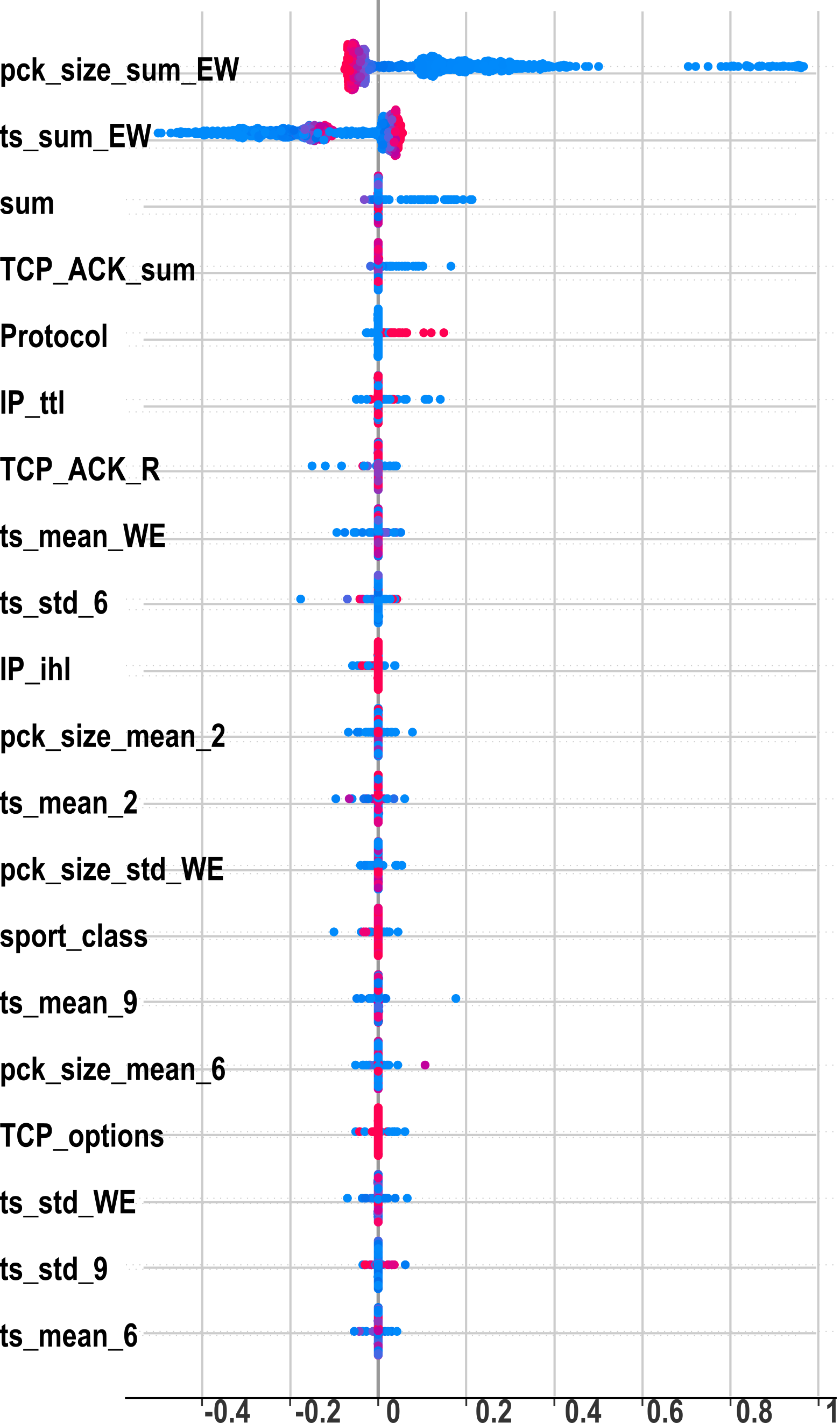}}\hspace{-0.1mm}
	\subfloat[\label{fig:shd}\centering SHD XGB.]{\includegraphics[width=25mm]{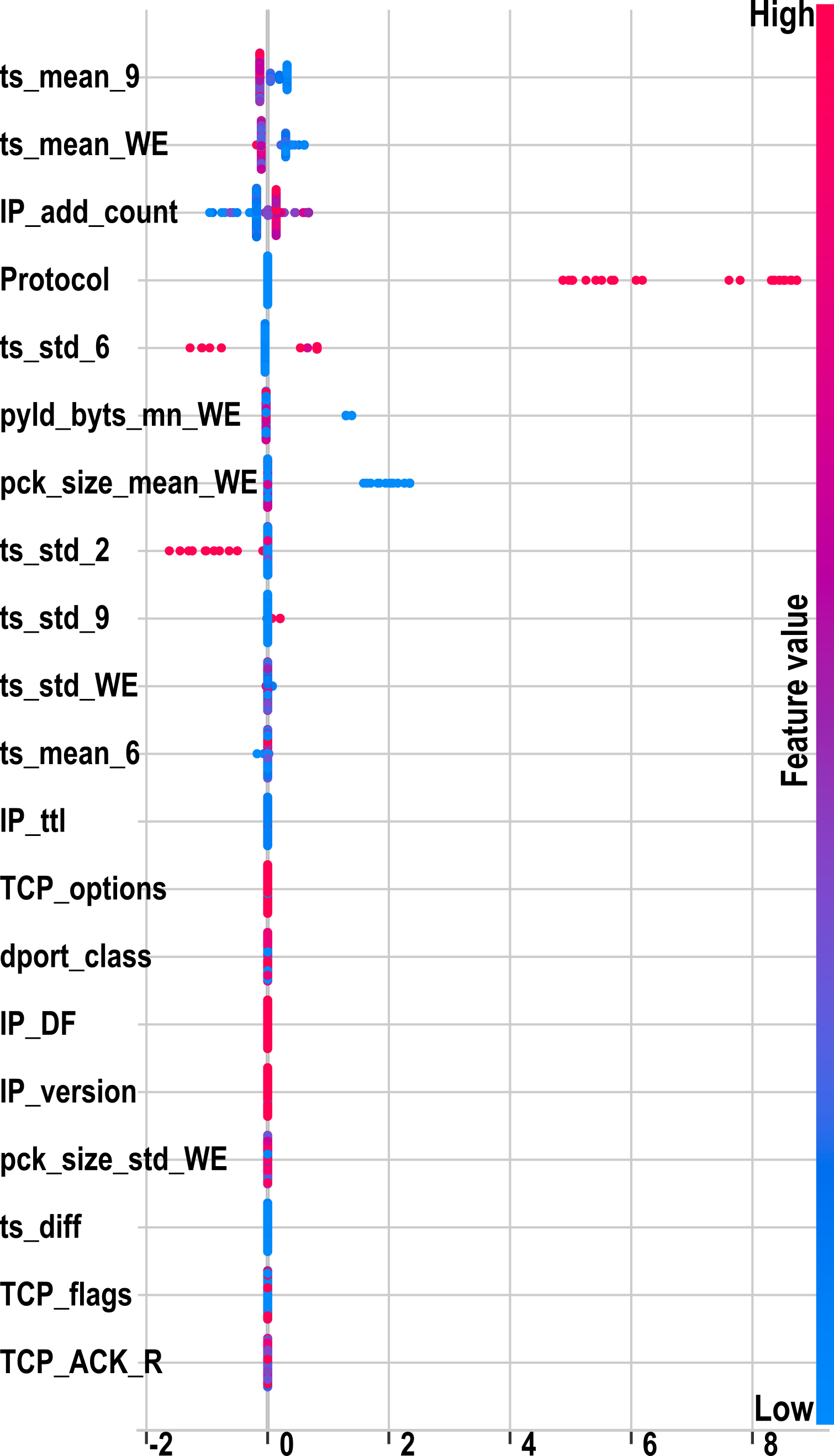}}\hfill
	\caption{SHAP graphs of models for MHD, SHD, OSD attacks.}
	\label{fig:unique}
\end{figure}

Fig.~\ref{fig:os} shows the characteristics given for the Operating System Version Discovery (OSD) attack. Since this attack consists of TCP packets with active SYN flags, TCP-based features are prominent. In addition, depending on the intensity of the attack, time-based features are also important. Even LR, which is the most successful model for detecting the OSD attack in our study, has not achieved significant success. 
This is likely due to the overfitting that occurred because the training dataset for this attack was very limited (see Table~\ref{tab:IoTID20} for sample numbers per attack).

For Mirai Host Discovery (MHD) attack, Fig.~\ref{fig:mhd} reveals that the Protocol, IP and TCP-related features are most significant, presumably because this attack does not have IP or TCP headers because it is composed of ARP packets. Considering that all network packets in the ARP protocol have the same size, we can say that time characteristics are effective in distinguishing between benign-attack packets. 

Fig.~\ref{fig:shd} shows the XGB model analysis for the  Scanning Host Discovery (SHD) attack. Since SHD is a version of MHD with a different tool, similar characteristics are evident. 

\subsubsection{Summary}
With the notable exception of ARPS attacks, the window approach is generally ahead in the detection of all attacks. The success of the flow approach is particularly low in cases where data from another dataset is used to evaluate the models. Based on the analysis of these attack models, one reason for this appears to be the extensive use of network-specific statistics in the features of the flow approach. Although models trained with these features may perform well in a specific network environment, the model is unable to sustain this performance when evaluated using data collected from a different network environment. Furthermore, the limitation of flow feature extraction to the IP level restricts adaptability to certain attack types, such as MHD and SHD.

\textcolor{black}{In summary, the superior generalization of the window-based approach, demonstrated quantitatively in Table~\ref{tab:compare2}, is directly attributable to the nature of the features learned. The SHAP analyses consistently show that our method identifies and prioritizes features related to the core mechanics of an attack, such as protocol flag anomalies, payload entropy, and port diversity. Conversely, the flow-based approach repeatedly overfits to features like Inter-Arrival Times (IATs). While powerful within a single network capture, these temporal features are highly sensitive to the specific dynamics of the network environment and thus fail to generalize, leading to a significant drop in performance on unseen datasets.}

\section{Limitations and Future Work}\label{Limitations}
This section outlines the key limitations of our current approach and highlights promising directions for future research and practical deployment enhancements.

\subsection{Dataset and Model Scope}
Our approach is limited by the availability and inherent characteristics of public datasets. While chosen for its attack diversity, our primary training dataset, IoT-NID, has limited device diversity. This can impact the generalizability of our results to more diverse real-world environments. Future work should focus on validating our methodology using newer, large-scale datasets with more heterogeneous IoT device profiles as they become available. 

Furthermore, this study focused on classical machine learning models to isolate the impact of our feature engineering pipeline. While effective in controlled experiments, this choice limits our ability to benchmark against more advanced models. A comprehensive comparison with state-of-the-art deep learning architectures—such as LSTMs or Transformers, is a critical next step in fully evaluating the performance and scalability of our approach.  

\subsection{Real-World Deployment Considerations}
The practical deployment of IoTGeM in live network environments introduces several performance-related challenges. Our current feature extraction prototype, developed using Python and the Scapy library, is functional but relatively slow. For real-time or large-scale production use, this component would require reimplementation in a more performant language, such as C++ or Go. 

\textbf{Resource Constraints}: While the trained models (e.g., Decision Trees) are lightweight and suitable for deployment on edge devices, the feature extraction step remains the most computationally intensive. Optimizing this stage is essential for operation on devices with limited CPU and memory resources. To fully understand and improve performance, a formal analysis of the time and space complexity (e.g., Big O notation), along with direct benchmarking against other state-of-the-art systems, represents a key future research direction.  

\textbf{Latency}: Our window-based detection approach enables earlier identification of malicious activity compared to traditional flow-based systems, which often wait for a flow to terminate. However, in the case of high-volume attacks such as DDoS, the requirement to fill the sliding window before classification can introduce unacceptable delays. Incorporating a timeout mechanism would allow the system to flag potential threats in real time, even if a complete window has not yet been observed.  

Future studies should also explore applying our methodology to analyze a single, sophisticated, and evasive attack scenario. This would allow for a deeper, feature-level understanding of complex threats that often evade traditional detection systems.  

\section{Conclusions}\label{Conclusions}

\textcolor{black}{This paper introduced IoTGeM, an approach that successfully creates generalizable models for behaviour-based IoT attack detection. The primary outcome of this work is the clear demonstration that our holistic framework—combining an improved window-based feature extraction with a novel Genetic Algorithm using external feedback for feature selection—produces models that are significantly more robust than conventional flow-based methods, particularly when generalizing to previously unseen datasets.}

\textcolor{black}{This superiority is proven in our most stringent, cross-dataset evaluations, where our final models achieved outstanding performance. We obtained F1 scores of 99\% for ACK, HTTP, SYN, MHD, and PS attacks, 94\% for UDP attacks, and 91\% for SHD attacks. Our analysis concludes that this success is because IoTGeM learns features that are intrinsically tied to the nature of the attacks themselves, avoiding the overfitting to network-specific statistics that plagues traditional approaches.}

In this context, we have analysed different feature sets, proposed a window-based approach to feature extraction that can provide packet-level attack detection, and compared it with the more conventional flow-based approach. We used a multi-step feature selection method, including a GA, to discover sets of features which are both predictive and generalizable. Since the feature selection method is based on external feedback from an independent dataset, it is very successful in eliminating features that are caused by data leakage and that lead to overfitting, which is a common problem in attack detection.

The resulting feature sets were then used by eight different machine learning algorithms to detect ten attack types. The resulting attack detection models were evaluated using three different scenarios, testing their ability to generalize beyond the training distribution. Particularly notable was the superior ability of our window-based models to generalize to previously unseen datasets, when compared to flow-based models. 

\textcolor{black}{While concepts like sliding windows and cross-dataset testing exist, the core contribution of this work is the holistic IoTGeM framework that synergistically combines them. In particular, our novel use of a Genetic Algorithm with external feedback for feature selection demonstrably produces more robust and generalizable models for IoT attack detection.} 
SHAP analysis of the most important features used by models showed that highly specific features such as inter-arrival times (IAT) in the flow characteristics are prominent in many attacks. However, since these features provide information about the dynamics of the network rather than the nature of the attacks, they are far from generalizable and lead to models that suffer from overfitting. 

On the other hand, we find that the features of our approach are more in line with the nature of the attack, thus producing generalizable models that are successful even for previously unseen attack patterns.

Overall, our results showed success in detecting 10 different isolated attacks. In this context, it achieved $\ge$99\% F1 score in ACK, HTTP, SYN, MHD, and PS attacks, 94\% in UDP, and 91\% in SHD. Although no significant success was achieved in ARPS, OSD and BF attacks, the reasons for the failure of the models were analysed. For ARPS, our feature set may not be well-suited; this is one case where the host of IAT features used by flow-based approaches are more effective. For OSD, the data contains few attacks on which to reliably train a model, and this appears to have led to overfitting features that reflect properties of the benign class. For BF, poor discrimination appears to be due to using a different attack sub-type for testing, rather than being due to deficiency in the learned model.

\textcolor{black}{Future research will focus on extending this methodology by integrating state-of-the-art deep learning architectures and further testing our approach on new, large-scale, and more diverse IoT datasets}. 

\bibliographystyle{cas-model2-names}
\bibliography{cas-refs}

\bio{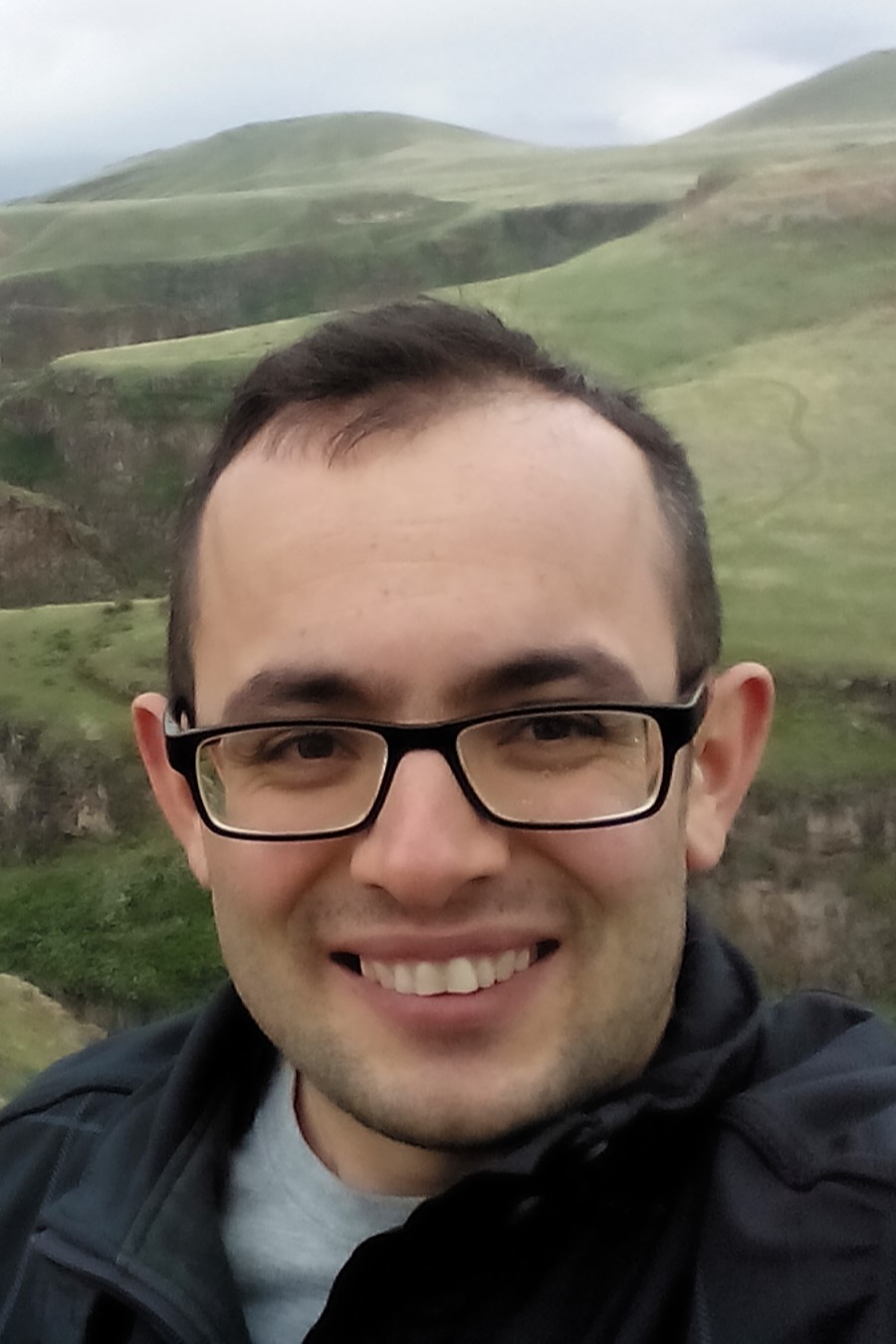} 
Kahraman Kostas	received  the MSc degree  in  Computer Networks and Security from  the  University of Essex, Colchester, U.K., in 2018. He is a PhD candidate in Computer  Science at  Heriot-Watt University, Edinburgh, U.K. His research focuses on the security of computer networks and Internet of Things.  You can find more information at \url{https://kahramankostas.github.io/}.\endbio
\vspace{.5cm}
\bio{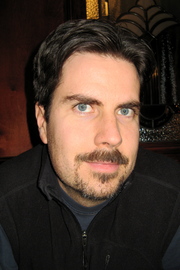} Mike Just earned his Ph.D. in Computer Science from Carleton University in 1998 and is currently an Associate Professor at Heriot-Watt University. He is primarily interested in computer security, and in applying human-computer interaction and machine learning techniques to solve computer security problems. You can find more information at \url{https://justmikejust.wordpress.com/}.\endbio
\vspace{.5cm}
\bio{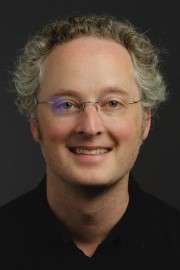} Michael A. Lones	(M’01—SM’10) is a Professor of Computer Science at Heriot-Watt University. He received both MEng and PhD degrees from the University of York. He carries out research in the areas of machine learning and optimisation, where he has a particular interest in biologically-inspired approaches and issues of replicability. Application areas of his work include medicine, robotics and security. You can find more information at \href{http://www.macs.hw.ac.uk/\%7Eml355}{http://www.macs.hw.ac.uk/$\sim$ml355}.\endbio

\onecolumn
\appendix

\section{Appendices}\label{app}

\begin{table}[H]
	\centering
	\caption{Publicly available datasets relevant to attack detection}
	\begin{tabular}{lp{0.6\textwidth}} 
		\hline
		Datasets &Web address\\\hline
		AWID2 &\href{https://icsdweb.aegean.gr/awid/awid2 }{icsdweb.aegean.gr/awid/awid2 }\\
		BoT-IoT &\href{https://research.unsw.edu.au/projects/bot-iot-dataset }{research.unsw.edu.au/projects/bot-iot-dataset }\\
		CICDDoS2019 &\href{www.unb.ca/cic/datasets/ddos-2019.html }{www.unb.ca/cic/datasets/ddos-2019.html }\\
		CICIDS2017 &\href{www.unb.ca/cic/datasets/ids-2017.html }{www.unb.ca/cic/datasets/ids-2017.html }\\
		CICIDS2018 &\href{www.unb.ca/cic/datasets/ids-2018.html }{www.unb.ca/cic/datasets/ids-2018.html }\\
		CIDDS-01 &\href{https://www.hs-coburg.de/forschung/forschungsprojekte-oeffentlich/informationstechnologie/cidds-coburg-intrusion-detection-data-sets.html}{https://www.hs-coburg.de/forschung/forschungsprojekte-oeffentlich/informationstechnologie/cidds-coburg-intrusion-detection-data-sets.html}\\
		DS2OS &\href{www.kaggle.com/datasets/francoisxa/ds2ostraffictraces?resource=download }{www.kaggle.com/datasets/francoisxa/ds2ostraffictraces?resource=download }\\
		ECU-IoHT &\href{https://github.com/iMohi/ECU-IoFT }{github.com/iMohi/ECU-IoFT }\\
		Edge-IIoTset &\href{https://ieee-dataport.org/documents/edge-iiotset-new-comprehensive-realistic-cyber-security-dataset-iot-and-iiot-applications }{ieee-dataport.org/documents/edge-iiotset-new-comprehensive-realistic-cyber-security-dataset-iot-and-iiot-applications }\\
		EMBER &\href{https://github.com/elastic/ember }{github.com/elastic/ember }\\
		InSDN &\href{https://aseados.ucd.ie/datasets/SDN }{aseados.ucd.ie/datasets/SDN }\\
		IoT-23 &\href{www.stratosphereips.org/datasets-iot23 }{www.stratosphereips.org/datasets-iot23 }\\
		IoTNID &\href{https://ocslab.hksecurity.net/Datasets/iot-network-intrusion-dataset }{ocslab.hksecurity.net/Datasets/iot-network-intrusion-dataset }\\
		ISCXIDS2012 &\href{www.unb.ca/cic/datasets/ids.html }{www.unb.ca/cic/datasets/ids.html }\\
		KDD99 &\href{https://kdd.ics.uci.edu/databases/kddcup99/kddcup99.html }{kdd.ics.uci.edu/databases/kddcup99/kddcup99.html }\\
		Kitsune &\href{https://archive.ics.uci.edu/dataset/516/kitsune+network+attack+dataset }{archive.ics.uci.edu/dataset/516/kitsune+network+attack+dataset }\\
		MQTT-IoT-IDS20 &\href{https://ieee-dataport.org/open-access/mqtt-iot-ids2020-mqtt-internet-things-intrusion-detection-dataset }{ieee-dataport.org/open-access/mqtt-iot-ids2020-mqtt-internet-things-intrusion-detection-dataset }\\
		MQTTset &\href{www.kaggle.com/datasets/cnrieiit/mqttset}{www.kaggle.com/datasets/cnrieiit/mqttset}\\
		N-BaIoT &\href{https://archive.ics.uci.edu/dataset/442/detection+of+iot+botnet+attacks+n+baiot }{archive.ics.uci.edu/dataset/442/detection+of+iot+botnet+attacks+n+baiot }\\
		NSL-KDD &\href{www.kaggle.com/datasets/hassan06/nslkdd }{www.kaggle.com/datasets/hassan06/nslkdd }\\
		TON\_IoT &\href{https://research.unsw.edu.au/projects/toniot-datasets }{research.unsw.edu.au/projects/toniot-datasets }\\
		UNSW-NB15 &\href{https://research.unsw.edu.au/projects/unsw-nb15-dataset }{research.unsw.edu.au/projects/unsw-nb15-dataset }\\\hline
	\end{tabular}%
	\label{tab:datasetlist}%
\end{table}%

\subsection{Literature Review Criteria}\label{LRC}

We applied the following criteria while searching the literature:
\begin{itemize}
	\item Published in 2019 and later,
	\item IoT focused,
	\item Including anomaly detection or intrusion detection,
	\item Used supervised machine learning methods.
\end{itemize}

We did not include Wireless Sensor Networks (WSN), which is included under IoT in many studies. This is because WSN studies generally focus on radio frequency-based attacks at the physical layer, which is beyond our scope of research. 
We searched \href{https://scholar.google.com/}{Google Scholar} using the conditions mentioned above and the following queries: ``(IoT OR internet of things) AND (“anomaly detection” OR “Intrusion detection”  OR IDS ) AND (machine learning  OR deep learning) AND (Supervised)''. This identified about 350 articles. After reviewing these, we eliminated unrelated ones (articles not including IoT device, intrusion/attack detection, or ML methods). The remaining articles are listed in Table~\ref{tab:LR1} with summary information such as dataset, ML algorithm(s) and publication year.

	\begin{table}[htbp]
		\centering
		\caption{Comparison of related work. Only the best scores are included in the evaluation section.}
		
		\setlength{\tabcolsep}{1pt}
		
		\resizebox{.9\textwidth}{!}{
\begin{tabular}{|c|c!{\vrule width 2pt}r|r|r|r|r|r|r|r|l|r|r|r|r|r|r|r|r|r!{\vrule width 2pt}r|l|r|r|l|r|r|l|r|r|r|l!{\vrule width 2pt}r|r|r|r|r|l|r|r|r!{\vrule width 2pt}l|c|c|c|c|c|}
				\hline
				&  & \multicolumn{18}{c!{\vrule width 2pt}}{Dataset}                                                                                                                  & \multicolumn{12}{c!{\vrule width 2pt}}{Machine Learning}                                                         & \multicolumn{9}{c!{\vrule width 2pt}}{Feature Type}                                      & \multicolumn{6}{c|}{Evaluation} \\
\hline
		\end{tabular}}
		\label{tab:LR1}%
	\end{table}%
	
	\begin{table}[htbp]
		\ContinuedFloat
		\centering
		\caption{Comparison of related work. Only the best scores are included in the evaluation section (continued). }
		\setlength{\tabcolsep}{1pt}
	
	\resizebox{.9\textwidth}{!}{\begin{tabular}{|c|c!{\vrule width 2pt}r|r|r|r|r|r|r|r|l|r|r|r|r|r|r|r|r|r!{\vrule width 2pt}r|l|r|r|l|r|r|l|r|r|r|l!{\vrule width 2pt}r|r|r|r|r|l|r|r|r!{\vrule width 2pt}l|c|c|c|c|c|}
			\hline
			&  & \multicolumn{18}{c!{\vrule width 2pt}}{Dataset}                                                                                                                  & \multicolumn{12}{c!{\vrule width 2pt}}{Machine Learning}                                                         & \multicolumn{9}{c!{\vrule width 2pt}}{Feature Type}                                      & \multicolumn{6}{c|}{Evaluation} \\
\hline
				
		\end{tabular}}%
		\label{tab:LR2}%
	\end{table}%
	
\begin{table}[htbp]
	\ContinuedFloat
	\centering
	\caption{Comparison of related work. Only the best scores are included in the evaluation section (continued). }
	\setlength{\tabcolsep}{1pt}
	
	\resizebox{.9\textwidth}{!}{\begin{tabular}{|c|c!{\vrule width 2pt}r|r|r|r|r|r|r|r|l|r|r|r|r|r|r|r|r|r!{\vrule width 2pt}r|l|r|r|l|r|r|l|r|r|r|l!{\vrule width 2pt}r|r|r|r|r|l|r|r|r!{\vrule width 2pt}l|c|c|c|c|c|}
			\hline
			&  & \multicolumn{18}{c!{\vrule width 2pt}}{Dataset}                                                                                                                  & \multicolumn{12}{c!{\vrule width 2pt}}{Machine Learning}                                                         & \multicolumn{9}{c!{\vrule width 2pt}}{Feature Type}                                      & \multicolumn{6}{c|}{Evaluation} \\
\end{tabular}}%
	\label{tab:LR3}%
\end{table}%

\begin{table}[htbp]
	\centering
	\caption{Summary feature list, avoiding repetition. The complete list of features and descriptions are available at: \href{https://github.com/kahramankostas/HerIoT/blob/main/featurelist.md}{github.com/kahramankostas/HerIoT/blob/main/featurelist.md} } 
	\resizebox{\textwidth}{!}{  \begin{tabular}{rl|ll}
    \toprule
    \multicolumn{1}{l}{Feature} & Explanation  & Feature & Explanation  \\
    \midrule
    \multicolumn{1}{l}{ts} &    Time Stamp   &  ID     &  Source + Destination   \\
    \multicolumn{1}{l}{Ether\_dst} &    Destination MAC Address   &  pck\_size\_diff  &  the size difference of consecutive packets   \\
    \multicolumn{1}{l}{Ether\_src} &    Source MAC Address   &  pck\_size\_mean\_WE  &  EW (expanding windows) packet size mean   \\
    \multicolumn{1}{l}{IP\_src} &    Source IP Address   &  pck\_size\_std\_WE  &  EW packet size Std.   \\
    \multicolumn{1}{l}{IP\_dst} &    Destination IP Address   &  pck\_size\_sum\_of\_EW  &  EW packet size total.   \\
    \multicolumn{1}{l}{pck\_size} &    Packet (Frame) Size   &  ts\_diff  &  the time difference of consecutive packets   \\
    \multicolumn{1}{l}{Ether\_type} &    Ethernet Type   &  ts\_mean\_WE  &  EW packet time mean   \\
\bottomrule
    \end{tabular}}%
	\label{tab:indF}%
\end{table}%

\subsection{Data Leakage in the Use of Individual Features}\label{use-ind}
\begin{figure}[htbp]
	\centering
	\includegraphics[width=155mm]{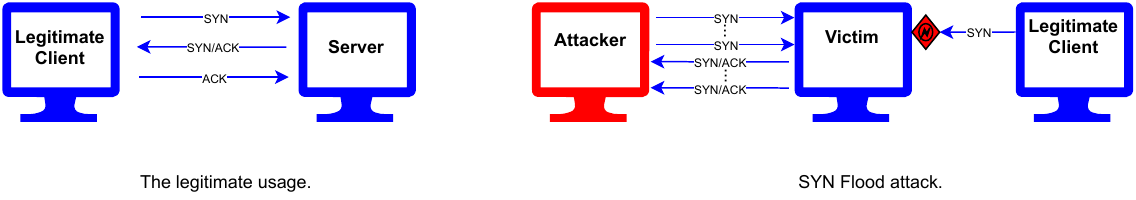}
	\caption{Legitimate and malicious use of 3-way handshake.}
	\label{fig:synfig}
\end{figure}

It is not possible to detect an attack using individual packet features alone. 
This situation can be explained with a simple example: SYN Flood~\cite{gulihar2020cooperative}, which is a DoS attack based on the exploit of 3-handshake in the TCP protocol. During the 3-way handshake process, the client that wants to establish a connection sends a synchronization (SYN) packet to the server. The server receiving this packet sends a synchronization and acknowledgment (SYN-ACK) packet to the client. Finally, the client responds with the acknowledgment (ACK) packet, and the TCP connection is established (see Fig.~\ref{fig:synfig}).

In the SYN Flood attack, the attacker sends a large number of SYN packets to the server. The server sends the SYN-ACK packet in response and waits for the ACK packet from the other side to establish the connection. The attacker does not send these packets, resulting in a large number of unfinished handshakes. The server whose resources are exhausted cannot then serve legitimate users.
If we look at these SYN packets individually, we cannot see any features that distinguish them from each other (except for identifying features such as IP or MAC addresses. This information is useless in attack detection because it is impossible to know beforehand). It is impossible to tell whether SYN packets are viewed individually as benign or malicious. However, if we look at the whole, we see that few or no ACK packets are received despite thousands of SYN packets from the same source. So while looking at the individual packet is useless for detection, it is possible to detect this attack by looking at the whole. This is roughly the same as other types of attacks. However, like with any generalizations, there are exceptions. One example is single-packet attacks that exploit malformed packets. These attacks typically involve sending malformed network packets that the device cannot handle, leading to malfunctions or crashes in the receiving device~\cite{singlepackets}. However, addressing these attacks usually involves packet filtering or firewalls, not machine learning behavioural analysis, since they can be effectively managed using signature/rule-based approaches.

\begin{figure}[htbp]
	\centering
	
	\centering{\includegraphics[width=168mm]{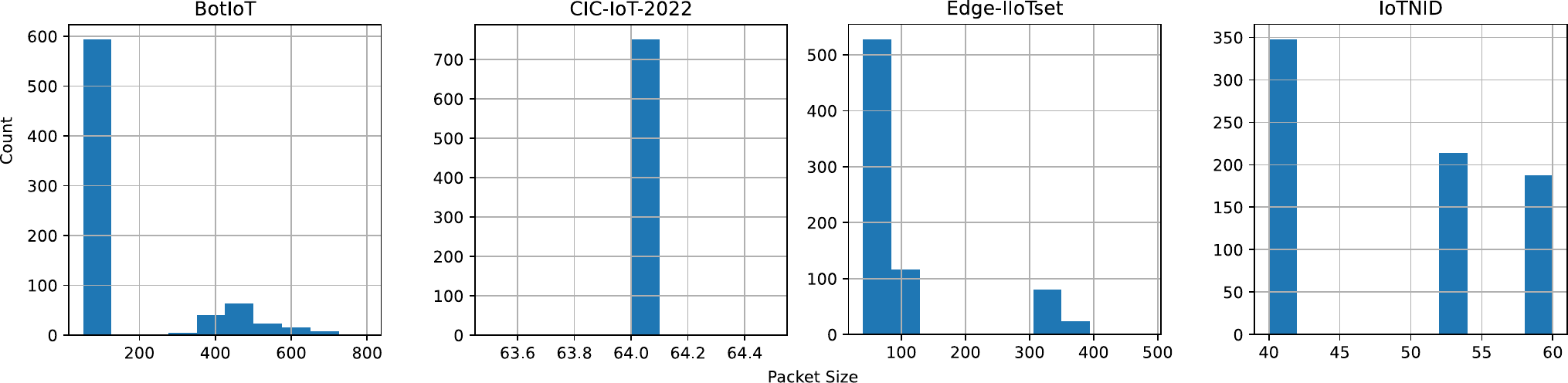}}%
	\caption{The distribution of packet sizes of malicious data (HTTP Flood attack) in four different datasets.}
	\label{fig:httphist}
\end{figure}

The high scores obtained by using individual packet features can be explained in two different ways. Identifying features: they can be specific features that identify a target or an attacker such as IP address, MAC address, or in some cases port number. They can be features that identify or give hints about a session. The use of these features will lead to over-fitting of the model, resulting in systems that are not useful in real-life data.

\begin{figure}[htbp]
	\centering

	\subfloat[\label{fig:comp_model2}\centering In Kitsune dataset - Mirai attack, the labels follow a certain time pattern.]{{\includegraphics[width=100mm]{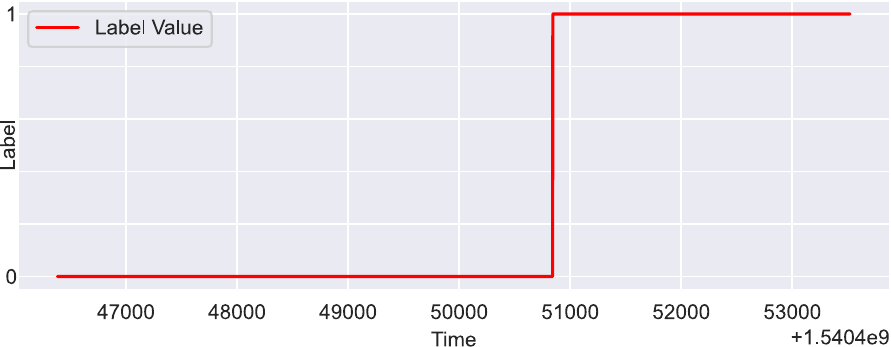}}}
	\subfloat[\label{fig:comp_model1}\centering In the IoT-NID dataset - UDP flood attack, attack packets have a specific size.]{{\includegraphics[width=70mm]{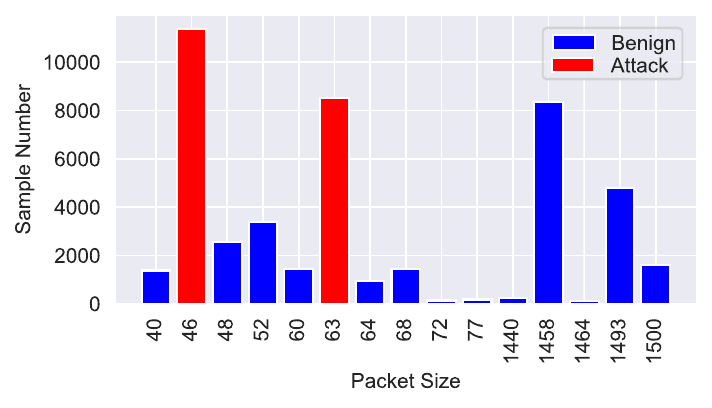}}}
	\caption{Distribution of labels in Kitsune and IoT-NID datasets.}
	\label{fig:comp_model}	
\end{figure}

\begin{figure}[htbp]
	\centering
	
	\subfloat[\label{fig:comp_1}\centering Mirai Attack Time Complexity]{{\includegraphics[width=43mm]{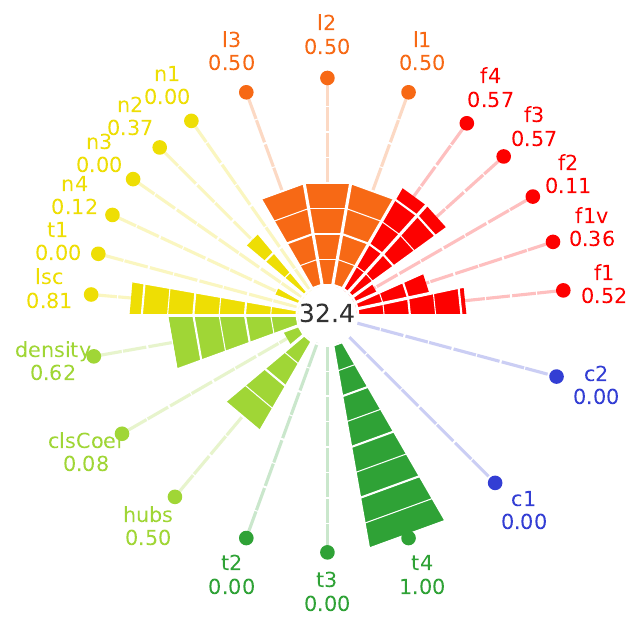}}}%
	\subfloat[\label{fig:comp_2}\centering UPD Flood Attack Time Complexity]{{\includegraphics[width=43mm]{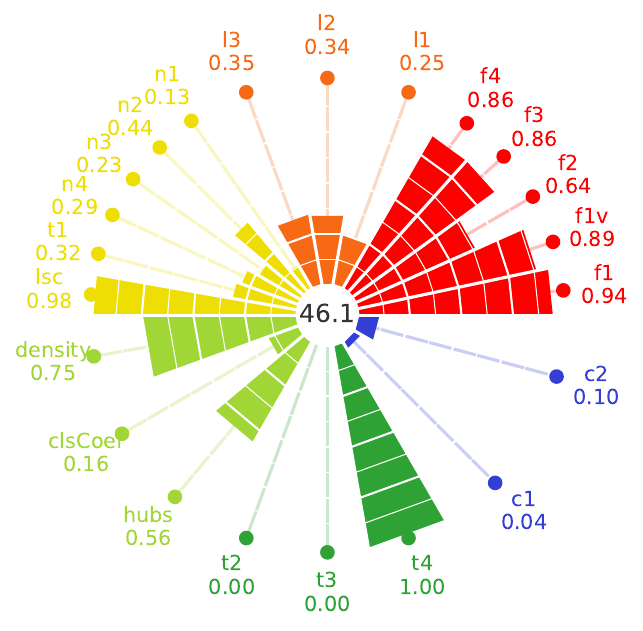}}}
	\subfloat[\label{fig:comp_3}\centering Mirai Attack Size Complexity]{{\includegraphics[width=43mm]{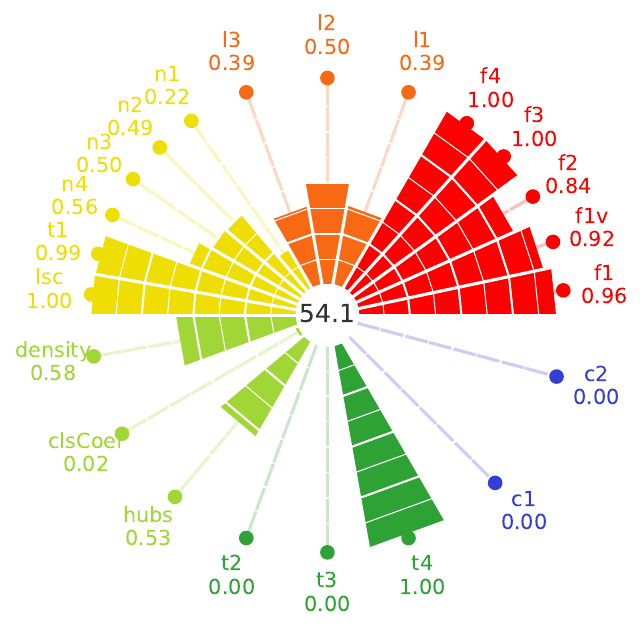}}}%
	\subfloat[\label{fig:comp_4}\centering UPD Flood Attack Size Complexity]{{\includegraphics[width=43mm]{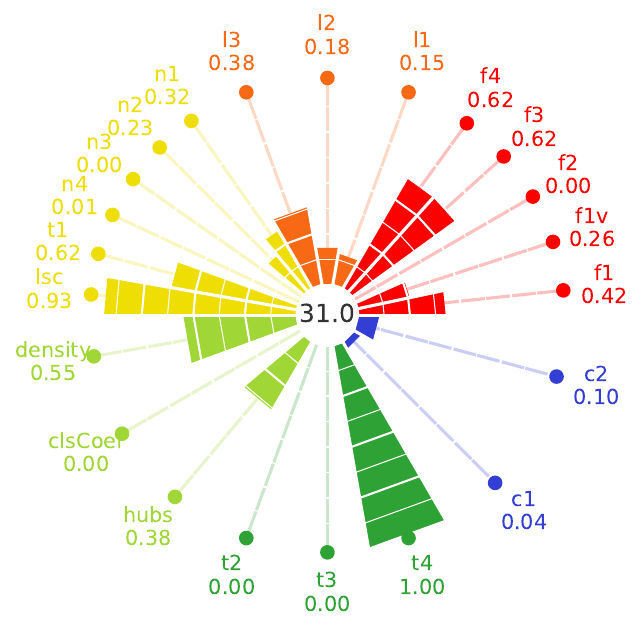}}}\\
	\caption{Complexity analysis of UDP and Mirai attacks according to time and size characteristics. The central score depicted in the figures represents the comprehensive complexity score, which is derived as the average of 22 distinct analytical methods (The complexity score ranges from 0 to 100, where higher values indicate increasing complexity). These 22 methods stem from six different approaches to assessing complexity, each distinctly colour-coded for clarity of categorization~\cite{lorena2019complex}: red for feature-based, orange for linearity-based, yellow for neighbourhood-based, light green for network-based, dark green for dimensionality-based, and blue for class imbalance-based.
		When these figures are analysed, it can be seen that the UDP flood attack has low size complexity and high time complexity. On the other hand, Mirai attack show low size complexity and high time complexity.}
	\label{fig:comp_graph}	
\end{figure}

\begin{figure}[htbp]
	
	\includegraphics[width=85mm]{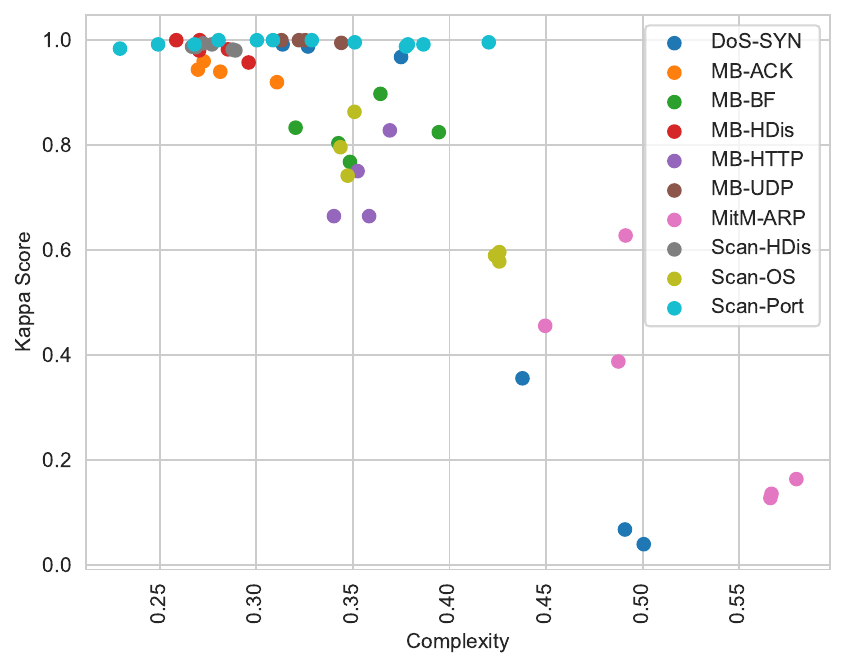}
	\centering	\caption{Reflection of how dataset complexity affects the predictive value of individual features.}
	\label{fig:compexity_graph}
\end{figure}

On the other hand, apart from some specific attacks, most attacks follow a pattern and produce uniform outputs. For example, the distribution of the size of the attack packets in the HTTP flood attack on four different datasets is shown in Fig.~\ref{fig:httphist}. Because of this uniform nature of the attacks, if the complexity level of the dataset is low in studies using individual packets, basic characteristics, such as size or time, can become identifying features. For example, by using only packet size in the IoT-NID UDP dataset or only time in the Kitsune Mirai dataset, a near-perfect classification can be made (see Table~\ref{tab:ind-data}). This is only because the complexity of these datasets in terms of these features is quite low (see the complexity analyses in Fig.~\ref{fig:comp_graph}). However, these classification successes will be specific to this dataset, and so the model will not be successful on other datasets, since they have different time and size patterns. Fig.~\ref{fig:compexity_graph} shows the results of another experiment on complexity. In this experiment, a basic feature (size) was used to separate the attack data. Especially if the complexity is below the critical level (45), a high rate of discrimination is achieved. As complexity increases, success decreases.

\begin{table}[htbp]
	\centering
	\caption{Near-perfect classification of 2 datasets afflicted by low data complexity. In Kitsune data, the labels follow a certain time pattern. In the IoT-NID dataset, attack packets have a specific size.}
	\begin{tabular}{llllrrrrr}
		\toprule
		Dataset & Attack & Feature & ML    & \multicolumn{1}{l}{Acc} & \multicolumn{1}{l}{Prec} & \multicolumn{1}{l}{Rec} & \multicolumn{1}{l}{F1} & \multicolumn{1}{l}{Kappa}  \\
		\midrule
		Kitsune & Mirai & Time Stamp & ExtraTree & 0.952 & 0.885 & 0.972 & 0.920 & 0.841  \\
		IoT-NID & UDP   & Packet Size & ExtraTree & 1.000 & 1.000 & 0.999 & 1.000 & 0.999  \\
		\bottomrule
	\end{tabular}%
	\label{tab:ind-data}%
\end{table}%

\newpage

\begin{algorithm}
\caption{IoTGeM Feature Selection Process}
\begin{algorithmic}[1]
    \Require $Training\_Data$, $HPO\_Data$, $Validation\_Data$, $All\_Features$
    \Ensure $Best\_Feature\_Set$
    
    \State \textbf{Step 1: Feature Elimination via Voting}
    \State $Initial\_Features \leftarrow \emptyset$
    \For{each feature $f$ in $All\_Features$}
        \State $votes \leftarrow 0$
        
        \State \textbf{// Scenario 1: Cross-validation on training data}
        \State $\kappa_1 \leftarrow Evaluate\_Feature(f, Training\_Data)$
        \If{$\kappa_1 > 0$}
            \State $votes \leftarrow votes + 1$
        \EndIf
        
        \State \textbf{// Scenario 2: Train on Training\_Data, test on HPO\_Data}
        \State $\kappa_2 \leftarrow Evaluate\_Feature(f, Training\_Data, HPO\_Data)$
        \If{$\kappa_2 > 0$}
            \State $votes \leftarrow votes + 1$
        \EndIf
        
        \State \textbf{// Scenario 3: Train on Training\_Data, test on Validation\_Data}
        \State $\kappa_3 \leftarrow Evaluate\_Feature(f, Training\_Data, Validation\_Data)$
        \If{$\kappa_3 > 0$}
            \State $votes \leftarrow votes + 1$
        \EndIf
        
        \If{$votes \geq 2$ \textbf{and} $\kappa_3 > 0$}
            \State Add $f$ to $Initial\_Features$
        \EndIf
    \EndFor
    
    \State \textbf{Step 2: GA with External Feedback}
    \State $Population \leftarrow Initialize\_Population(Initial\_Features)$
    \State $Best\_Global\_Solution \leftarrow null$
    \State $Best\_Global\_Fitness \leftarrow -1$
    
    \For{$i = 1$ to $num\_generations$}
        \For{each solution $s$ in $Population$}
            \State \textbf{// Fitness is evaluated on separate validation data}
            \State $model \leftarrow Train\_Model(Training\_Data, features=s)$
            \State $fitness \leftarrow Evaluate\_Model(model, Validation\_Data)$ \textbf{// e.g., F1 Score}
            \State $s.fitness \leftarrow fitness$
        \EndFor
        
        \If{$\max(Population.fitness) > Best\_Global\_Fitness$}
            \State $Best\_Global\_Solution \leftarrow Get\_Best\_Solution(Population)$
            \State $Best\_Global\_Fitness \leftarrow Best\_Global\_Solution.fitness$
        \EndIf
        
        \State $Population \leftarrow Evolve\_New\_Population(Population)$ \textbf{// Crossover \& Mutation}
    \EndFor
    
    \State $Best\_Feature\_Set \leftarrow Best\_Global\_Solution$
    \Return $Best\_Feature\_Set$
\end{algorithmic}
\end{algorithm}

\subsection{Issues with the IoTID20 Dataset}\label{IoTID20}
We performed the feature extraction process by applying CICFlowMeter to all the pcap files we used. Although a dataset (IoTID20) has already been produced using CICFlowMeter from IoT-NID raw data, we preferred to redo the feature extraction process instead of using it. We used the  \href{https://github.com/ahlashkari/CICFlowMeter}{CICFlowMeter}\footnote{\url{https://github.com/ahlashkari/CICFlowMeter}} tool to extract features from the dataset and labeled them in parallel with the defined rules in IoT-NID dataset\cite{IoT-NID} using our  \href{https://github.com/kahramankostas/IoTGeM/blob/main/0001-Feature-Extraction-PCAP2CSV/000-FLOW-LABELLER.ipynb}{flow-labeller script}\footnote{\url{https://github.com/kahramankostas/IoTGeM/blob/main/0001-Feature-Extraction-PCAP2CSV/000-FLOW-LABELLER.ipynb}}. We believe that there may have been issues with the feature extraction process of the IoTID20 dataset. Some clues leading to this conclusion are as follows. Although the raw data contains 10 different attacks, there are 8 attacks in the dataset. It is conceivable that some pcap files are labelled as if they contain a single attack, ignoring the fact that they contain more than one type of attack.

When examining the content generator directives in the IoT-NID dataset~\cite{IoT-NID}, distinct pcap files are apparent, each corresponding to a specific attack and session. However, these files encompass not only malicious packets but also benign ones within each attack file. 
This becomes most apparent when considering that the benign file contains 137,000 packets, whereas the entire dataset comprises over 15,000,000 benign packets. Notably, this dataset lacks pre-defined labels; solely raw data and filtering rules for packet-level labelling are provided. Employing the rules from content creators facilitates the separation of benign and malicious segments within all files. Yet, in the context of IoTID20, it appears that these rules were disregarded, resulting in the mislabelling of all pcap files as attacks. Substantiating this assertion, we observe the following: a flow comprises either a single packet or a combination of several. In this regard, the number of flows derived from a pcap file must consistently be less than or equal to the number of packets. However, a scrutiny of IoTID20 reveals numerous instances where this principle is violated, with the count of attacks often surpassing the expected count (refer to Table~\ref{tab:IoTID20}). For many attacks such as HTTP, BF, OSD, PS, MHD, and SHD the number of samples provided in IoTID20 is much higher than the number of samples in the raw data.

\begin{table}[htbp]
	\centering
	\caption{The comparison of the number of packets in the IoT-NID raw data with the number of flows in the IoTID20 dataset}
\begin{tabular}{llrr}
		\toprule
		Type  & Label & \multicolumn{1}{l}{Packets} & \multicolumn{1}{l}{Flows} \\
		\midrule
		\multicolumn{1}{p{10.835em}}{Normal} & Normal & 1,756,276 & 40,073 \\
		\midrule
		Mirai Botnet & Host Discovery (MHD)& 673   & \multirow{2}[2]{*}{22,192} \\
		Scanning & Host Discovery (SHD) & 2,454 &  \\
		\midrule
		Scanning & Port Scanning (PS) & 20,939 & \multirow{2}[2]{*}{53,073} \\
		Scanning & OS/Version Detection (OSD) & 1,817 &  \\
		\midrule
		Man in the Middle (MITM) & ARP Spoofing (ARPS)& 101,885 & 35,377 \\
		Denial of Service (DoS) & SYN Flooding (SYN) & 64,646 & 59,391 \\
		Mirai Botnet & Telnet Bruteforce (BF) & 1,924 & 121,181 \\
		Mirai Botnet & UDP Flooding (UDP)& 949,284 & 183,554 \\
		Mirai Botnet & ACK Flooding (ACK) & 75,632 & 55,124 \\
		Mirai Botnet & HTTP Flooding(HTTP)& 10,464 & 55,818 \\
		\bottomrule
	\end{tabular}%
	\label{tab:IoTID20}%
\end{table}%

\begin{table}[htbp]
  \centering
  \caption{Attack and benign sample distribution for each binary classification model across different datasets (CV, Different Session, and Different Dataset). This table supports our per-attack modeling strategy by detailing the number of attack and benign samples used in training each model, along with their corresponding attack-to-benign ratios. This information enhances transparency regarding class imbalance considerations.}
    \begin{tabular}{llrrr}
    \toprule
    \multicolumn{1}{c}{\textbf{Dataset}} & \multicolumn{1}{c}{\textbf{Attack Type}} & \multicolumn{1}{p{3.72em}}{\textbf{Attack Samples}} & \multicolumn{1}{p{3.78em}}{\textbf{Benign Samples}} & \multicolumn{1}{p{6.72em}}{\textbf{Attack/Benign Ratio}} \\
    \midrule
    CV    & SYN Flooding & 9160  & 10835 & 0.85 \\
    CV    & HTTP Flooding & 800   & 19195 & 0.04 \\
    CV    & ACK Flooding & 1400  & 18595 & 0.08 \\
    CV    & UDP Flooding & 7835  & 12155 & 0.64 \\
    CV    & ARP   & 14995 & 4995  & 3.00 \\
    CV    & SP    & 1240  & 10000 & 0.12 \\
    CV    & BF    & 505   & 10000 & 0.05 \\
    CV    & OS    & 605   & 10000 & 0.06 \\
    CV    & SCHD  & 485   & 10000 & 0.05 \\
    CV    & MHDis & 205   & 9995  & 0.02 \\
    \midrule
    Session & SYN Flooding & 8526  & 11474 & 0.74 \\
    Session & HTTP Flooding & 764   & 13000 & 0.06 \\
    Session & ACK Flooding & 13897 & 6103  & 2.28 \\
    Session & UDP Flooding & 19378 & 622   & 31.15 \\
    Session & ARP   & 13211 & 6703  & 1.97 \\
    Session & SP    & 1230  & 6094  & 0.20 \\
    Session & BF    & 500   & 10000 & 0.05 \\
    Session & OS    & 50    & 9221  & 0.01 \\
    Session & SCHD  & 101   & 10000 & 0.01 \\
    Session & MHDis & 477   & 6094  & 0.08 \\
    \midrule
    Dataset & SYN Flooding & 10000 & 10000 & 1.00 \\
    Dataset & HTTP Flooding & 10000 & 10000 & 1.00 \\
    Dataset & ACK Flooding & 10000 & 10000 & 1.00 \\
    Dataset & UDP Flooding & 10000 & 10000 & 1.00 \\
    Dataset & ARP   & 10000 & 10000 & 1.00 \\
    Dataset & SP    & 4048  & 4818  & 0.84 \\
    Dataset & BF    & 312   & 261   & 1.20 \\
    Dataset & OS    & 10000 & 10000 & 1.00 \\
    Dataset & SCHD  & 210   & 10000 & 0.02 \\
    Dataset & MHDis & 488   & 10000 & 0.05 \\
    \bottomrule
    \end{tabular}%
  \label{tab:distribution}%
\end{table}%

\newpage
\subsection{Full ML Results}\label{MLR}
\begin{table}[htbp]
	\centering
		\vspace{-1.5cm} 
	\caption{Window approach CV results. T is time}

	\resizebox{.8\textwidth}{!}{

	\begin{tabular}{ccllllllllrr}
		\cmidrule{2-12}    \multirow{81}[22]{*}{\begin{sideways}Window approach CV results.\end{sideways}} & \multicolumn{1}{l}{Attack} & ML    & Accuracy & BalancedAcc & Precision & Recall & F1-Score & kappa & ROC   & \multicolumn{1}{l}{Train-T} & \multicolumn{1}{l}{Test-T} \\
		\cmidrule{2-12}          & \multirow{8}[2]{*}{ACK} & DT    & 1.000±0.000 & 1.000±0.000 & 1.000±0.000 & 1.000±0.000 & 1.000±0.000 & 1.000±0.000 & 1.000±0.000 & 0.106 & 0.000 \\
		&       & KNN   & 1.000±0.000 & 1.000±0.000 & 1.000±0.000 & 1.000±0.000 & 1.000±0.000 & 0.999±0.000 & 1.000±0.000 & 0.130  & 0.128 \\
		&       & LR    & 1.000±0.000 & 1.000±0.000 & 1.000±0.000 & 1.000±0.000 & 1.000±0.000 & 1.000±0.000 & 1.000±0.000 & 0.437 & 0.000 \\
		&       & MLP   & 1.000±0.000 & 1.000±0.001 & 1.000±0.000 & 1.000±0.001 & 1.000±0.001 & 1.000±0.001 & 1.000±0.001 & 16.252 & 0.132 \\
		&       & NB    & 1.000±0.000 & 0.999±0.001 & 1.000±0.000 & 0.999±0.001 & 1.000±0.000 & 0.999±0.001 & 0.999±0.001 & 0.031 & 0.006 \\
		&       & RF    & 1.000±0.000 & 1.000±0.000 & 1.000±0.000 & 1.000±0.000 & 1.000±0.000 & 0.999±0.000 & 1.000±0.000 & 5.534 & 0.103 \\
		&       & SVM   & 0.948±0.005 & 0.631±0.011 & 0.974±0.002 & 0.631±0.011 & 0.694±0.014 & 0.398±0.027 & 0.631±0.011 & 16.365 & 1.069 \\
\end{tabular}}%
	\label{tab:wcv}%
\end{table}%
\pagenumbering{gobble}
\begin{table}[htbp]
	\centering
	\vspace{-1.5cm} 
	\caption{Window approach Session results. T is time}

	\resizebox{.8\textwidth}{!}{

	\begin{tabular}{ccllllllllrr}
		\cmidrule{2-12}    \multirow{81}[22]{*}{\begin{sideways}Window approach Session results.\end{sideways}} & \multicolumn{1}{l}{Attack} & ML    & Accuracy & BalancedAcc & Precision & Recall & F1-Score & kappa & ROC   & \multicolumn{1}{l}{Train-T} & \multicolumn{1}{l}{Test-T} \\
		\cmidrule{2-12}          & \multirow{8}[2]{*}{ACK} & DT    & 1.000±0.000 & 1.000±0.000 & 0.999±0.000 & 1.000±0.000 & 0.999±0.000 & 0.999±0.001 & 1.000±0.000 & 0.021 & 0.002 \\
		&       & KNN   & 1.000±0.000 & 1.000±0.000 & 1.000±0.000 & 1.000±0.000 & 1.000±0.000 & 0.999±0.000 & 1.000±0.000 & 0.032 & 0.190 \\
		&       & LR    & 1.000±0.000 & 1.000±0.000 & 0.999±0.000 & 1.000±0.000 & 1.000±0.000 & 0.999±0.000 & 1.000±0.000 & 0.098 & 0.003 \\
		&       & MLP   & 1.000±0.000 & 1.000±0.000 & 0.999±0.000 & 1.000±0.000 & 0.999±0.000 & 0.999±0.000 & 1.000±0.000 & 26.160 & 0.814 \\
		&       & NB    & 1.000±0.000 & 1.000±0.000 & 1.000±0.000 & 1.000±0.000 & 1.000±0.000 & 0.999±0.000 & 1.000±0.000 & 0.003 & 0.012 \\
		&       & RF    & 0.999±0.000 & 1.000±0.000 & 0.999±0.000 & 1.000±0.000 & 0.999±0.000 & 0.998±0.000 & 1.000±0.000 & 1.113 & 0.169 \\
		&       & SVM   & 0.306±0.000 & 0.501±0.000 & 0.653±0.000 & 0.501±0.000 & 0.235±0.000 & 0.001±0.000 & 0.501±0.000 & 29.912 & 6.743 \\
\end{tabular}}%
	\label{tab:ws}%
\end{table}%
\pagenumbering{gobble}
\begin{table}[htbp]
	\centering
		\vspace{-1.5cm} 
	\caption{Window approach Test results. T is time}

	\resizebox{.8\textwidth}{!}{

	\begin{tabular}{ccllllllllrr}
		\cmidrule{2-12}    \multirow{81}[22]{*}{\begin{sideways}Window approach Test results.\end{sideways}} & \multicolumn{1}{l}{Attack} & ML    & Accuracy & BalancedAcc & Precision & Recall & F1-Score & kappa & ROC   & \multicolumn{1}{l}{Train-T} & \multicolumn{1}{l}{Test-T} \\
		\cmidrule{2-12}          & \multirow{8}[2]{*}{ACK} & DT    & 0.920±0.229 & 0.945±0.156 & 0.900±0.269 & 0.945±0.156 & 0.912±0.246 & 0.882±0.310 & 0.945±0.156 & 0.018 & 0.006 \\
		&       & KNN   & 0.269±0.000 & 0.500±0.000 & 0.134±0.000 & 0.500±0.000 & 0.212±0.000 & 0.000±0.000 & 0.500±0.000 & 0.029 & 1.555 \\
		&       & LR    & 0.989±0.000 & 0.992±0.000 & 0.980±0.000 & 0.992±0.000 & 0.986±0.000 & 0.972±0.000 & 0.992±0.000 & 0.084 & 0.001 \\
		&       & MLP   & 0.500±0.000 & 0.500±0.000 & 0.250±0.000 & 0.500±0.000 & 0.333±0.000 & 0.000±0.000 & 0.500±0.000 & 20.468 & 0.613 \\
		&       & NB    & 0.269±0.000 & 0.500±0.000 & 0.134±0.000 & 0.500±0.000 & 0.212±0.000 & 0.000±0.000 & 0.500±0.000 & 0.006 & 0.011 \\
		&       & RF    & 0.991±0.000 & 0.994±0.000 & 0.984±0.000 & 0.994±0.000 & 0.989±0.000 & 0.978±0.000 & 0.994±0.000 & 0.941 & 0.296 \\
		&       & SVM   & 0.500±0.000 & 0.500±0.000 & 0.250±0.000 & 0.500±0.000 & 0.333±0.000 & 0.000±0.000 & 0.500±0.000 & 30.356 & 6.798 \\
\end{tabular}}%
	\label{tab:wt}%
\end{table}%

\pagenumbering{gobble}
\begin{table}[htbp]
	\centering
		\vspace{-1.5cm} 
	\caption{Flow approach CV results. T is time}

	\resizebox{\textwidth}{!}{

	\begin{tabular}{ccllllllllrr}
		\cmidrule{2-12}    \multirow{81}[22]{*}{\begin{sideways}Flow approach CV results.\end{sideways}} & \multicolumn{1}{l}{Attack} & ML    & Accuracy & BalancedAcc & Precision & Recall & F1-Score & kappa & ROC   & \multicolumn{1}{l}{Train-T} & \multicolumn{1}{l}{Test-T} \\
		\cmidrule{2-12}          & \multirow{8}[2]{*}{ACK} & DT    & 0.894±0.015 & 0.910±0.013 & 0.897±0.015 & 0.910±0.013 & 0.893±0.016 & 0.789±0.030 & 0.910±0.013 & 0.000     & 0.000 \\
		&       & KNN   & 0.588±0.020 & 0.498±0.002 & 0.294±0.010 & 0.498±0.002 & 0.370±0.008 & 0.004±0.004 & 0.498±0.002 & 0.002 & 0.010 \\
		&       & LR    & 0.896±0.017 & 0.911±0.015 & 0.898±0.015 & 0.911±0.015 & 0.894±0.017 & 0.791±0.033 & 0.911±0.015 & 0.043 & 0.002 \\
		&       & MLP   & 0.896±0.012 & 0.912±0.009 & 0.898±0.012 & 0.912±0.009 & 0.895±0.012 & 0.792±0.024 & 0.912±0.009 & 1.808 & 0.003 \\
		&       & NB    & 0.799±0.028 & 0.830±0.021 & 0.836±0.020 & 0.830±0.021 & 0.799±0.028 & 0.613±0.050 & 0.830±0.021 & 0.001 & 0.002 \\
		&       & RF    & 0.894±0.011 & 0.910±0.007 & 0.897±0.014 & 0.910±0.007 & 0.893±0.011 & 0.788±0.022 & 0.910±0.007 & 0.330  & 0.025 \\
		&       & SVM   & 0.895±0.019 & 0.911±0.017 & 0.898±0.016 & 0.911±0.017 & 0.894±0.019 & 0.790±0.037 & 0.911±0.017 & 0.031 & 0.019 \\
\end{tabular}}%
	\label{tab:fcv}%
\end{table}%

\begin{table}[htbp]
	\centering
		\vspace{-1.5cm} 
	\caption{Flow approach Session results. T is time}

	\resizebox{\textwidth}{!}{

	\begin{tabular}{ccllllllllrr}
		\cmidrule{2-12}    \multirow{81}[22]{*}{\begin{sideways}Flow approach Session results.\end{sideways}} & \multicolumn{1}{l}{Attack} & ML    & Accuracy & BalancedAcc & Precision & Recall & F1-Score & kappa & ROC   & \multicolumn{1}{l}{Train-T} & \multicolumn{1}{l}{Test-T} \\
		\cmidrule{2-12}          & \multirow{8}[2]{*}{ACK} & DT    & 0.997±0.000 & 0.939±0.001 & 0.995±0.001 & 0.939±0.001 & 0.966±0.001 & 0.931±0.002 & 0.939±0.001 & 0.001 & 0.000 \\
		&       & KNN   & 0.022±0.000 & 0.500±0.000 & 0.511±0.000 & 0.500±0.000 & 0.022±0.000 & 0.000±0.000 & 0.500±0.000 & 0.002 & 0.295 \\
		&       & LR    & 0.997±0.000 & 0.940±0.000 & 0.995±0.000 & 0.940±0.000 & 0.966±0.000 & 0.931±0.000 & 0.940±0.000 & 0.060  & 0.001 \\
		&       & MLP   & 0.997±0.000 & 0.940±0.000 & 0.995±0.000 & 0.940±0.000 & 0.966±0.000 & 0.931±0.000 & 0.940±0.000 & 2.264 & 0.014 \\
		&       & NB    & 0.992±0.000 & 0.816±0.000 & 0.996±0.000 & 0.816±0.000 & 0.886±0.000 & 0.771±0.000 & 0.816±0.000 & 0.001 & 0.004 \\
		&       & RF    & 0.997±0.000 & 0.940±0.000 & 0.995±0.000 & 0.940±0.000 & 0.966±0.000 & 0.931±0.000 & 0.940±0.000 & 0.340 & 0.119 \\
		&       & SVM   & 0.997±0.000 & 0.940±0.000 & 0.995±0.000 & 0.940±0.000 & 0.966±0.000 & 0.931±0.000 & 0.940±0.000 & 0.035 & 0.407 \\
\end{tabular}}%
	\label{tab:fs}%
\end{table}%

\begin{table}[htbp]
	\centering
		\vspace{-1.5cm} 
	\caption{Flow approach Test results. T is time}

	\resizebox{\textwidth}{!}{

 \begin{tabular}{ccllllllllrr}
\cmidrule{2-12}    \multirow{81}[22]{*}{\begin{sideways}Flow  approach Test results.\end{sideways}} & \multicolumn{1}{l}{Attack} & ML    & Accuracy & BalancedAcc & Precision & Recall & F1-Score & kappa & ROC   & \multicolumn{1}{l}{Train-T} & \multicolumn{1}{l}{Test-T} \\
\cmidrule{2-12}          & \multirow{8}[2]{*}{ACK} & DT    & 0.116±0.000 & 0.443±0.000 & 0.375±0.000 & 0.443±0.000 & 0.113±0.000 & -0.024±0.00 & 0.443±0.000 & 0.000  & 0.000 \\
          &       & KNN   & 0.102±0.000 & 0.500±0.000 & 0.051±0.000 & 0.500±0.000 & 0.093±0.000 & 0.000±0.000 & 0.500±0.000 & 0.003 & 0.021 \\
          &       & LR    & 0.116±0.000 & 0.443±0.000 & 0.375±0.000 & 0.443±0.000 & 0.113±0.000 & -0.024±0.00 & 0.443±0.000 & 0.044 & 0.000 \\
          &       & MLP   & 0.116±0.000 & 0.443±0.000 & 0.375±0.000 & 0.443±0.000 & 0.113±0.000 & -0.024±0.00 & 0.443±0.000 & 2.851 & 0.002 \\
          &       & NB    & 0.118±0.000 & 0.444±0.000 & 0.383±0.000 & 0.444±0.000 & 0.115±0.000 & -0.024±0.00 & 0.444±0.000 & 0.001 & 0.001 \\
          &       & RF    & 0.116±0.000 & 0.443±0.000 & 0.375±0.000 & 0.443±0.000 & 0.113±0.000 & -0.024±0.00 & 0.443±0.000 & 0.357 & 0.036 \\
          &       & SVM   & 0.116±0.000 & 0.443±0.000 & 0.375±0.000 & 0.443±0.000 & 0.113±0.000 & -0.024±0.00 & 0.443±0.000 & 0.024 & 0.022 \\
\end{tabular}}%
	\label{tab:ft}%
\end{table}%

\end{document}